\documentclass[epj,nopacs]{svjour}
\usepackage{epsfig,multicol,multirow}
\usepackage{amssymb}
\usepackage{hep,hyperref}
\usepackage{graphics}
\usepackage{amsmath,latexsym}
\newcommand{\jump}{\vspace{0.3cm}}
\usepackage{color}

\newcommand{\hzero}{\ensuremath{\PHiggslightzero}} 
\newcommand{\Hzero}{\ensuremath{\PHiggsheavyzero}} 
\newcommand{\Azero}{\ensuremath{\PHiggspszero}} 
\newcommand{\Hpm}{\ensuremath{\PHiggspm}} 
\newcommand{\CP}{\ensuremath{\mathcal{C}\mathcal{P}}}
\newcommand{\lag}{\ensuremath{\mathcal{L}}}

\newcommand{\retildehat}{\ensuremath{\hat{\Sigma}}}

\newcommand{\self}{\ensuremath{\Sigma}}
\newcommand{\sw}{\ensuremath{s_{\PW}}}
\newcommand{\cw}{\ensuremath{c_{\PW}}}
\newcommand{\swd}{\ensuremath{s_{\PW}^2}}
\newcommand{\cwd}{\ensuremath{c_{\PW}^2}}

\newcommand{\mhzero}{\ensuremath{M_{\hzero}}}
\newcommand{\mHzero}{\ensuremath{M_{\Hzero}}}
\newcommand{\mAzero}{\ensuremath{M_{\Azero}}}
\newcommand{\mHp}{\ensuremath{M_{\PHiggs^{\pm}}}}
\newcommand{\aew}{\ensuremath{\alpha_{\rm ew}}}

\newcommand{\thdmcalc}{Eriksson:2009ws}
\newcommand{\citeFH}{Frank:2006yh}
\newcommand{\hollikrev}{Langacker:1996qb,Hollik:1988ii,Hollik:2003cj,Hollik:2006hd}
\newcommand{\hollikold}{Hollik:1986gg,Hollik:1987fg}

\begin{document}
\title{$\Delta\,r$ in the Two-Higgs-Doublet Model at full one loop level -- and beyond}
\author{David L\'opez-Val \inst{1} \and Joan Sol\`a \inst{2}
}                     
\institute{Institut f\"ur Theoretische Physik, Universit\"at Heidelberg \\
Philosophenweg 16, D-69120 Heidelberg, Germany. \\
\email{lopez@thphys.uni-heidelberg.de} \and
 Dept. Estructura i
Constituents de la Mat\`eria, Universitat de Barcelona, and Institut
de Ci\`encies del Cosmos\\ Av. Diagonal 647, E-08028
    Barcelona, Catalonia, Spain. \\ \email{sola@ecm.ub.edu}}
\date{Received: date / Revised version: date}
%
\abstract{ After the recent discovery of a Higgs-like boson particle at
the CERN LHC-collider, it becomes more necessary than ever to prepare
ourselves for identifying its standard or non-standard nature. The
fundamental parameter $\Delta r$, relating the values of the electroweak
gauge boson masses and the Fermi constant, is the traditional observable
encoding high precision information of the quantum effects. In this work
we present a complete quantitative study of $\Delta r$ in the framework of
the general Two-Higgs-Doublet Model (2HDM). While the one-loop analysis of
$\Delta r$ in this model was carried out long ago, in the first part of
our work we consistently incorporate the higher order effects that have
been computed since then for the SM part of $\Delta r$. Within the
on-shell scheme, we find typical corrections leading to shifts of $\sim
20-40 \,\MeV$ on the $\PW$ mass, resulting in a better agreement with its
experimentally measured value and in a degree no less significant than in
the MSSM case. In the second part of our study we devise a set of
effective couplings that capture the dominant higher order genuine 2HDM
quantum effects on the $\delta\rho$ part of $\Delta r$ in the limit of
large Higgs boson self-interactions. This limit constitutes a telltale
property of the general 2HDM which is unmatched by e.g. the MSSM.}
%
\maketitle

\section{Introduction}

After the recent announcement of the discovery of a Higgs-like boson
candidate at the CERN
LHC-collider\,\cite{incandela12,gianotti12,cmshiggstwo,atlashiggstwo}, we
might be closer than ever at unveiling the ultimate architecture of the
Electroweak Symmetry Breaking (EWSB) mechanism. Since the idea of
spontaneous symmetry breaking (SSB) was incorporated in the structure of
the Standard Model (SM), it remained a most pressing and unsettled
conundrum in our understanding of High Energy Physics. The ideas pioneered
by Higgs\,\cite{Higgs:1964ia,Higgs:1964pj}, Englert and
Brout\,\cite{Englert:1964et}, and Guralnik and
Kibble\cite{Guralnik:1964eu} crystallized  into the present-day paradigm,
in which we assume the existence of one fundamental scalar $SU_L(2)$
doublet whose particle remnant after SSB of the gauge symmetry corresponds
to the physical Higgs boson of the SM. The quest for an experimental
confirmation of this picture has hitherto ranked very high in the wishing
list of the experimental efforts first conducted at LEP, and later on at
the Tevatron and the LHC. The tantalizing Higgs-event candidates reported
by the ATLAS and CMS collaborations during the last year have apparently
been confirmed by the recent analysis of the new
data~\cite{incandela12,gianotti12,cmshiggstwo,atlashiggstwo}, which have
led them to conclude (at a $\sim 5\sigma$ confidence level) that a new
particle has been discovered carrying most of the ingredients to be
expected from a Higgs-like boson with a mass close to $125-126 \,\GeV$.
The high evidence has been gathered by both the ATLAS and CMS
collaborations from the detected signals in the diphoton
($\gamma\gamma$) and weak gauge boson ($\PZ^0\PZ^{0*}$ and $\PW\PW^*$)
decay modes\,\cite{incandela12,gianotti12,cmshiggstwo,atlashiggstwo}. If
confirmed by subsequent searches and final identification, this could be
the greatest achievement of Particle Physics since the discovery of the
top quark 17 years ago at Fermilab. And in that case it would represent
the most impressive and significant success of Particle Physics ever,
since it would constitute the confirmation of the physical reality of the
SSB, i.e. of the most subtle and far reaching quantum field theoretical
(QFT) structure of the SM. Not surprisingly it would at the same time
raise many other problems, even outside the strict domain of Particle
Physics, such as for example in cosmology through the certified existence
of the huge electroweak vacuum energy. However, we understand that
obtaining a consistent overall picture of our world may still take quite
some more time.

\smallskip{}
In the meanwhile, even after that phenomenal discovery, it is difficult to
ascertain the nature of the found Higgs-like particle, whether it is the
SM Higgs boson or a member of an extended Higgs sector whose remaining
constituents are yet to be found, among other possibilities. There is no reason
\emph{a priori} for scalar particles not to appear within multiple
families, as fermions or gauge bosons do. Both theoretically and
phenomenologically there are many motivations for Higgs physics beyond the
SM. For example, the Vacuum Expectation Value (VEV) of the Higgs field, $v
\simeq 246 \,\GeV$, is well known to be unstable under radiative
corrections as ultraviolet (UV) contributions to this VEV must be
fine-tuned to insure its stability at low energies.
This naturalness puzzle has constituted a primary driving force for the
study of extensions of the Standard Model (SM), and of the Higgs sector in
particular. As a simple, yet very attractive example of such an extension,
we find the Two-Higgs-Doublet Model (2HDM) \cite{Hunter,Djouadi:2005gi}.
Here, by introducing a second $SU_L(2)$ scalar doublet $\Phi_2$ we meet a
compelling phenomenological profile \cite{Hunter}-\cite{Branco:2011iw}.
On a theoretical basis, there is indeed strong support for a multi-Higgs
doublet structure. To begin with, a generic 2HDM furnishes a suitable
low-energy description of multifarious EWSB models, such as the Minimal
Supersymmetric Standard Model (MSSM) \cite{Ferrara87}, Composite Higgs
\cite{composite} and Little Higgs models
\cite{little}. Despite its simplicity, it may
accommodate a plethora of new mechanisms either for spontaneous or
explicit \CP-violation as well as a rich vacuum structure
\cite{Gunion:2005ja}. This sets the ground for novel
scenarios in which to address a wide variety of unsolved riddles in very
different areas, from neutrino mass generation \cite{Ma:2006km} to
Electroweak Baryogenesis \cite{baryogenesis}. In
particular, we cannot exclude more exotic possibilities in the framework
of Grand Unified Theories, which could also adapt to the properties of the
purportedly found Higgs boson\,\cite{Stech:2012zr}.

The simplicity and nevertheless very rich potentiality of the 2HDM
qualifies as an excellent starting point for a broad class of
model-building studies. For instance, if one posits additional discrete
symmetries in the scalar potential, the generic 2HDM structure becomes
further restricted and may lead to the so-called \text{Inert Doublet
Models} (IDM) \cite{Deshpande:1977rw}, in which at least one of the scalar
degrees of freedom becomes a stable WIMP, and thus provides a natural
candidate for Dark Matter. These models have been portrayed at length in
the literature, both from the perspective of collider observables (cf.
e.g. \cite{inert-collider}) and
astrophysics
\cite{inert-astro}.
A 2HDM structure can also realize a \textit{Higgs portal} scenario
\cite{Schabinger:2005ei}. Here, one entertains the possibility of an
additional Higgs field in a hidden sector, with no couplings to the SM
particles -- with the exception of the standard Higgs boson. The Higgs
self-interactions
 thus constitute the only link (the ``portal'') between the visible and the hidden domains, with
a dramatic impact on the expected Higgs boson widths -- and so on their
foreseeable collider signatures
\cite{portal}. Further extensions of
the minimal 2HDM include vectorphobic \cite{Cervero:2012cx} and
scale-invariant formulations \cite{Lee:2012jn}; or combinations with
additional gauge bosons \cite{Panotopoulos:2011xb}, fermionic generations
\cite{BarShalom:2011zj} or heavy neutrinos \cite{Arhrib:2011uy}, among
others.

Numerous studies have scrutinized the prospects for pinning down evidences
of 2HDM physics at the LHC, see. e.g. Refs.\cite{2hdm-collider}. In this
vein, the potential Higgs-like candidates identified by ATLAS and CMS have
already endorsed respective analyses on the corresponding implications for
the model \cite{2hdm-current,Altmannshofer:2012ar,Pich13}. This task is certainly not easy, as long as
the studies are restricted to direct collider observables. But the
situation can improve when counting also on the information conveyed by EW
precision analyses. In this respect it is fair to say that the LHC will
soon be able to broaden its capabilities from direct discovery to
precision physics measurements. This is of foremost importance since it
may provide virtual access to possible new degrees of freedom coupled to
the SM, and hence to new -- or modified -- interaction patterns which can
induce departures of the EW precision observables from the pure SM
expectations\,\footnote{Cf. for instance
\cite{Hollik:1988ii,Hollik:2003cj,Hollik:2006hd,Langacker:1996qb,Wells:2005vk}
for a comprehensive exposition of the method.}.

A most important example of precision observable sensitive to virtual
effects from new physics is the mass of the charged EW gauge boson
[$M_{\PW^{\pm}}$ ]. The current uncertainty in its theoretical
determination within the SM is estimated to be  $\Delta\,M_{\PW}^{\rm
th}\,\simeq\, 4\,\MeV$\,\cite{LEPEWWG}. On the other hand the present
world-average of the experimental measurements renders the value
$M_{\PW}^{\rm{exp}} = 80.385 \pm 0.015$~\GeV\,\cite{pdg}, thus carrying an
accuracy at a remarkable level  $\lesssim 0.02\,\%$, i.e. better than 2
parts in ten thousand. But not less remarkable is the fact that the
experimental error still gives room for non-negligible non-SM
contributions. Indeed, for a SM Higgs boson mass of $M_{\PHiggs} = 125
\,\GeV$ (cf. the forthcoming discussions in Section~\ref{sec:analysis}),
the corresponding prediction for the mass of the $\PW$-boson renders
$M_{\PW}^{\rm SM} = 80.363 \,\GeV$. Despite the current discrepancy
$|M_{\PW}^{\rm SM}-M_{\PW}^{\rm{exp}}| \simeq 20 \,\MeV$ lies to within
one-sigma level of the experimental measurement, it is as big as five
times the estimated theoretical error in the SM. Since the latter falls
within the reach of the most accurate planned measurements of the
$\PW$-mass in the future there is little doubt that a deviation from the
SM of that sort should eventually be accessible to observation. This is
strongly hinted by the expectation that with the upcoming LHC data the
uncertainty on the W-boson mass can be narrowed down to $\Delta\,M^{\rm
exp}_{\PW} \simeq 10\,\MeV$ \cite{Bozzi:2011ww}, while a high-luminosity
linear collider running in a low-energy mode at the $\PW^+\PW^-$ threshold
should be able to reduce it even further, namely at the
$\Delta\,M_{\PW}^{\rm exp} \simeq 6$~MeV level or less. The profound
impact that a high precision measurement of $M_{\PW}$ as well as its
correlation to the $\PZ$-mass, $M_Z$, could have both as a precision test
of the SM and as a probe of new physics should not be underestimated. This
is not new of course, what is new is the fact that we are now much closer
than ever to exploit this feature at the LHC. {Let us recall} that the
$M_W-M_Z$ correlation is usually parameterized in terms of the quantity
$\Delta r$, defined as $G_F/\sqrt{2}=(g^2/8\,M_W^2)(1+\Delta r)$ where
$G_F$ is Fermi's constant and $g$ is the weak $SU(2)$ gauge coupling-- see
Sec. \ref{Sec:Deltarandallthat} for a more detailed definition in the
on-shell scheme. It suffices to say here that $g^2$ in this scheme is to
be replaced by $e^2/\sw^2\equiv4\pi\alpha/\left(1-M_W^2/M_Z^2\right)$,
where $\alpha$ is the e.m. fine structure constant.

The history of $\Delta r$ and its companion parameter $\delta\rho$ (i.e.
that part of $\Delta r$ which parametrizes the breaking of custodial
symmetry) is extensive and already quite
old\,\cite{\hollikrev,Wells:2005vk}; it has at present more than thirty
years. Let us recall that $\Delta r$ was first computed in the SM context
in 1980 by A. Sirlin and coworkers\,\cite{Sirlin:1980nh,Marciano:1980pb},
whereas the $\delta\rho$ parameter was defined earlier by Veltmann and
collaborators\,\cite{Ross:1975fq,Veltman:1976rt,Veltman:1977kh,Einhorn:1981cy}.
Since then the calculations of these parameters in the SM became improved
over the last three decades by important QCD and electroweak higher order
effects, hence establishing a powerful relation which allows to perform
accurate predictions of the W-boson mass in high-precision tests of the
standard model. {Not only so}, the calculations were soon extended to
physics beyond the SM, mainly from Supersymmetry (SUSY). In this regard,
an exhaustive coverage of $\Delta r$ and $\delta\rho$ within the MSSM is
available in the literature.  The first preliminary calculations
(including only the so-called oblique contributions) within the MSSM were
presented some twenty years ago in \cite{Barbieri:1989dc}, and then
shortly afterwards at full one-loop level in \cite{Garcia:1993sb} and
\cite {Chankowski:1993eu} -- see also \cite{Gosdzinsky:1990ga},
\cite{Gosdzinsky:1990sk}\,\footnote{Let us point out that to the best of
our knowledge the oldest full one-loop MSSM calculation of the electroweak
gauge boson masses existing in the literature was provided quite earlier
in references \cite{Grifols:1983gu} and \cite{Grifols:1984xs}. Although it
was presented in a renormalization framework slightly different from the
usual one, it was later adapted to the standard on-shell scheme and
this resulted in the first full one-loop MSSM calculation of $\Delta r$
reported in the literature\,\cite{Garcia:1993sb}, followed shortly after
by the similar analysis of \cite {Chankowski:1993eu}. }. Subsequently a
lot of refinements which include higher order effects have been performed
up to the present days, see the comprehensive studies
\cite{Heinemeyer:2006px}, including dedicated work e.g. on two-loop
effects \cite{Freitas:2002ve} or flavor violation
\cite{Heinemeyer:2004by}.

In spite of the generous literature on the $\Delta r$ parameter in various
contexts of physics beyond the SM -- mainly focused on the MSSM, as we
have just seen -- it is a bit surprising the scarce attention that has
been paid to this topic from the viewpoint of the general 2HDM, except for
some works presented long
ago\,\cite{Frere:1982ma,Bertolini:1985ia,\hollikold,Froggatt:1991qw,He:2001tp}.
In these old papers the one-loop calculation of $\Delta r$ was
first presented. However, for a modern numerical prediction of the
$W$-mass within the 2HDM at a level comparable to the SM, a consistent
estimate of the higher order effects in the 2HDM is necessary. Filling
such gap in the literature is our main task herewith. To that aim we
provide a fully-fledged updated analysis of the 2HDM contributions to the
parameter $\Delta\,r$  by consistently including the known higher order
effects from the SM as a part of the full 2HDM contribution in the
currently allowed region of the parameter space. From here we obtain
a theoretical determination of $M_{\PW}$ in the 2HDM at a level comparable
to the SM, taking $G_F$, $\alpha$ and $M_Z$ as experimental inputs. In
addition to that, we extend the previous analyses in an attempt to
estimate the maximum impact on $\delta\rho$ from the genuine 2HDM higher
order corrections associated to the Higgs boson self-interactions,
namely in the limit where these self-interactions become very large
(bordering the perturbative unitarity bounds). It is known that these
scenarios can strongly modify the Higgs/gauge boson couplings, owing to
the enhanced quantum effects driven by the Higgs self-interactions.
Interestingly, multi-Higgs doublet structures with strongly coupled Higgs
particles are well motivated from the theory side, in particular in view
of strongly-interacting realizations of the EWSB. Our main focus in this
second part of the paper will be to determine whether such augmented Higgs
self-couplings may be able to stamp any sensible fingerprint on the EW
precision observables under scope.

\bigskip
The paper is organized as follows. In Section~\ref{sec:theory} we
succinctly review the 2HDM setup and relevant constraints; we also
consider a preview of the EW precision observables to be examined in
detail thereafter, and set our definitions and notation.
Section~\ref{sec:analysis} is devoted to present the results of our
detailed analysis of the predictions for $\Delta r$ and $M_{\PW}$ at
full one loop level in the 2HDM in which the known higher order SM
effects are consistently incorporated. In Section~\ref{sec:higher}
we explore the maximum size of the genuine 2HDM contributions beyond
one-loop level by introducing a set of effective couplings
or form factors for the Higgs-gauge boson interactions that enable
us to estimate the leading higher order effects on $\delta\rho$ in
the large Higgs self-coupling limit. Conclusions and closing
remarks are finally delivered in Section~\ref{sec:conclusions}.
Additional analytical details of the calculation are quoted in the
Appendix.

\section{Theoretical setup}
\label{sec:theory}

\subsection{The general Two-Higgs-Doublet-Model in a nutshell}

The 2HDM \cite{Hunter} canonically extends the SM Higgs sector with a
second $SU_L(2)$ doublet of weak hypercharge $Y=+1$, so that it contains
$4$ complex scalar fields. The most general form of a gauge invariant,
renormalizable, \CP-conserving potential that one may construct out of two
doublets $\Phi_i\, (i=1,2)$  can be cast as follows
%
\begin{eqnarray}
 V(\Phi)&=&\lambda_1 \left( \Phi_1^\dagger \Phi_1 - \tfrac{v_1^2}2 \right)^2 +
         \lambda_2 \left( \Phi_2^\dagger \Phi_2 - \tfrac{v_2^2}2 \right)^2 \nonumber \\
& &+  \lambda_3 \left( \Phi_1^\dagger \Phi_1 - \tfrac{v_1^2}2 +
                          \Phi_2^\dagger \Phi_2 - \tfrac{v_2^2}2 \right)^2 +
 \nonumber \\ &+&
         \lambda_4 \left[ (\Phi_1^\dagger \Phi_1) (\Phi_2^\dagger \Phi_2) -
                          (\Phi_1^\dagger \Phi_2) (\Phi_2^\dagger \Phi_1) \right] \nonumber \\
&& +
         \lambda_5 \left[ \Re e(\Phi_1^\dagger \Phi_2) - \tfrac{v_1 v_2}2  \right]^2 +
         \lambda_6 \left[ \Im m(\Phi_1^\dagger \Phi_2) \right]^2
 \label{eq:potential}
\end{eqnarray}

\noindent where the self-couplings $\lambda_i$ may be rewritten in terms
of the masses of the physical Higgs particles ($M_{h^0}$, $M_{H^0}$,
$M_{A^0}$, $M_{H^\pm}$); $\tan \beta=v_2/v_1$ (the ratio of the two VEV's
$\langle \phi_i^0\rangle$ giving masses to the up- and down-like quarks);
the mixing angle $\alpha$ between the two $\CP$-even states; and,  last
but not least, the self-coupling $\lambda_5$, which cannot be absorbed in
the previous quantities. Therefore we end up with a $7$-parameter set:
$(M_{h^0}$, $M_{H^0}$, $M_{A^0}$, $M_{H^{\pm}}$, $\sin\alpha$,
$\tan\beta$, $\lambda_5)$. An additional discrete symmetry $\Phi_i \to
(-1)^i\, \Phi_i (i = 1,2)$ -- which is exact up to soft-breaking terms of
dimension 2  -- is canonically assumed as a warrant of Flavor-Changing
Neutral-Current (FCNC) suppression\footnote{Let us note in passing that
such a symmetry is automatically preserved in the MSSM.}. Alternative
constructions with no explicit $Z_2$ symmetry are described e.g. in
\cite{Mahmoudi:2009zx} and references therein.

As for the Yukawa sector involving the Higgs/quark interactions, the
absence of tree-level flavor changing neutral currents (FCNC) leads to two
main canonical realizations: 1) type-I 2HDM, in which just one Higgs
doublet couples to all quarks, whereas the other doublet does not; 2)
type-II 2HDM, where one doublet couples only to down-like quarks and the
other doublet just to up-like quarks. Other flavor structures are also
conceivable and have indeed attracted a growing attention in the recent
years \cite{2hdm-flavor,Altmannshofer:2012ar}. That said, we will
see that the evaluation of $\Delta r$ is barely influenced by the
particular form of the Yukawa couplings after we impose the various
phenomenological and theoretical restrictions. For a more comprehensive
exposition within our notation, including a detailed list of Higgs boson
couplings to fermions and bosons, and a discussion of the on-shell
renormalization of the unconstrained 2HDM Higgs sector, see
Ref.\,\cite{LopezVal:2009qy}.

It is also worth recalling that the Higgs sector of the MSSM corresponds
to a particular (constrained) realization of the general two-Higgs doublet
structure. The underlying SUSY invariance restricts the form of the
potential \eqref{eq:potential} in a way that has far-reaching
phenomenological implications. While in the MSSM the Higgs
self-interactions, and thereby also the Higgs mass spectrum, are dictated
by the EW gauge couplings, in the 2HDM we no longer have such a dynamical
restriction \emph{a priori}. This implies that the triple, as well as the
quartic, 2HDM Higgs self-interactions are fundamentally unconstrained --
the influence of the latter being comparatively milder. In practice, they
can be boosted as much as permitted by perturbative unitarity and vacuum
stability, giving rise to trademark signatures which are completely
foreign to the MSSM \footnote{In contrast, the core of the enhancement
capabilities of the MSSM Lagrangian resides in the richer pattern of
Yukawa-like couplings between the Higgs bosons and the quarks, as well as
between quarks, squarks and charginos/neutralinos. Their implications for
collider and EW precision physics have been object of dedicated attention
in the past for a plethora of varied processes, see
e.g.\cite{mssm-pheno}.
For reviews on the subject, see e.g.
\cite{Carena:2002es,Heinemeyer:2008tw,Hunter,Djouadi:2005gj}.}.

\jump A rather extensive set of bounds restricts the regions of the 2HDM
parameter space with potential significance to Phenomenology. Dedicated
accounts on these topics can be found e.g. in
Refs.\cite{elkaffas, 2hdm-bounds,Mahmoudi:2009zx,superiso},
and in particular also in Section II of Ref.\cite{LopezVal:2009qy}, which
very closely follows the notation and conventions employed herewith.
It goes without saying that the recent identification of a $\sim$ 125 GeV
SM-like Higgs boson candidate by ATLAS and CMS raises a number of
very significant implications, which we must take into
account for realistic studies. A comprehensive updated
analysis of the 2HDM parameter space constraints in the light of these novel results is not yet available.
Nevertheless, we can rely on model-independent approaches \cite{Azatov:2012wq}
that spell out the general conditions to be satisfied by
phenomenologically viable extensions of the standard Higgs sector.

\subsection{Phenomenological
restrictions}\label{subsec:phenorestrictions}

The basic restrictions to be imposed on the 2HDM parameter space can be
outlined as follows: i) first, we need to account for an upper limit on
the loop-induced breaking of the (approximate) $SU(2)$ custodial symmetry,
which one can precisely trade through $\delta\rho$, that is, one of the EW
precision quantities under scrutiny in this paper. Its numerical value
should lie below $|\delta\rho_{\rm 2HDM}|\lesssim {\cal O}{(10^{-3})}$
\cite{pdg}, if we allow up to $3\sigma$ deviations from the current
best-fit value. In practice, this translates into restrictions on the mass
splitting among the different Higgs bosons. Tight requirements ensue also
from the radiative $B$-meson decay $\mathcal{B}(b \to s\gamma)$ and the
$B_d^0 - \bar{B}_d^0$ mixing. Generically, the former process sets a lower
bound on the charged Higgs mass of $M_{H^{\pm}} \gtrsim 300$ GeV for $\tan
\beta \ge 1$ \cite{Mahmoudi:2009zx} in the case of type-II 2HDM, while the
latter strongly disfavors the $\tan\beta \lesssim 1$ regions (for both
type-I and type-II) and tends to enforce $\tan\beta$ to be roughly above
$1.5 -  2$ if the charged Higgs boson is kept relatively light (viz.
$M_{\PHiggs^{\pm}} \sim 100-150 \,\GeV$, which is allowed for type-I
models). Perturbative unitarity, as well as vacuum stability, impose as
well very severe limitations. These translate into wide excluded areas
across the $\tan\beta - \lambda_5$ plane. Unitarity places an upper limit
of $|\lambda_5| \sim \mathcal{O}(10)$ for Higgs boson masses of few
hundred GeV, and vacuum stability excludes the $\lambda_5 > 0$ region up
to a very narrow band\,\cite{LopezVal:2009qy}. In short, in order to
satisfy these restrictions we are confined to regions where
$\tan\beta\simeq 1$ and $|\lambda_5| \simeq 5-10\ (\lambda_5 < 0)$ {for
maximum Higgs self-coupling enhancements}. Additionally, any chosen Higgs
mass spectrum ought to satisfy all the current limits from direct searches
at LEP, Tevatron and LHC.

In practice, all these constraints are systematically included in our
calculation by combining the latest version of the public codes { \sc
2HDMCalc-1.1} \cite{\thdmcalc}, {\sc SuperISO-3.1} \cite{superiso} and
{\sc HiggsBounds-2.2} \cite{higgsbounds}, hand in hand with several
alternative and/or complementary in-house routines. Last but not least, we
must deal with the phenomenological implications of the $5\sigma$ Higgs
boson candidate recently unveiled at the
LHC\,\cite{incandela12,gianotti12,cmshiggstwo,atlashiggstwo}.
A most natural choice when embedding the current experimental picture into
a concrete realization of the 2HDM is to identify the lightest neutral
CP-even Higgs boson [$\hzero$] with the $\sim 125$ GeV resonance. Under
this assumption, the Higgs couplings to the gauge bosons become severely
constrained, as the current data show no substantial departure with
respect to the SM-like decay patterns. Consequently, we are left with very
tight restrictions on the trigonometric factors $\tan\beta$ and
$\sin\alpha$. It follows that we are essentially restrained to the
so-called \emph{decoupling} regime $\alpha = \beta-\pi/2$ (with
$\alpha<0$) or, equivalently, $\alpha = \beta+\pi/2$ (with $\alpha>0$) --
both cases featuring $g^2_{\hzero VV} \sim \sin^2(\beta-\alpha)\simeq
\,1$. If we instead identify the $\sim 125$ GeV resonance with the heavy
neutral \CP-even Higgs boson [$\Hzero$], the observed decay rates into
gauge bosons enforce $\alpha \simeq \beta$ since then the
corresponding coupling to gauge bosons yields $g^2_{\Hzero VV} \sim
\cos^2(\beta-\alpha)\simeq \,1$. Independently, these conditions also
disallow some specific choices of $\sin\alpha$. For instance, $\sin\alpha
\simeq 0$ cannot be realized within a type-II 2HDM. In the decoupling
limit, this choice would imply $\beta \simeq -\pi/2$, thus rendering an
unduly enhanced $\hzero b\bar{b}$ interaction,
$\sim|\sin\alpha/\cos\beta|\gg 1$, incompatible with the fermionic modes
of the current Higgs candidate observations. Conversely, for $\alpha =
\beta$ the choice $\alpha \simeq 0$ would spoil the perturbativity of the
Higgs/top Yukawa coupling. By similar arguments, one can prove that
$\alpha \simeq \pi/2$ is not permitted within type-II realizations of the
2HDM. Notice, however, that $\alpha=0$ (resp. $\alpha = \pi/2$), namely
the fermiophobic limit for $\Hzero$ (resp. $\hzero$) is still viable
within type-I models owing to the different Higgs/quark interaction
strengths $\sim \sin\alpha/\sin\beta$ (resp. $\sim \cos\alpha/\sin\beta$).

Further constraints can be imposed over $\tan\beta$. For example, we must
comply with the aforementioned B-physics limits on both the low
($\tan\beta \lesssim 1$) and high ($\tan\beta \gtrsim 10$)
$\tan\beta$-ranges. Also the non-observation of enhanced $h\to b\bar{b}$
decays rules out the large $\tan\beta$ regimes of a type-II 2HDM. In
addition, recent studies (cf. Ref.~\cite{Azatov:2012wq}) have concluded
that the fermionic decay signatures of the $\sim 125$ GeV Higgs-like
resonance, even if fully compatible with a SM-like Yukawa sector, exhibit
a mild statistical tilt towards slightly enhanced (resp. suppressed)
Higgs/top (resp. Higgs/bottom) couplings. These scenarios would be
realized for $\tan\beta \lesssim 1$. Let us recall, finally, that moderate
choices of $\tan\beta \sim \mathcal{O}(1)$ are particularly appealing and
well motivated for the purposes of our analysis, as they enable to
maximize the triple (3H) self-couplings through a relatively large value
of the parameter $|\lambda_5|$ -- therefore by resorting only to the
intrinsic structures of the 2HDM Higgs potential.

A variety of processes can probe the potentially enhanced 2HDM Higgs boson
self-interactions, and have indeed been intensively analysed over the past
years -- mostly in the context of linear colliders. Available studies
include, on the one hand, the tree-level production of triple Higgs-boson
final states \cite{Ferrera:2007sp}; the double
Higgs-strahlung channels $hh\PZ^0$ \cite{Arhrib:2008jp}; and the inclusive
Higgs-pair production via gauge-boson fusion \cite{Hodgkinson:2009uj}. In
the same vein, also the $\Pphoton\Pphoton$ mode of a linac has been
explored, in particular the loop-induced production of a single neutral
Higgs boson \cite{Bernal:2009rk} and of a
Higgs boson pair
\cite{Cornet:2008nq}. In all
the above mentioned cases, promising signatures were pinpointed, which
could be revealing of an unconstrained multi-Higgs doublet pattern.
Similar genuine 2HDM effects might also manifest as large radiative
corrections to a number of Higgs production channels. One loop studies of
pairwise Higgs boson final states were first addressed for charged Higgs
bosons $\APelectron\Pelectron \to \PHiggs^{+}\PHiggs^{-}$
\cite{Arhrib:1998gr} and later on carried to completion by a
full-fledged study of the neutral Higgs sector $\APelectron\Pelectron \to
\hzero\Azero, \Hzero\Azero$, including also the more traditional
Higgs-strahlung events $\APelectron\Pelectron \to \hzero\,\PZ^0,
\Hzero\,\PZ^0$
\cite{LopezVal:2009qy,LopezVal:2010yf}.
All these studies reveal the possible existence of:  i) sizable Higgs
boson production rates, typically in the ballpark of $\mathcal{O}(10-100)$
fb for $\sqrt{s} = 500$ GeV at a future linear collider; ii) sizable quantum effects, up to $\delta
\sim \pm 50\%$; and iii) a very characteristic complementarity of the
dominant Higgs production modes at different center-of-mass energies --
these properties being correlated, once more, to significant Higgs boson
self-interactions, and so to a non-supersymmetric multi-Higgs doublet
structure.

\subsection{Electroweak Precision Quantities from muon decay}\label{Sec:Deltarandallthat}

The relation between the EW gauge boson masses ($M_{\PW}$, $M_Z$) in terms
of the Fermi constant ($G_F$) and the fine structure constant ($\alpha$)
is an essential tool for testing the quantum effects within the SM as well
as to place bounds on its manifold conceivable extensions. Such relation
can be derived in terms of the muon lifetime  $\tau_{\mu}$, whose decay
rate is precisely defined by the Fermi constant, $G_F$, via the expression
\cite{\hollikrev}
\begin{eqnarray}
\tau_{\mu}^{-1} = \frac{G_F^2 \, m_\mu^5}{192 \pi^3} \;
F\left(\frac{m_{\mathrm{e}}^2}{m_\mu^2}\right)
\left(1 + \frac{3}{5} \frac{m_\mu^2}{M_{\PW}^2} \right)
\left(1 + \Delta_{\rm QED} \right) ,
\label{eq:fermi}
\end{eqnarray}
\noindent where $F(x)=1-8x-12x^2\ln x+8x^3-x^4$. Following the standard
conventions in the literature, the above defining equation for $G_F$
includes the expression $\Delta_{\rm QED}$, i.e., the finite QED
contribution obtained within the Fermi Model, which is known up to
two-loop order. Calculating the muon lifetime within the SM at the quantum
level and comparing with ~\eqref{eq:fermi} yields the relation:
\begin{eqnarray}
M_{\PW}^2 \left(1 - \frac{M_{\PW}^2}{M_Z^2}\right)=
\frac{\pi \alpha}{\sqrt{2} G_F} \left(1 + \Delta r\right)\,,
\label{eq:deltar_def1}
\end{eqnarray}
in which
\begin{equation}\label{eq:Deltar}
\Delta r\equiv\frac{\hat\Sigma_{\PW}(0)}{M_W^2}+\Delta\,r^{[{\rm vert, box}]}\,.
\end{equation}
These expressions define the quantity $\Delta r$ in a precise way. Let us
notice that $\hat\Sigma_{\PW}(k^2)$ is the on-shell renormalized
self-energy of the $\PW$-boson; it accounts for the universal
(``oblique'') part of the electroweak radiative corrections to the muon
decay. The non-universal (i.e. process-dependent) corrections -- which
stem from the vertex and box contributions to the muon decay -- are
encoded in the subleading term $\Delta\,r^{[{\rm vert, box}]}$. The
explicit expression for $\Delta r$ consists of a combination of loop
diagrams and counter terms. For the renormalization details, see
e.g.\,\,\cite{\hollikrev}. Here we will only remind the reader of some
basic facts which can be helpful to contextualize our 2HDM computation. To
start with, let us write down the explicit structure of $\Delta r$ after
renormalization:

\begin{eqnarray}
\Delta\,r &=& \Pi_{\gamma}(0)- \frac{\cwd}{\swd}\,\left(\frac{\delta\,M_Z^2}{M_Z^2}
- \frac{\delta\,M^2_{\PW}}{M_{\PW}^2}\right) + \frac{\self_{\PW}(0)-\delta\,M_{\PW}^2}{M_{\PW}^2} \nonumber \\
&& + 2\,\frac{\cw}{\sw}\,\frac{\Sigma_{\gamma\PZ}(0)}{M_Z^2}+\Delta\,r^{[{\rm vert, box}]}\,.\nonumber\\
\label{eq:deltar_def2}
\end{eqnarray}
%
%
%
We stress that this expression is finite because its original definition
\eqref{eq:Deltar} depends only on the on-shell renormalized self-energy of
the $\PW$-boson  and the remainder $\Delta\,r^{[{\rm vert, box}]}$ --
which is also finite by virtue of the Ward-Takahashi identities of the EW
theory. In the previous expression
$\Pi_{\gamma}(k^2)\equiv\partial\Sigma_{\gamma}(k^2)/\partial k^2$ is the
photon vacuum polarization, and we are using the notation $\swd \equiv
1-M_{\PW}^2/M_{\PZ}^2$, and $\cwd \equiv M_{\PW}^2/M_{\PZ}^2$.
Furthermore, since the renormalization is in the on-shell scheme the weak
gauge boson mass counter terms read: $\delta\,M^2_V = \Re
e\Sigma_{V}(M_V^2)$, with  $V =\PW^{\pm}, \PZ^0$. By $\Sigma_{V}(q^2)
\equiv \Sigma^T_{V}(q^2)$ we will hereafter denote the (transverse parts
of the) unrenormalized gauge boson self-energies, which we conventionally
extract from the corresponding vacuum polarization tensor:
\begin{eqnarray}
\Pi^{\mu\nu}_{V}(q^2) = g^{\mu\nu}\,\Sigma^T_V(q^2) + q^\mu\,q^\nu\,\Sigma^L_V\,(q^2)
\label{eq:polarization_tensor}.
\end{eqnarray}
After introducing the renormalized photon vacuum polarization
$\hat{\Pi}_{\gamma}(k^2)={\Pi}_{\gamma}(k^2)-{\Pi}_{\gamma}(0)$ it is
convenient to consider its fermionic part at $k^2=M_Z^2$, i.e.
$\hat{\Pi}^{\rm ferm}_{\gamma}(M_Z^2)$. As it is well-known, for the light
fermions (quarks and leptons, excluding the top quark) the quantity
\begin{equation}\label{eq:Deltaalpha}
\Delta\alpha= - \Re e \,\hat{\Pi}^{\rm ferm}_{\gamma}(M_Z^2)
\end{equation}
is independent of the EW part of the SM and goes into a finite
renormalization of the QED fine structure constant: $\alpha\to\alpha
(1+\Delta\alpha)$. This can be resummed according to the renormalization
group to provide the value of $\alpha$ at the scale of the $Z$-mass:
$\alpha(M_Z^2)=\alpha/(1-\Delta\alpha)\simeq 1/128$ which is $\sim 6\%$
larger than its low energy (Thomson limit) value $\alpha\simeq 1/137$.
Such resummation takes into account all the leading logarithms of the type
$\alpha^n\,\ln^n{(M_Z/m_l)}$ which enter the renormalization of $\alpha$
from the leptonic sector $l=e,\mu,\tau$. The light quark contribution,
instead, is computed more precisely via a dispersion relation from the
experimental hadronic data collected in low energy $e^{+}e^{-}$
scattering. Finally, the top quark gives a negligible (decoupling-like,
$\sim M_Z^2/m_t^2$) contribution to \eqref{eq:Deltaalpha}.

Let us now emphasize one more type of effect which will have special
relevance for our analysis of the 2HDM contributions to $\Delta r$. We are
referring to the so-called $\delta\rho$
parameter\,\cite{Ross:1975fq,Veltman:1976rt,Veltman:1977kh,Einhorn:1981cy}.
By inspecting \eqref{eq:deltar_def2}, such contribution comes from the
term
\begin{equation}\label{eq:deltarho}
\frac{\delta\,M_{\PZ}^2}{M_{\PZ}^2} -
\frac{\delta\,M^2_{\PW}}{M_{\PW}^2}\to
\frac{\Sigma_{\PZ}(0)}{M_{\PZ}^2} -
\frac{\Sigma_{\PW}(0)}{M_{\PW}^2}\equiv\delta\rho \, .
\end{equation}
The $\delta\rho$ effect is finite for the matter fermions of the SM (for
each doublet separately) and provides a very important (non-decoupling)
contribution when large mass splitting are present within a given fermion
family. Its main source comes of course from the top quark, or to be more
precise from the top quark and bottom quark doublet. Even though the
bottom quark contributes a negligible finite correction, it is essential
to make the quark doublet contribution to $\delta\rho$ perfectly finite on
its own. The corresponding impact on $\Delta r$ is not just
$-(\cwd/\swd)\,\delta\rho^t$ as it actually contains additional terms
(encapsulated in the so-called ``remainder'' $\Delta r_{\rm rem}$, cf.
Eq.~\eqref{eq:DeltarGeneral} and the discussion further down) which are
numerically significant:
\begin{eqnarray}\label{eq:DeltarTop}
\Delta r^{\rm top}&=&-\frac{\sqrt{2}G_FM_W^2}{16\pi^2}\left[3\frac{\cwd}{\swd}\,\frac{m_t^2}{M_W^2}+2\,
\left(\frac{\cwd}{\swd}-\frac13\right)\ln\frac{m_t^2}{M_W^2} \right. \nonumber \\
&& \left. +\frac43\ln\cwd+\frac{\cwd}{\swd}-\frac79\right]\,.
\end{eqnarray}
For a more physical interpretation, let us remind the reader that
$\delta\rho$ stands for the possible deviations of the value of the Fermi
constant in neutral current processes ($G_F^{NC}$) -- typically induced by
neutrino interactions -- from the corresponding Fermi constant in charged
processes ($G_F$) such as muon decay. Both in the SM and in the 2HDM we
have
\begin{equation}\label{eq:GNCGF}
\rho\equiv \frac{G_F^{NC}}{G_F}=\frac{M_W^2}{M_Z^2\,\cwd}=1+\delta\rho\,.
\end{equation}
In the absence of weak hypercharge interaction $g'\to 0$ ($M_{\PZ}\to
M_{\PW}$, $\swd\to 0$) we would have $\delta\rho=0$ and the Fermi
constants in both kind of processes would have equal strength.  In the SM,
as well as in the 2HDM, the only sources of $\delta\rho$ come from quantum
effects. These deviations are bound to satisfy $|\delta\rho|\lesssim {\cal
O}{(10^{-3})}$\,\cite{pdg}. But this bound still gives a substantial
margin for physics beyond the SM, as we shall see.

After clarifying the physical significance of the main terms in
\eqref{eq:deltar_def2} we may now write down the general structure of
$\Delta r$ in the following traditional form\cite{\hollikrev}:
\begin{equation}\label{eq:DeltarGeneral}
\Delta r=\Delta\alpha-\frac{\cwd}{\swd}\,\delta\rho+\Delta r_{\rm rem} = \Delta\alpha + \Delta r^{[\delta\rho]}+\Delta r_{\rm rem}\,,
\end{equation}
where the leading contributions $\Delta\alpha$ and $\delta\rho$ have been
defined above, and we have introduced $\Delta r^{[\delta\rho]} \equiv
-(\cwd/\swd)\delta\rho$. The so-called ``remainder'' piece $\Delta r_{\rm
rem}$ collects the remaining effects, which in the SM entail subleading
(which should \emph{not} be taken as synonymous of negligible)
contributions \footnote{Beyond the one-loop order, resummations of the
leading one-loop contributions have been derived, see e.g.
\cite{Consoli:1989fg}, and further contrasted to exact higher-order
calculations \cite{Freitas:2002ja,Freitas:2002ve}.}. For example, while
$\Delta\alpha\simeq 0.06$, we have $\Delta r_{\rm rem}\simeq 0.01$. For
comparison, the top quark gives a contribution to \eqref{eq:DeltarGeneral}
of around $\Delta r^{[\delta\rho] {\rm (top)}}\simeq -0.03$, or $-0.04$ if
using the more accurate expression \eqref{eq:DeltarTop}. Although $\Delta
r_{\rm rem}$ is smaller than the leading terms, all these contributions
are in fact quite significant and produce an important numerical shift of
the $\PW$-mass, lowering its zeroth order value ($M_W^{\rm tree}\simeq
80.9$ GeV) by about $0.8\%$, i.e. more than $0.6$ GeV. In particular,
$\Delta r_{\rm rem}$ renders a non-negligible contribution of $\sim 160$
MeV to that total (see below). The additional effects that might come from
physics beyond the SM are generally much smaller but the current high
precision electroweak physics has improved sufficiently so as to make
possible to detect shifts of $\Delta r$ at the level of less than one per
mil. At the end of the next subsection we shall further illustrate this
point.

At present, the calculation of $\Delta r$ in the SM is complete
up to two loops
\cite{Freitas:2002ja,Freitas:2002ve,Awramik:2002wn,Awramik:2003rn,Onishchenko:2002ve}
and includes also the leading three and four-loop pieces
\cite{vanderBij:2000cg}.
As we have mentioned, the bulk of the contributions to $\Delta r$ is
linked to the renormalization of the fine structure constant. The
next-to-leading source of contributions comes from $\delta\rho$, which in
the SM is finite and dominated by the aforementioned $\mathcal{O}(m_t^2)$
terms from the top-quark loops.  In the SM, however, the Higgs
contribution to $\delta\rho$ is neither gauge invariant nor UV-finite,
only the sum with the remaining bosonic part is finite and gauge
independent. Numerically it is not very relevant as compared to the
fermionic contribution to $\Delta r$ and it increases only logarithmically
with the Higgs boson mass. This feature reflects the so-called screening
behavior of the SM Higgs boson\,\cite{Veltman:1976rt}, a property which
does not generally hold in extended Higgs sectors, such as e.g. in the
unconstrained 2HDM. As this issue is important for our considerations, let
us quote explicitly the leading effect of the SM Higgs boson in the limit
of large $M_H$. The contribution being not finite, it also depends on the
choice of the dimensional regularization scale $\mu$ used in the 't
Hooft-Veltman procedure (see the Appendix for more details). The splitting of
terms in the full bosonic contribution is therefore somewhat arbitrary,
but it is natural to set that scale at the EW value $\mu=M_W$. Finally,
omitting the UV-divergent pieces that cancel with the remaining bosonic
terms one arrives at
\begin{equation}\label{eq:deltarhoHSM}
\delta\rho^{H}\simeq -\frac{3\sqrt{2}\,G_F\,M_{\PW}^2}{16\,\pi^2}\,\frac{\swd}{\cwd}\,\left\{\ln\frac{M_H^2}{M_W^2}-\frac56\right\}+...
\end{equation}
and
\begin{equation}\label{eq:DeltarHSM}
\Delta r^{H}\simeq \frac{\sqrt{2}\,G_F\,M_{\PW}^2}{16\,\pi^2}\,\frac{11}{3}\,\left\{\ln\frac{M_H^2}{M_W^2}-\frac56\right\}+...
\end{equation}
As we can see the dominant term $-(\cwd/\swd)\delta\rho^H$  is corrected
by non-negligible additional finite parts from $\Delta r_{\rm rem}$ having
the same structure. The neat contribution of Eq.~(\ref{eq:DeltarHSM}) is
subsumed again in $\Delta r_{\rm rem}$. Let us indeed note that
numerically $\Delta r^{H}$ is rather irrelevant as compared to, say, the
top quark contribution and the overall $\Delta r$ value ($\simeq 0.04$)
within the SM; we find $\Delta r^{H}={\cal O}(10^{-3})$ for $M_H=200$ GeV
and ${\cal O}(10^{-4})$ for $M_H=125$ GeV.
In spite of this meager yield, we
wish to stress that the particular $\delta\rho^H$ piece in
Eq.~(\ref{eq:deltarhoHSM}) is formally very important because it measures
the departure from custodial symmetry, which is that global $SU(2)$
symmetry of the Higgs SM Lagrangian which is only broken by the weak
hypercharge interaction $g'=g\,s_W/c_W$. The ``custodial symmetry limit''
thus corresponds to $g'\to 0$ ($M_{\PZ}\to M_{\PW}$, $\swd\to 0$). In this
limit the gauge bosons $W^{\pm}$ and $Z$ form a degenerate triplet of a
global $SU(2)$ symmetry \cite{Einhorn:1981cy}. We
may indeed confirm from the above expressions that the $\delta\rho^{H}$
effect takes on the form $\delta\rho^{H}\simeq
(-3g'^2/16\pi^2)\,\ln{M_H^2/M_W^2}$ and hence it vanishes in the custodial
symmetry limit $g'^2\equiv g^2\swd/\cwd\to 0$, as expected.

A natural question to ask is if there are significant (non-screening)
custodial-breaking effects from Higgs physics beyond the SM, i.e. effects
not just growing logarithmically with the Higgs boson masses but as powers
of the masses themselves. Let us briefly mention the Higgs sector of the
MSSM. Although it features a type II 2HDM, it is of a constrained nature
owing to the underlying SUSY\,\cite{Hunter,Djouadi:2005gj}. As a result
there are no conspicuous non-screening effects and in this sense the
situation does not differ significantly from the SM.  There are
notwithstanding alternative contributions to $\delta\rho$ in the MSSM
which come from the stop/sbottom-mediated quantum effects and are
potentially very relevant
\cite{Barbieri:1983wy,Barbieri:1989dc,Gosdzinsky:1990ga,Garcia:1993sb,Chankowski:1994tn}.
Finally, in the general (unconstrained) 2HDM case the situation changes
dramatically. The corresponding $\Delta r$ contributions have
been studied at one-loop \cite{Frere:1982ma,Bertolini:1985ia,\hollikold,
Froggatt:1991qw,He:2001tp,Hunter,grimus} and have been employed in studies of
combined 2HDM parameter space constraints \cite{elkaffas}. However,
to the best of our knowledge no systematic study that includes a
complete account of $\Delta r$ and $M_{\PW}$ at a level comparable to the
SM is available in the literature. We believe that in the light of the
present experimental situation at the LHC it is highly desirable to cover
this gap.

\subsection{$\Delta r$, $\delta\rho$ and $M_{\PW}$ in the 2HDM. Preliminary considerations}
\label{subsec:Deltardeltarho}

It is important to emphasize that $\Delta r$ is sensitive (through quantum
effects) to all the SM parameters (couplings and masses) as well as to the
full list of parameters involved in any possible extension of the SM. In the
relevant 2HDM case under consideration,
\begin{equation}\label{eq:DeltarbeyondSM}
\Delta r=\Delta r(e,M_{\PW},M_{\PZ},m_f; M_i, \sin\alpha, \tan\beta, \lambda_5)\,,
\end{equation}
where $m_f$ and $M_i=M_{h^0}, M_{H^0}, M_{A^0}, M_{H^\pm}$ are
respectively the masses of the fermions and of the Higgs bosons. Finally,
$\lambda_5$ is the parameter of the Higgs potential \eqref{eq:potential}
which cannot be related to a physical mass of the 2HDM. That parameter
enters $\Delta r$ beyond one loop only, but we will see that it can still
furnish significant quantum effects. In general these effects will enter
Eq.~\eqref{eq:DeltarGeneral} through $\delta\rho$ and $\Delta r_{\rm
rem}$.
Let us clarify that the contributions to $\Delta r$
from the charged Higgs bosons come only from $\delta\rho$ and
$\Delta r_{rem}$ in Eq.\,(2.11), as the charged Higgs boson loop
effects on $\Delta\alpha$ leave no finite yield to the photon
self-energy nor to the photon-Z mixing, as expected from gauge
invariance.

In this work we revisit the calculation of $\delta\rho$ in the general
2HDM and shall consider also the full one-loop calculation of $\Delta r$,
i.e. the complete quantity \eqref{eq:deltar_def2} or
\eqref{eq:DeltarGeneral}. As advertised, we will subsequently include
higher order effects on $\delta\rho$ related with the self-couplings of
the Higgs bosons, mainly governed by the $\lambda_5$ parameter. These
effects are non-existent in the MSSM owing to the pure gauge nature of the
self-couplings in the tree-level MSSM Higgs potential\,\cite{Hunter}.
\begin{figure*}
\begin{center}
\includegraphics[scale=1.0]{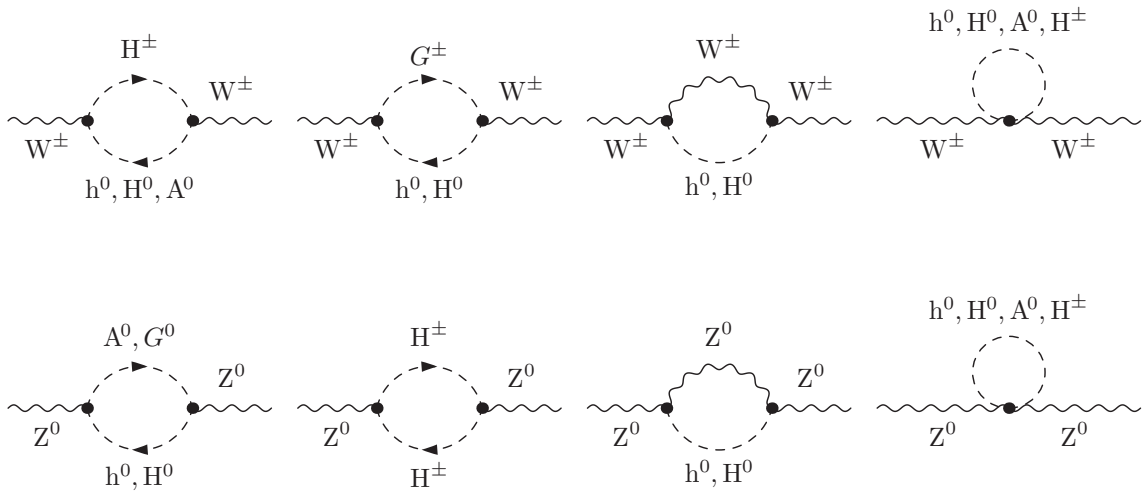}
\caption{Set of Feynman diagrams describing the pure 2HDM one-loop
contributions to the $\PW^{\pm}$ and $\PZ^0$ self-energies in the 't
Hooft-Feynman gauge. All Feynman diagrams in this paper have been generated
by means of {\sc FeynArts} \cite{feynarts}. \label{fig:self}}
\end{center}
\end{figure*}
The dominant part of the 2HDM corrections comes from the one-loop
contributions to $\delta\rho$ mediated by the various 2HDM Higgs
bosons\,\cite{Frere:1982ma,Bertolini:1985ia,\hollikold,
Froggatt:1991qw,He:2001tp}. As in the case of the  SM we perform the calculation in
the Feynman gauge (cf. Fig.\,\ref{fig:self}) and we call the result
$\delta\rho_{\rm 2HDM}$. It amounts to a compact and finite expression
whose full form is provided in the Appendix, see
Eq.\,\eqref{eq:deltarho_an}. Here we single out only the portion of
$\delta\rho_{\rm 2HDM}$ that depends on the Higgs boson mass splittings,
which may be called $\delta\rho_{{\rm 2HDM}}^{*}$. It can be cast as
follows:
\begin{eqnarray}\label{drho2HDM}
\delta\rho_{{\rm
2HDM}}^{*}&&=\frac{-G_F}{8\sqrt{2}\,\pi^2}\Big\{M_{H^{\pm}}^2
\left[1-\frac{M_{A^0}^2}{M_{H^{\pm}}^2-M_{{A^0}}^2}\,
\ln\frac{M_{H^{\pm}}^2}{M_{A^0}^2}\right]\nonumber\\
&&+\cos^2(\beta-\alpha)\,M_{h^0}^2\left[\frac{M_{A^0}^2}{M_{A^0}^2-M_{h^0}^2}\,
\ln\frac{M_{A^0}^2}{M_{h^0}^2} \right. \nonumber \\
&& \; -\left.\frac{M_{H^{\pm}}^2}{M_{H^{\pm}}^2-M_{h^0}^2}\,
\ln\frac{M_{H^{\pm}}^2}{M_{h^0}^2}\right]\nonumber\\
&&+\sin^2(\beta-\alpha)\,M_{H^0}^2\left[\frac{M_{A^0}^2}{M_{A^0}^2-M_{H^0}^2}\,
\ln\frac{M_{A^0}^2}{M_{H^0}^2} \right. \nonumber \\
&& \; - \left.\frac{M_{H^{\pm}}^2}{M_{H^{\pm}}^2-M_{H^0}^2}\,
\ln\frac{M_{H^{\pm}}^2}{M_{H^0}^2}\right]\Big\}\,. \nonumber \\
\end{eqnarray}
From this expression, and by comparison with Eq.\,\eqref{eq:deltarhoHSM},
it is clear that in the unconstrained 2HDM the contributions to
$\delta\rho$ do not follow the screening theorem of the SM Higgs boson.
Indeed, from the quadratic dependence on the various Higgs boson masses
outside the logarithms and the fact that arbitrary mass splittings between
these Higgs bosons are possible, one can expect that they could easily
overshoot the limits on $\delta\rho$ within the 2HDM, namely the $3\sigma$
bound $|\delta\rho|\lesssim {\cal O}{(10^{-3})}$ obtained from the EW
precision fits\,\cite{pdg}. However, we note that if $M_{\Azero}\to
M_{H^{\pm}}$ then $\delta\rho_{{\rm 2HDM}}^{*}\to 0$. Hence if the mass
splitting between $M_{A^0}$ and $M_{H^{\pm}}$ is not too large
$\delta\rho_{{\rm 2HDM}}^{*}$ can be kept under control. This is also true
for the full $\delta\rho_{{\rm 2HDM}}$, if at the same time we are not far
away from the decoupling limit $\beta-\alpha=\pi/2$ -- where the lightest
CP-even state  $h^0$ behaves SM-like (see the Appendix for more details).

Let us now provide some discussion on the general strategy we will
follow to estimate $\Delta r$ in the unconstrained 2HDM at one-loop
and beyond. More detailed considerations will be made in sections 3
and 4, together with the full quantitative analysis of course. Let
us assume that $\Delta r$ in the defining equation
\eqref{eq:deltar_def1} collects the expanded form of the various
contributions up to some order of perturbation theory where they are
presumably computed. We can split $\Delta r_{\rm 2HDM} $ as follows:
\begin{equation}
\Delta r_{\rm 2HDM} =\Delta r_{\rm 2HDM}^{[1]({\rm RG})} +
\Delta r_{\rm 2HDM}^{({\rm boson})}+\Delta r_{\rm 2HDM}^{({\rm Ykw-QCD})}\,.
\label{eq:deltar-def5}
\end{equation}
This structure assumes that all of the leading (and some
next-to-leading) effects on $\Delta r$ are involved within the 2HDM.
There are various subclasses of contributions in it that will
deserve some particular comment. At the moment we note that the
first term on the \textit{r.h.s.} of \eqref{eq:deltar-def5} is the
improved one-loop correction, namely the total one-loop correction
plus the renormalization group (RG) resummed effects of
$\Delta\alpha$, which go into the important renormalization of the
e.m. fine structure constant; the second term represents the
complete 2-loop bosonic part, i.e. the pure EW contribution from
gauge bosons, Goldstone bosons and Higgs bosons in all possible
combinations; finally, the third term stands for the joint
Yukawa-coupling and QCD effects to ${\cal O}(\aew\alpha_s)$,
including some resummed contributions. Needless to say all these
terms are rather complicated but many of them exhibit a structure
very similar to that of the known SM
case\,\cite{deltar-corrections}.
There is, however, a subset of higher order diagrams in the 2HDM
which provide a qualitatively (and maybe also a quantitatively important)
new contribution. The latter is buried in the
$\Delta r_{\rm 2HDM}^{({\rm boson})}$ piece of
(\ref{eq:deltar-def5}) and for this reason it proves convenient to
further split it as follows:
\begin{equation}
\Delta r_{\rm 2HDM}^{({\rm boson})} =\Delta r_{\rm 2HDM}^{({\rm boson *})}+ \Delta r_{\rm 2HDM}^{({\rm \lambda_5})}\,.
\label{eq:DeltaBoson}
\end{equation}
\begin{figure*}
\begin{center}
\includegraphics[scale=1.0]{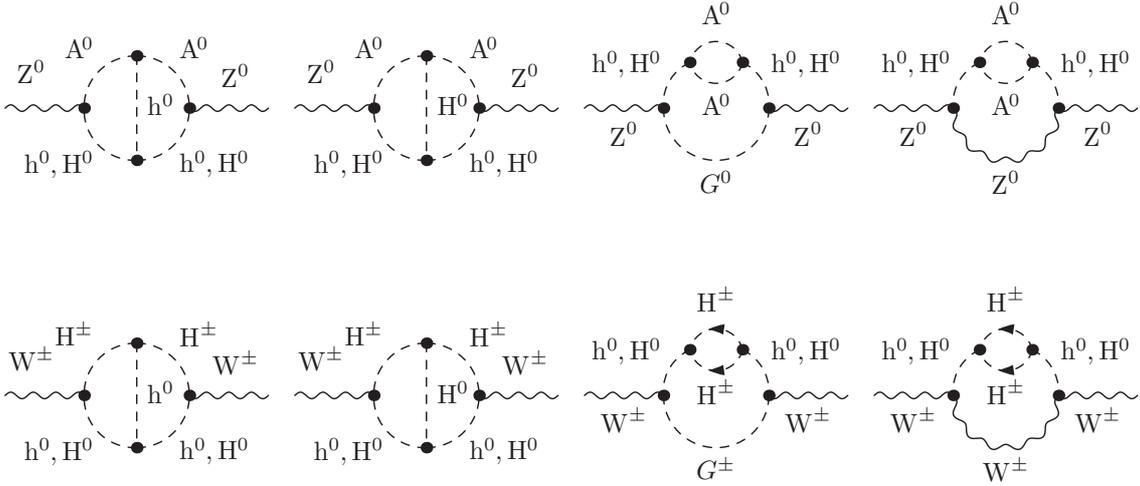}
\caption{Small representative sample of Feynman diagrams describing
potentially significant 2-loop contributions to $\Delta r$ within the 2HDM
in the 't Hooft-Feynman gauge. The complete list of 2-loop diagrams of
this class is much more extensive, and the full list of 2-loop graphs of
all classes within the 2HDM (involving bosons and/or fermions) is huge.
However, even with the displayed sample we can immediately appreciate the
presence of trilinear Higgs boson couplings and their foreseeable
significance. In the limit of large $\lambda_5$, all these diagrams are of
${\cal O}(\aew\lambda_5^2)$ and constitute the (gauge invariant and
finite) leading 2-loop contribution to $\Delta r$.} \label{fig:2self}
\end{center}
\end{figure*}
The special term here is the second one on the \textit{r.h.s.} of the
above expression, $\Delta r_{\rm 2HDM}^{({\rm \lambda_5})}$, whose
relevance will soon become apparent. We start by explaining the meaning of
the first term, $\Delta r_{\rm 2HDM}^{({\rm boson *})}$, which stands in
part for the ordinary gauge boson and Golstone boson diagrams up to two
loops. Among these diagrams we find the 2-loop pieces contributing to
${\cal O}(\aew^2)$, which are not particularly important from the
numerical point of view, although they are technically difficult to
account for. Furthermore, $\Delta r_{\rm 2HDM}^{({\rm boson *})}$ involves
as well the 2-loop diagrams containing Higgs bosons only, and also those
graphs mixing gauge bosons and Higgs bosons; the corresponding amplitudes
can be constructed from the one-loop diagrams of Fig.~\ref{fig:self} after
bridging them with another Higgs internal line (propagator). See
Fig.\,\ref{fig:2self} for a detailed sample of 2-loop diagrams constructed
in this way. The graphs in that figure are proportional to the product of
two trilinear Higgs boson self-couplings, $\lambda_{hhh}$, and two
ordinary gauge couplings, i.e. the corresponding amplitudes are of ${\cal
O}(\aew\lambda_{hhh}^2)$. There are other 2-loop graphs constructed from
the aforementioned bridging procedure that involve three gauge couplings
and one trilinear Higgs coupling, and hence are of ${\cal
O}(g^3\lambda_{hhh})$. We can expect that some of these amplitudes,
especially those involving two trilinear couplings, can be very important
from the quantitative point of view because they can be enhanced even
after preserving all known bounds on perturbative unitarity, custodial
symmetry and vacuum stability. For a full display of the detailed
structure of the trilinear $\lambda_{hhh}$ self-couplings in the general
2HDM, see e.g. Ref.\,\cite{LopezVal:2009qy} and particularly its Table II.
As an example we quote the coupling of three light \CP-even Higgs bosons:
\begin{eqnarray}\label{eq:extrilinear}
\lambda_{\hzero\hzero\hzero}&=&-\frac{3 i e}{2 M_W\sin2\beta\, \sw}
\left[M_{h^0}^2 (2 \cos(\alpha+\beta) \right. \nonumber \\ && + \left. \sin2\alpha\sin(\beta-\alpha))\right.\nonumber\\
&& \left. -\cos(\alpha+\beta) \cos^2(\beta-\alpha)\,\ \frac{4 \lambda_5
   M_W^2\, \sw^2}{e^2}\right]\,. \nonumber \\
\end{eqnarray}
These structures can be enhanced, in principle, on three accounts: a)
assuming large $\tan\beta\gg 1$ or small $\tan\beta\ll 1$, b) arranging
for large mass splittings among the Higgs bosons, and c) pushing the value
of the characteristic $\lambda_5$ coupling (the free parameter which is
unrelated to the Higgs masses) in the general 2HDM Higgs potential
(\ref{eq:potential}). Of these possibilities in practice only the last one
can be used efficiently after enforcing all the known theoretical and phenomenological
bounds described in Sec. \ref{subsec:phenorestrictions}.

Let us note in passing that the $\lambda_{\PHiggs\PHiggs\PHiggs}^2$
contributions are not present in the SM at two loops (with $\PHiggs$ standing
here for the SM Higgs boson) because the $\PZ$ gauge boson cannot couple
to $\PHiggs\PHiggs$ in the SM and hence a double $\PHiggs\PHiggs\PHiggs$
vertex at two loops cannot be formed by the aforesaid bridging procedure.
Only a single vertex $\PHiggs\PHiggs\PHiggs$ can appear at this order in
the SM, thus giving rise to ${\cal O}(g^3\lambda_{\PHiggs\PHiggs\PHiggs})$
effects from the Higgs self-interaction in combination with pure gauge
boson couplings. But even in this case $\lambda_{\PHiggs\PHiggs\PHiggs}$
has no special enhancing property in the SM beyond the artificial raising
of the Higgs boson mass, recall that $\lambda_{\PHiggs\PHiggs\PHiggs}\sim
g\,M_{\PHiggs}^2/M_{\PW}$. For a relatively light Higgs boson,
$M_H\gtrsim125$ GeV the trilinear Higgs boson self-coupling in the SM is
not very substantial. In particular, for a 125 GeV Higgs boson we get

\begin{equation}
 \lambda_{\PHiggs\PHiggs\PHiggs} (M_{\PHiggs} = 125\,\GeV) =
 \left. \frac{3e\,M^2_{\PHiggs}}{2\,M_{\PW}\,\swd} \right|_{M_{\PHiggs} = 125\,\GeV} \simeq 188\, \GeV
\label{eq:triplesm}.
\end{equation}

\noindent To certify the meager impact of this numerical result notice that it
yields roughly $2\%$ of the Lee-Quick-Thacker bound only
\cite{lee}, which as we know represents the upper bound
on the SM Higgs boson self-interaction ensuing from the perturbative
unitarity requirement. In view of the fact that the presumed value of
$M_H$ is not much higher than $M_W$, the mentioned effects can be finally
subsumed into the fairly irrelevant pure bosonic ${\cal O}(\aew^2)$ part
in the context of the SM\,\cite{Awramik:2002wn,Awramik:2003rn}.

Let us now finally address the precise origin of the  $\Delta r_{\rm
2HDM}^{({\rm \lambda_5})}$ piece in Eq.\,\eqref{eq:DeltaBoson}. Such
contribution appears when we select the $\lambda_5^2$ tag in the subclass
of all the 2-loop diagrams containing two trilinear Higgs boson couplings.
Recall that the $\lambda_5$ term is present in \emph{all} the Higgs boson
self-couplings $\lambda_{hhh}$. (At this important point we refer the
reader once more to Table II of Ref.\cite{LopezVal:2009qy}). In
particular, see the last term on the \textit{r.h.s.} of
Eq.\,(\ref{eq:extrilinear}). It was precisely in order to single out such
peculiar and potentially relevant 2-loop $\lambda_5^2$ proportional
effects in the 2HDM computation of $\Delta r$ that we have introduced the
last term of Eq.\,\eqref{eq:DeltaBoson}. In our approach we wish to
explicitly detach this term from the 2-loop bosonic contribution. To do
this in practice we split all the trilinear Higgs boson self couplings of
the 2HDM as follows
\begin{equation}\label{eq:splithhh}
\lambda_{hhh}=\lambda_{hhh}^{*}+c_h\lambda_5
\end{equation}
where $\lambda_{hhh}^{*}$ is the part of the coupling not containing
$\lambda_5$, and $c_h$ is a coefficient which can be easily identified
from the list of trilinear couplings in Table II of
\cite{LopezVal:2009qy}. Let us now emphasize that the part of the
amplitudes which is proportional to $\lambda_{hhh}^{*}$ is also
incorporated into $\Delta r_{\rm 2HDM}^{({\rm boson *})}$ in
Eq.\,(\ref{eq:DeltaBoson}) whereas the part which carries the
$\lambda_5^2$ coefficient constitutes our precise definition of the piece
$\Delta r_{\rm 2HDM}^{({\rm \lambda_5})}$ in (\ref{eq:DeltaBoson}). In
practice, we shall project the $\mathcal{O}(\lambda_5^2)$ component for
all of the Higgs boson self-couplings from the generic structure
\eqref{eq:splithhh}, meaning that, at the end of the day, $\Delta r_{\rm
2HDM}^{({\rm \lambda_5})}$ will be built upon $\mathcal{O}(\lambda^2_5)$
contributions; these are the leading (Higgs-mediated) effects beyond
1-loop in the limit of large Higgs boson self-interactions -- for more
details cf. our discussion in Section~\ref{sec:higher}.

It is important to note that the formal separation of these two kind of
contributions is well-defined, and it is worthwhile performing it because
the $\lambda_5$-part is potentially the most significant one from the
phenomenologically point of view. We note that even though both
$\lambda_{hhh}^{*}$ and $c_h$ could also be enhanced at large or small
$\tan\beta$, we do not contemplate this possibility in view of the
perturbative unitarity constraints, alongside with the phenomenological
preference for $\tan\beta \sim \mathcal{O}(1)$ values. As a result the
entire enhancing power of the $\lambda_{hhh}$ couplings resides
exclusively in the $\lambda_5$ part, whereas the rest of the 2-loop
bosonic yield is subsumed in $\Delta r_{\rm 2HDM}^{({\rm boson}*)}$,
altogether quite irrelevant in practice as in the SM
case\,\cite{Awramik:2002wn,Awramik:2003rn}.

The sum of all the 2-loop bosonic diagrams involved in the computation of
$\Delta r_{\rm 2HDM}^{({\rm boson})}$ in Eq.\,\eqref{eq:DeltaBoson} is
well defined, namely it is gauge invariant and finite.  Most important for
our purposes is that we have checked explicitly the gauge invariance and
finiteness of the overall piece $\Delta r_{\rm 2HDM}^{({\rm \lambda_5})}$
constructed from the above procedure (see section 4 for more details).
This feature could be expected because the $\lambda_5$ coupling does not
appear at one-loop in this calculation and hence that parameter is not
needed for the renormalization of the gauge boson sector at 2-loops. As a
result we can exploit this property and test the impact of the 2-loop
diagrams involving one or two trilinear Higgs boson couplings in the limit
of large $\lambda_5$. Of course ``large'' means, in this context, as large
as allowed by the perturbative unitarity constraints, which are well-known
in the
literature\,\cite{unitarity}
-- see e.g. \cite{LopezVal:2009qy} for a discussion in our framework --
and we certainly impose them as an important restriction in our
calculation.

From the above considerations the following strategy is suggested to
obtain an estimate of $\Delta r$ in the general 2HDM,
Eq.\,\eqref{eq:deltar-def5}. First, the pure bosonic higher order effects
independent of $\lambda_5$ are as tiny in the 2HDM as they are in the SM;
and, second, in the regime $\tan\beta={\cal O}(1)$ we should have in the
2HDM at most the same Yukawa couplings and QCD loop corrections as in the
SM\,\cite{Freitas:2002ve,Freitas:2002ja,Awramik:2002wn,Awramik:2003rn}. We express these
two statements in a nutshell as follows:
\begin{equation}\label{eq:estimates}
\Delta r_{\rm 2HDM}^{({\rm boson}*)}\simeq \Delta r_{\rm SM}^{({\rm boson})};
\ \ \ \Delta r_{\rm 2HDM}^{({\rm Ykw-QCD})}\simeq \Delta r_{\rm SM}^{({\rm Ykw-QCD})}\,,
\end{equation}
where the bosonic part $\Delta r_{\rm 2HDM}^{({\rm boson}*)}$ is of course
the same quantity that we defined in (\ref{eq:DeltaBoson}) through the
splitting (\ref{eq:splithhh}), and $\Delta r_{\rm SM}^{({\rm boson})}$ is
the known 2-loop bosonic SM result. A practical recipe for a reasonable
estimate of $\Delta r$ in the general 2HDM should therefore be the
following effective quantity:
\begin{equation}
\Delta r_{\rm 2HDM}^{\rm eff} \simeq\Delta r_{\rm SM} +
\delta(\Delta r_{\rm 2HDM}^{[1]})+\Delta r_{\rm 2HDM}^{({\rm
\lambda_5})}\,. \label{eq:deltar-def4}
\end{equation}
The first term, $\Delta r_{\rm SM}$, stands for the full set of SM
contributions known to date, hence with a structure completely similar to
(\ref{eq:deltar-def5}) but within the SM:
\begin{equation}
\Delta r_{\rm SM} =\Delta r_{\rm SM}^{[1]({\rm RG})} +
\Delta r_{\rm SM}^{({\rm boson})}+\Delta r_{\rm SM}^{({\rm Ykw-QCD})}\,.
\label{eq:deltar-SM}
\end{equation}
It includes different sorts of one loop and higher order effects of
various kinds\,\cite{Freitas:2002ve,Freitas:2002ja,Awramik:2002wn,Awramik:2003rn}, as
well as some resummed contributions -- the most important one being of
course the one affecting $\Delta\alpha$. The overall SM contribution
(\ref{eq:deltar-SM}) is usually accounted for quantitatively with the help
of a detailed numerical parameterization (see section 3 for details). The
second term on the \textit{r.h.s.} of (\ref{eq:deltar-def4}), i.e.
$\delta(\Delta r_{\rm 2HDM}^{[1]})$, denotes the one-loop shift to $\Delta
r$ driven by the ``genuine'' (viz. non-standard) 2HDM contributions to the
weak gauge boson self-energies. To determine this characteristic one-loop
effect from the 2HDM we compute the set of (Higgs-mediated,
$h=\hzero,\Hzero,\Azero,\Hpm$) Feynman diagrams displayed in
Fig.~\ref{fig:self}, whose result we indicate by $\Delta r_{\rm
2HDM}^{h[1]}$, and then we subtract from it the one-loop contribution from
the SM Higgs boson, $H$, denoted as $\Delta r_{\rm SM}^{H[1]}$, namely
\begin{equation}
\delta(\Delta r_{\rm
2HDM}^{[1]})=\Delta r_{\rm 2HDM}^{h[1]}-\Delta r_{\rm SM}^{H[1]}\,,
  \label{eq:deltar-def42}
\end{equation}
see more details in the next section and in the Appendix.

Notice that the one-loop SM Higgs contribution is included as part of the
first term on the \textit{r.h.s.} of (\ref{eq:deltar-def4}), as we have
mentioned, and it would be counted twice if it was not subtracted.
Therefore, upon subtracting \eqref{eq:deltar-def4} from
\eqref{eq:deltar-def5} and using equations (\ref{eq:DeltaBoson}) and (\ref
{eq:estimates}) and \eqref{eq:deltar-SM}, we find
\begin{eqnarray}\label{eq:diffeff}
\Delta r_{\rm 2HDM}-\Delta r_{\rm 2HDM}^{\rm eff}&\simeq&\Delta r_{\rm 2HDM}^{[1]({\rm RG})} \nonumber \\ && -\Delta r_{\rm SM}^{[1]({\rm RG})} -
\delta(\Delta r_{\rm 2HDM}^{[1]})=0\,.\nonumber \\
\end{eqnarray}
From now on we will identify $\Delta r_{\rm 2HDM}$ with $\Delta r_{\rm
2HDM}^{\rm eff}$ and will use Eq.\eqref{eq:deltar-def4} for the practical
evaluation of the 2HDM contribution.

To summarize, Eq.\,\eqref{eq:deltar-def4} encodes a reasonable estimate of
the basic contributions to $\Delta r$ in the 2HDM under the presently
known perturbative unitarity constraints. It actually provides an upper
bound to the maximum value that this parameter can reach in the general
2HDM after taking into account the leading one loop and higher order
effects. This is because on the one hand it includes the corresponding SM
value, and on the other it collects the two distinctive sources of
potentially relevant 2HDM effects: i) the genuine 1-loop Higgs boson
effects beyond the SM, i.e. $\delta(\Delta r_{\rm 2HDM}^{[1]})$ ; ii) and
the leading 2-loop ${\cal O}(\aew\lambda_5^2)$ effects triggered by the
enhanced Higgs boson self-couplings in the limit of large $\lambda_5$,
i.e. $\Delta r_{\rm 2HDM}^{({\rm \lambda_5})}$. The latter furnishes, at
large $\lambda_5$, the biggest source of enhancement at higher order
within the known perturbative unitarity limits.

Let us finish this section with a simple estimate of the numerical impact
on the $\PW$-mass which follows from a generic shift of the parameter
$\Delta r$, which we denote $\delta(\Delta r)$. In the case under
consideration that shift may essentially receive the following two genuine
2HDM contributions at one loop and beyond, which we have discussed above:
\begin{equation}\label{eq:deltaDeltar}
\delta(\Delta r)=\delta(\Delta r_{\rm 2HDM}^{[1]})+\Delta r_{\rm 2HDM}^{({\rm \lambda_5})}\,.
\end{equation}
From Eq.~\eqref{eq:deltar_def1} and taking the values for $M_{\PZ}$ and
$G_F$ as experimental inputs our evaluation of $\Delta r$ can be
translated into a theoretical prediction for the W-boson mass, $[M^{\rm
th}_{\PW}]$, and further confronted to $[M_{\PW}^{\rm exp}]$. For this we
need to solve the equation
\begin{eqnarray}
M_{\PW}^2 = \frac12\,M_Z^2\,\left[1 + \sqrt{1 - \frac{4\,\pi\alpha}{\sqrt{2}\,G_F\,M_Z^2}
\,[1 + \Delta\,r(M_{\PW}^2)]}\, \right]. \nonumber \\
\label{eq:mwpred}.
\end{eqnarray}
For $\Delta r=0$ one obtains the tree-level value $M_W^{\rm tree}\simeq
80.94$ GeV. But as mentioned above the full theoretical result is smaller
because quantum effects imply $\Delta r>0$. Mind that since $\Delta r$
itself depends on $M_{\PW}$, Eq.~\eqref{eq:mwpred} must be worked out
iteratively if one aims at a precise prediction. To within first order,
Eq.\,\eqref{eq:mwpred} implies that a shift $\delta(\Delta r)$ in $\Delta
r$ translates into a small shift in the $\PW$ mass given by
\begin{equation}\label{eq:shiftMW}
\delta M_{\PW}\simeq -\frac12\,M_{\PW}\,\frac{\swd}{\cwd-\swd}\,\delta(\Delta r)\,,
\end{equation}
with $\swd\simeq 0.22$. In the SM, as well as in all known promising
extensions of it, the quantum effects yield $\Delta r$ of order few
percent and positive, hence improving the agreement of the theoretical
prediction with experiment. For example, in the SM $\Delta r\simeq
0.04>0$. The physically measured value of $M_W$ ($M_W^{\rm exp}\simeq
80.38$ GeV) is in fact smaller than the tree-level value by about $0.8\%$,
i.e. some $\sim 600$ MeV smaller. This is consistent with
(\ref{eq:shiftMW}) since $\Delta r$ itself is small and so that equation
applies reasonably well to the entire $\Delta r$ value.

Furthermore, Eq.\,\eqref{eq:shiftMW} tells us that even a tiny departure
$|\delta(\Delta r)|\simeq 10^{-3}$ from physics beyond the SM induces a
shift of $|\delta M_{\PW}|\simeq 16$ MeV in the $\PW$ mass. This shift is
significant since it is very close to the present experimental error in
the $\PW$ mass ($M^{\rm exp}_{\PW} = 80.385 \pm 0.015\, \GeV$). It follows
that further improvement (lessening) of that error will enable us to
discriminate very subtle quantum effects, perhaps digging already into
physics beyond the SM. Interestingly enough we have indicated above that
the 2HDM can induce genuine shifts of this order from the two sources
indicated on the \textit{r.h.s.} of \eqref{eq:deltaDeltar}. In the next
sections we shall confirm by explicit numerical analysis that the maximum
size of the total shift $\delta(\Delta r)$ can be of order of a few times
$10^{-3}$ and displace the value of $M_W$ a few tens of MeV -- it
can reach as high as $\delta M_W\simeq 35-40$ MeV in the optimal cases
presented here. This potentially significant thrust from the genuine
quantum effects in the 2HDM sector can bring the theoretical prediction on
$M_W$ larger and hence offers a better agreement with the experiment than
the SM prediction, which persistently tends to stay too low as compared to
the experimental value.

%
%

\section{2HDM prediction for $\Delta\,r$ and $M_{\PW}$ at full one loop. Detailed analysis}
\label{sec:analysis}

Hereafter we carry out a dedicated numerical analysis of the
different EW precision quantities under scope. To start with, we
focus on the pure one-loop evaluation of $\Delta r$, and thereby of
$M_{\PW}^{\rm th}$, in the framework of the general 2HDM. Let us
emphasize from the beginning the meaning of our denomination ``full
one loop'' in the title of this section. It refers to the order of
perturbation theory at which we compute the genuine 2HDM effects on
the complete $\Delta r$ quantity, and not just to $\delta\rho$.
However, we incorporate in our calculation of $\Delta r$ (following
the procedure explained in Sec. \ref{subsec:Deltardeltarho}) the
known higher order effects within the SM because otherwise it would
be meaningless to compare with the experiment in view of the current
high precision of the measurements. It is only in a second stage
(cf. Sec. \ref{sec:higher}), where we perform a closer look to the
genuine higher order effects emerging from the 2HDM. Specifically we
will focus there on the parameter $\delta \rho$ within the 2HDM and
examine the Higgs boson self-interactions and their enhancement
capabilities therein. In doing this we will find appropriate to
resort  to a Born-improved Lagrangian approach for a first
reliable approximation to these higher order genuine 2HDM
contributions.

\begin{table*}
\begin{center}
\begin{tabular}{|l||c|c|c|c|} \hline
 & $M_{\PHiggslightzero}\,[\GeV]$ & $M_{\PHiggsheavyzero}\,[\GeV]$ & $M_{\PHiggspszero}\,[\GeV]$ & $M_{\PHiggs^\pm}\,[\GeV]$  \\ \hline
Set I & 125 & 135 & 200 & 220 \\
Set II & 115 & 125 & 200 & 220 \\
Set III & 125 & 200 & 200 & 215  \\
Set IV & 125 & 300 & 300 & 310 \\ \hline
Set V & 125 & 126 & 260 & 300   \\
Set VI & 125 & 126 & 210 & 250 \\
Set VII & 125 & 126 & 160 & 200  \\
Set VIII & 125 & 126 & 110 & 150 \\ \hline
Set V' ($M_{\PHiggs^{\pm}} < M_{\Azero}$) & 125 & 126 & 300 & 260  \\
Set VI' ($M_{\PHiggs^{\pm}} < M_{\Azero}$) & 125 & 126 & 250 & 210 \\
Set VII' ($M_{\PHiggs^{\pm}} < M_{\Azero}$) & 125 & 126 & 200 & 160  \\
Set VIII' ($M_{\PHiggs^{\pm}} < M_{\Azero}$) & 125 & 126 & 150 & 110
\\ \hline
\end{tabular}
\caption{Higgs boson mass sets employed throughout our computation. Owing
to the phenomenological restrictions from $B$-physics observables (cf.
Sec.~\ref{subsec:phenorestrictions}), Sets IV and V can be realized both
for type I and type II 2HDM's, while the other scenarios would mostly be
suitable for a type-I 2HDM setup only. Sets III and IV have been specially
devised to reproduce the characteristic mass splittings of the MSSM Higgs
bosons (see the text for details). Sets V-VIII feature two neutral,
$\CP$-even states with masses around $125$ GeV, and a constant splitting
[$|M_{\Azero} - M_{\PHiggs^{\pm}}|$ = 40 GeV] between their \CP-odd
[$\Azero$] and charged [$\PHiggs^{\pm}$] companions. The two possible mass
hierarchies, which we refer to as \textit{direct} [$M_{\PHiggs^{\pm}} >
M_{\Azero}$] and \emph{inverted} [$M_{\PHiggs^{\pm}} < M_{\Azero}$] (sets
marked with a prime) are analysed separately. \label{tab:masses}}
\end{center}
\end{table*}

The unrenormalized gauge boson self-energies $\self_{V}^{\rm 2HDM}$, as
defined in Eq.~\eqref{eq:polarization_tensor}, reunite all the information
we need to compute the 2HDM contributions to $\Delta r$ at one-loop: in
our notational setup (cf. Eqs.~\eqref{eq:deltar-def4} and
~\eqref{eq:deltar-def42}) the full one-loop payoff is denoted
\begin{equation}\label{eq:Deltar2HDM1loop}
\Delta r^{\rm [1]}_{\rm 2HDM} =
\Delta r_{\rm SM}^{\rm [1]} + \delta(\Delta\, r^{[1]}_{\rm 2HDM})\,,
\end{equation}
where the second term on the \textit{r.h.s.} of this equation was defined
in \eqref{eq:deltar-def42}.

The various contributions are described by the collection of Feynman
diagrams we display in Fig.~\ref{fig:self}. For a consistent calculation
of $\Delta r_{\rm 2HDM}^{\rm [1]}$ we must remove the existing overlap
between the one-loop 2HDM effects carried by the light, neutral \CP-even
Higgs boson ($\hzero$), and those driven by the SM Higgs boson
($\PHiggs_{\rm SM}$) -- which are contained in $\Delta r_{\rm SM}$ itself.
This is achieved, in practice, by subtracting from $\self_{\PW,\PZ}$ the
$\hzero$-mediated one-loop diagrams in the SM-like limit $\beta =
\alpha-\pi/2$. What remains is the genuine 2HDM effect encoded in
$\delta(\Delta\, r^{[1]}_{\rm 2HDM})$. Explicit details are given in the
Appendix. We perform our calculation in the 't Hooft-Feynman gauge and
regularize the UV divergences using standard dimensional
regularization\cite{'tHooft:1972fi}, which is both Lorentz and gauge
invariant. Analytical manipulations at this stage are carried out with the
help of the packages {\sc FeynArts} and {\sc FormCalc}
\cite{feynarts,formcalc}.

Let us now describe our numerical setup. As for the Higgs boson
spectra, we define several mass choices (cf. Table~\ref{tab:masses})
that span over representative parameter space regions, featuring in
particular the $\sim$125 GeV Higgs-like resonance spotlighted by
ATLAS and CMS and in agreement with all the phenomenological limits
discussed in Sec.~\ref{subsec:phenorestrictions}. Set I describes a
light, neutral \CP-even Higgs state [$\hzero$] taking on the role of
the 5$\sigma$ Higgs-candidate. It is thus preferably studied in the
decoupling regime $\alpha = \beta-\pi/2$. Notice that this scenario
can in principle accommodate a fermiophobic heavy Higgs [$\Hzero$]
($\alpha = 0$) for type-I 2HDM's. Further on we shall entertain this
possibility, even though a more accurate determination of the
fermionic decay modes of the putative Higgs boson candidate might
soon rule out this region of the parameter space. In Set II we still
assume a type-I 2HDM, but in contrast to the previous set we
identify the $\sim$125 GeV resonance with the heavier neutral
\CP-even Higgs field [$\Hzero$], whereas the lighter $\hzero$ is
assumed to have escaped detection so far. In this latter case we are
of course setting $\alpha\simeq\beta$, since this choice insures
SM-like couplings of $\Hzero$ to gauge bosons, and suppressed
couplings for $\hzero$. Sets I and II, thus, represent alternative
choices of Higgs bosons, featuring a non-constrained mass spectrum
and in full compliance with the current experimental picture. Sets
III and IV, on the other hand, consider (as in the case of Set I)
$\alpha = \beta-\pi/2$ and are devised to mimic typical MSSM-like
spectra -- say a relatively light $\hzero$ state playing the role of
the $\sim$125 GeV resonance along with heavier (almost
mass-degenerate) companions [$\Hzero,\Azero,\PHiggs^{\pm}$]. The
mass spectrum in both sets has been numerically derived from selected
MSSM parameter configurations with the help of {\sc FeynHiggs}
\cite{\citeFH}. In particular, Set III corresponds to the so-called
\emph{decoupling regime} of the MSSM for \emph{maximal mixing} (cf.
the recent analysis of Ref.~\cite{Arbey:2012dq}). Similarly, we
define Set IV, now with heavier masses for the scalar companions of
the light Higgs boson, as a realization of the \emph{decoupling
regime} for \emph{typical mixing}~\cite{Arbey:2012dq} of the MSSM.
Let us note that owing to the phenomenological restrictions
described in Sec.~\ref{subsec:phenorestrictions}, Set III must
correspond to type-I while Set IV can be either type-I or type-II
2HDM.

Finally, we construct the two groups V-VIII and V'-VIII' of mass
sets. In both groups we have two neutral, \CP-even, almost
mass-degenerate Higgs bosons with $M_h \sim 125\,\GeV$ while the
charged Higgs and the neutral \CP-odd states are typically heavier
and with a constant mass splitting $|M_{\Azero} - M_{\PHiggs^{\pm}}|
= 40$ GeV. The distinction of the two groups lies in the two
possible mass hierarchies: $M_{\PHiggs^{\pm}}
> M_{\Azero}$ (which we conventionally dub \emph{direct} hierarchy);
and $M_{\PHiggs^{\pm}} < M_{\Azero}$ (\emph{inverted} hierarchy --
corresponding to the sets V'-VIII' marked with a prime). The fact
that the two \CP-even states lie around $\sim 125\,\GeV$ is
particularly interesting, as they enable the two complementary
regions $\alpha \simeq \beta -\pi/2$ and $\alpha \simeq \beta$ to be
phenomenologically viable -- alongside with part of the intermediate
regimes. This is so because, regardless of whether the trigonometric
couplings are arranged as in the decoupling (SM-like) limit
[$\sin(\beta-\alpha) \simeq 1$]; or, conversely, if they are
anchored such that [$\cos(\beta-\alpha) \simeq 1$], in either case
we find a neutral, \CP-even state with SM-like couplings to the
gauge bosons. The latter can thus take on the role of the $\sim 125$
resonance unveiled at the LHC -- while its mass-degenerate companion
will remain virtually decoupled.


As reflected in Eq.~\eqref{eq:deltar-def4}, the 2HDM prediction for
$\Delta r$ is generated in our approach from  the combination of the SM
result $\Delta r_{\rm SM}$ and the genuinely new pieces sourced by the
2HDM degrees of freedom. Since we are interested here to evaluate the full
one-loop 2HDM correction beyond the conventional SM effects [$\Delta
r^{\rm [1]}_{\rm 2HDM}$], we will include in this section the
$\delta(\Delta r_{\rm 2HDM}^{[1]})$ piece from \eqref{eq:deltar-def4}
only. Recall
that the other piece in \eqref{eq:deltar-def4}, $\Delta r_{\rm
2HDM}^{({\rm \lambda_5})}$, first enters at two loops. Its quantitative
impact will be assessed in the next section, as it requires a careful
theoretical digression before we can tackle it.

In practice we shall evaluate $\Delta r_{\rm SM}$ from
Eq.~\eqref{eq:deltar_def1} upon extracting the SM theoretical value
[$M^{\rm SM}_W$] from the compact numerical parametrization of
Ref.\cite{Freitas:2002ja,Freitas:2002ve,Awramik:2003rn},
\begin{eqnarray}
M^{\rm SM}_{\PW} &=& M_{\PW}^{0} - d_1 \, \mathrm{dH} - d_2 \, \mathrm{dH}^2 + d_3 \, \mathrm{dH}^4 + d_4\,(dh-1)
\nonumber \\
       && - d_5 \, \mathrm{d}\alpha + d_6 \, \mathrm{dt}  - d_7 \, \mathrm{dt}^2
       - d_8 \, \mathrm{dH} \, \mathrm{dt} \nonumber \\ && + d_9\, dh\,dt - d_{10} \, \mathrm{d}\alpha_s
       + d_{11} \, \mathrm{dZ} ,
\label{eq:cplxpar}
\end{eqnarray}


\noindent The coefficients introduced thereof read
\begin{equation}
\begin{array}{lcllcl}
M_{\PW}^{0} &=& 80.3800 \,\GeV;&
d_6 &=& 0.5270\, \GeV;
\\
d_1 &=& 0.05253\, \GeV; &
d_7 &=& 0.0698 \,\GeV;
\\
d_2 &=& 0.010345 \,\GeV; &
d_8 &=& 0.004055 \,\GeV,
\\
d_3 &=& 0.001021 \,\GeV, \quad &
d_9 &=& 0.000110 \,\GeV,
\\
d_4 &=& -0.0000070 \,\GeV, &
d_{10} &=& 0.0716 \,\GeV,  \\
d_5 &=& 1.077 \,\GeV, & d_{11} &=& 115.0 \,\GeV; \\
\mathrm{dH} &=& \ln\left(\frac{M_{\PHiggs}}{100 \,\GeV}\right);
&
\mathrm{dt} &=& \left(\frac{m_t}{174.3 \,\GeV}\right)^2 - 1; \\
\mathrm{d}\alpha &=& \frac{\Delta\alpha}{0.05907} - 1;
& \mathrm{d}\alpha_s &=& \frac{\alpha_s(M_{\PZ})}{0.119} - 1; \\
\mathrm{dZ} &=& M_{\PZ}/(91.1875 \,\GeV) -1;
& \mathrm{dh} &=& \left(\frac{M_{\PHiggs}}{100\,\GeV}\right)^2;
\end{array}
\label{eq: MWparameterization}
\end{equation}

\noindent with the top-quark mass [$m_t = 173.5$ GeV] and the Z-boson mass  [$M_{\PZ} = 91.1875$ GeV]
evaluated
at their current best averaged values \cite{pdg}; while
$\alpha_s(M_{\PZ}) = 0.118$ and $\Delta\alpha = 0.05911$.
Notice that $M_{\PW}^{0}= 80.3800 \,\GeV$ in the above parameterization is
{\emph{not}} meant to be the zeroth order value of $M_{\PW}$ (which we
already indicated previously as $M_{\PW}^{\rm tree}$), but a fiducial mass
value which is chosen very close to the central value of the
experimentally measured $M_{\PW}$
--- cf. Eq.\,\eqref{eq:sminput} below. Therefore all the correcting terms on the \textit{r.h.s.} of \eqref{eq:cplxpar}
are expected to be small for any reasonable model, and some of them are
very much close to zero because the involved physical quantity  is known
with high accuracy (e.g. $\mathrm{dZ}\simeq 0$).

\noindent The parametrization \eqref{eq:cplxpar}-(\ref{eq:
MWparameterization}) reproduces the full set of available quantum
corrections within the SM. It relies on the following decomposition of
$\Delta r$ in all the one-loop and higher order effects computed to
date\,\cite{Freitas:2002ja,Freitas:2002ve}:
\begin{eqnarray}
 \Delta r_{\rm SM} &=& \Delta r^{(\alpha)} + \left[\Delta r^{(\alpha^2)}_{\rm bos} + \Delta r^{(\alpha^2)}_{\rm ferm}+\Delta r^{(\alpha\alpha_s)}+
\Delta r^{(\alpha\alpha^2_s)} \right. \nonumber \\
&&  \left. + \Delta r^{(G_F^2\alpha_s m_t^4)} + \Delta r^{(G_F^3\,m_t^6)}\right]
\label{eq:deltaR-estim}.
\end{eqnarray}
\noindent For more clarity we keep in \eqref{eq:deltaR-estim} the notation
of the aforesaid references, and hence the first term $\Delta
r^{(\alpha)}$ denotes the complete one-loop contribution; in our notation,
it corresponds to $\Delta r_{\rm SM}^{[1]}$ in \eqref{eq:Deltar2HDM1loop}.
The remaining terms are higher order contributions which have been
explicitly computed in the literature over the years\,\footnote{Recall
that $\Delta r$ in the defining equation \eqref{eq:deltar_def1} collects
the expanded form of the various contributions up to the order of
perturbation theory where the calculation has been carried out. In
contrast to the form sketched in \eqref{eq:deltar-SM}, no resummation is
assumed here of the $\Delta\alpha$ and $\delta\rho$ effects since the
terms obtained e.g. at two loop order from the resummation of leading
one-loop effects are already contained in the explicit two-loop
contribution. }; thus $\Delta r^{(\alpha^2)}_{\rm
bos}$\,\,\cite{Onishchenko:2002ve,Awramik:2002wv}
 and  $\Delta r^{(\alpha^2)}_{\rm
ferm}$\cite{Freitas:2000gg,Freitas:2002ja,Awramik:2002wv} stand for the
respective bosonic and fermionic electroweak two-loop corrections; $\Delta
r^{(\alpha\alpha_s)}$ and $\Delta r^{(\alpha\alpha^2_s)}$ for the
corresponding two-loop and three-loop QCD corrections
\cite{deltar-corrections}; and finally we have the pure electroweak
$\mathcal{O}(G_F^3\,m_t^6)$ and mixed electroweak-QCD
$\mathcal{O}(G_F^2\alpha_s m_t^4)$ terms, which track the leading 3-loop
contributions \cite{Faisst:2003px}. The remaining theoretical uncertainty
from the unknown higher order corrections is estimated to lie around
$\Delta\,M_{\PW}^{\rm SM} \simeq 4\,\MeV$ \cite{LEPEWWG}.

\begin{figure*}
\begin{center}
  \includegraphics[scale=0.5]{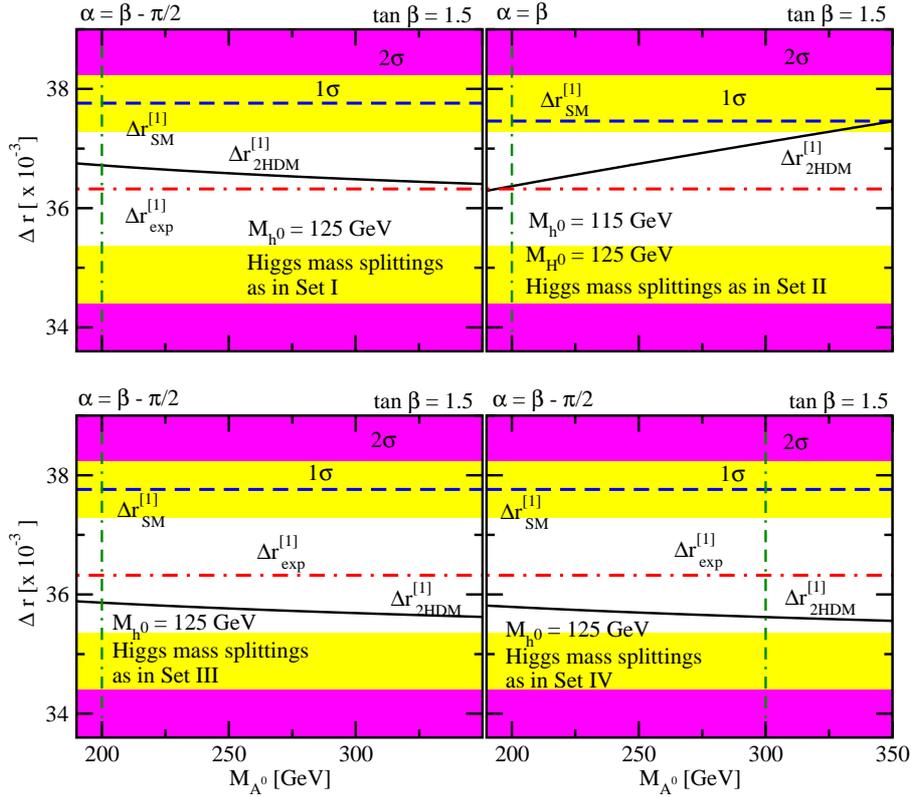}
 \caption{ Full one-loop evaluation of the parameter $\Delta r$ in the
2HDM, cf. Eq.\,\eqref{eq:Deltar2HDM1loop} (solid line, in black), as a
function of the neutral, \CP-odd Higgs boson mass [$M_{\Azero}$]. The
masses of the remaining heavy Higgs bosons
($M_{\Hzero},M_{\PHiggs^{\pm}}$) are also varied alongside, {with the mass
splittings fixed according to the mass sets from Table~\ref{tab:masses}
as indicated in each panel}. The vertical line signals the point in the Higgs
boson mass range that corresponds to the exact mass values quoted in
Table~\ref{tab:masses}. In the particular case of Set II, in which the
SM-like Higgs boson is identified with the heavy \CP-even Higgs boson
[$\Hzero$], its mass $M_{\Hzero}$ is also kept constant, so that in
practice we scan over $M_{\Azero}$ and $M_{\PHiggs^{\pm}}$ only. The
calculation is performed at fixed $\tan\beta = 1.5$, with $\alpha =
\beta-\pi/2$ (resp. $\alpha=\beta$) for Sets I,III and IV (resp. Set II),
in agreement with the SM-like decay patterns of the $125$ GeV Higgs boson
candidate. As a reference, we also display the SM prediction $\Delta
r_{\rm SM}$ (dashed line, in blue) and the experimental value $\Delta
r^{\rm exp}$ (dotted-dashed line, in red). The associated $1\sigma$ and
$2\sigma$ C.L. bands are explicitly indicated. \label{fig:deltaR} }
\end{center}
\end{figure*}

\begin{figure*}
 \begin{center} \includegraphics[scale=0.5]{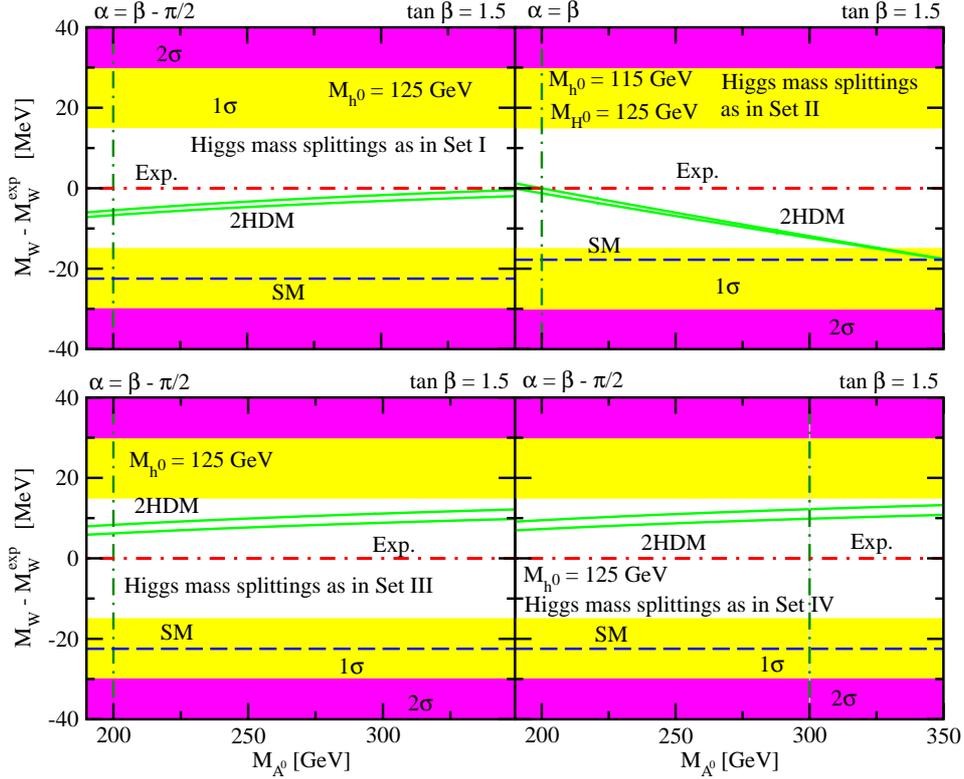}
 \caption{ Full one-loop prediction for the $\PW$-boson mass $[M^{\rm
th}_{\PW} \equiv M_{\PW}^{\rm SM} + \delta\,M_{\PW}^{\rm 2HDM}]$ in the
general 2HDM as a function of the neutral, \CP-odd Higgs boson mass
[$M_{\Azero}$]. The results are presented as a deviation $[\delta M_{\PW}]
= M_{\PW}^{\rm th} - M_{\PW}^{\rm exp}$ with respect to the experimental
value $M^{\rm exp}_{\PW} = 80.385 \pm 0.015$ GeV. The masses of the
remaining heavy Higgs bosons ($M_{\Hzero},M_{\PHiggs^{\pm}}$) are also
varied alongside, preserving the mass splitting {of the mass sets from
Table~\ref{tab:masses} as indicated in each panel}. The vertical line
signals the point in the Higgs boson mass range that corresponds to the
precise mass values quoted in Table~\ref{tab:masses}. In the particular
case of Set II, in which the SM-like Higgs boson is identified with the
heavy \CP-even Higgs boson [$\Hzero$], its mass $M_{\Hzero}$ is also kept
constant, so that in practice we scan over $M_{\Azero}$ and
$M_{\PHiggs^{\pm}}$ only. The calculation is performed at fixed $\tan\beta
= 1.5$, with $\alpha = \beta-\pi/2$ (resp. $\alpha=\beta$) for Sets I,III
and IV (resp. Set II), in agreement with the SM-like decay patterns of the
$125$ GeV Higgs boson candidate. The green band provides an estimate on
the theoretical uncertainty. As a reference, we also display the SM
prediction $M_{\PW}^{\rm SM}$ (dashed line, in blue). The associated
$1\sigma$ and $2\sigma$ C.L. bands for the measured value $M_{\PW}^{\rm
exp}$ are explicitly indicated. \label{fig:mw} } \end{center}
\end{figure*}

\begin{figure*}
 \begin{center}
  \includegraphics[scale=0.5]{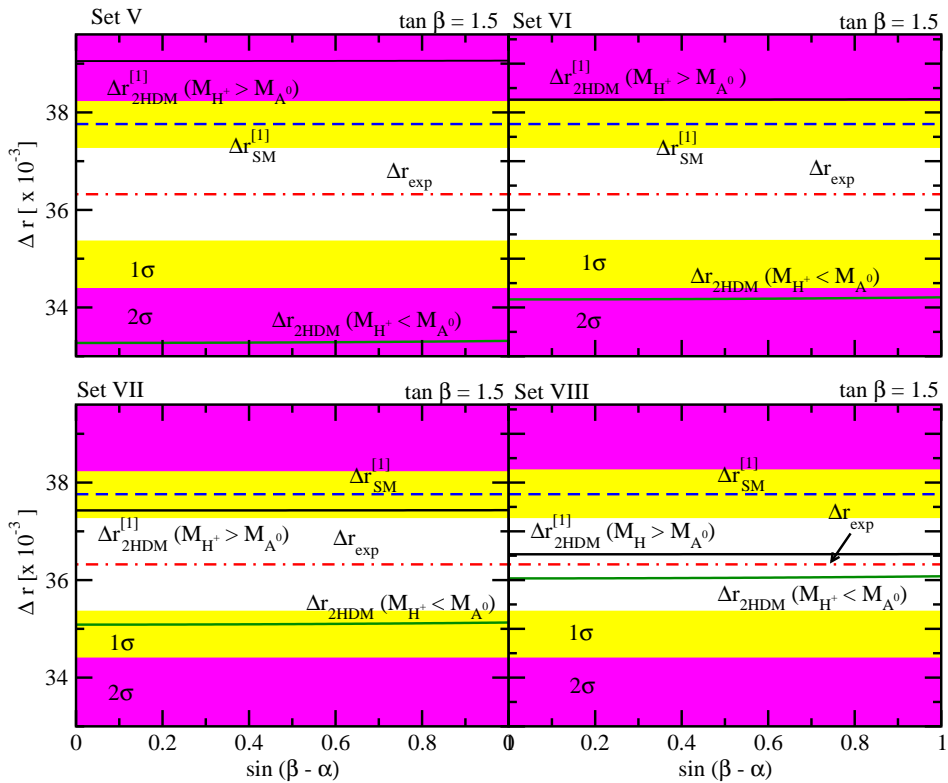}
\caption{Full one-loop evaluation of the parameter $\Delta r$ in the
general 2HDM, cf. Eq.\, \eqref{eq:Deltar2HDM1loop} (solid line, in black)
as a function of $\sin(\beta-\alpha)$. The calculation is performed at
fixed $\tan\beta = 1.5$, for sets V to VIII and V' to VIII' of Higgs boson
masses, and for the two different $M_{\Azero}/M_{\PHiggs^{\pm}}$ mass
hierarchies. As a reference, we also display the SM prediction $\Delta
r_{\rm SM}$ (dashed line, in blue) and the experimental value $\Delta
r^{\rm exp}$ (dotted-dashed line, in red). The associated $1\sigma$ and
$2\sigma$ C.L. bands are explicitly indicated.} \label{fig:deltaR-newsets}
 \end{center}
\end{figure*}

\noindent In accordance with the prescription \eqref{eq:deltar-def4}, we
then add up to the previous result the genuine 2HDM piece $\delta(\Delta
r_{\rm 2HDM}^{[1]})$ from the 2HDM Higgs boson-mediated self energies
[$\Sigma_{\PW,\PZ}$], as displayed in Fig.~\ref{fig:self} (see the
Appendix for explicit analytical details). Finally, we rephrase the
overall $\Delta r$ yield obtained in this way in terms of $M_{\PW}^{\rm
th}$ by iteratively solving Eq.~\eqref{eq:mwpred}. This is our estimate of
$\Delta r_{\rm 2HDM}$ at this point. As explained in the beginning of this
section, we call it the ``full one-loop 2HDM result'', in the sense that
all of the genuine one-loop effects from the 2HDM degrees of freedom have
been taken into account on the $\Delta r$ quantity within the 2HDM, but at
the same time we have incorporated to this result the complete current
knowledge of the leading SM quantum effects at various orders of
perturbation theory following the above method.

The upshot of our numerical analysis is summarized in
Figs.\,\ref{fig:deltaR}-\ref{fig:deltaR-newsets}. Figures~\ref{fig:deltaR}
and \ref{fig:mw} display the full one-loop 2HDM predictions for $\Delta r$
and $M^{\rm th}_{\PW}$, respectively, as a function of the neutral, CP-odd
Higgs boson mass [$M_{\Azero}$]. {To start with, we fix all the Higgs
masses as in Sets I-IV from Table~\ref{tab:masses}; then we sweep over a
range $190 < M_{\Azero} < 350$ [GeV] while varying the masses of the
remaining Higgs fields accordingly, so that the mass splittings among the
heavier 2HDM Higgs fields are mantained as in each of the considered mass
sets}. At the same time we keep fixed the mass of $\hzero$ at the value
$M_{\hzero} = 125 $ GeV. For Set II we actually anchor both $M_{\hzero}$
and $M_{\Hzero}$ at the constant values {indicated in the figure} --
recall that, in this scenario, the heavy \CP-even neutral field [$\Hzero$]
is meant to describe the putative $\sim 125$ GeV Higgs boson. We carry out
the calculation at a fixed value of $\tan\beta = 1.5$, in compliance with
the stringent conditions dictated by $B_d^0 - \bar{B}_d^0$ and
$\mathcal{B}(b \to s\gamma)$ physics, including the perturbative unitarity
and vacuum stability bounds (see Sec. \ref{subsec:phenorestrictions}). On
the same grounds we enforce SM-like Higgs/gauge boson couplings for the
corresponding $\sim 125$ GeV Higgs boson in our numerical analysis --
which amounts to work in the \emph{decoupling limit} $\alpha = \beta -
\pi/2$ for Sets I, III and IV, and $\alpha = \beta$ for Set II. Finally,
Sets V -- VIII and V' -- VIII', with two neutral, \CP-even, $\sim 125$ GeV
states, interpolate between both scenarios, and are examined in
Figure~\ref{fig:deltaR-newsets}. We shall comment on them later on.

The different panels of Fig.~\ref{fig:deltaR} present the $\Delta r^{\rm [1]}_{\rm 2HDM}$
results for Sets I - IV of Higgs bosons masses.
Therewith we also display the SM value $\Delta r_{\rm SM}$, computed as
described above, together with the experimental value $\Delta r_{\rm exp}$
obtained from Eq.~\eqref{eq:deltar_def1} using the experimental inputs
%

\begin{equation}
\begin{array}{ccc} M^{\rm exp}_{\PW} = 80.385 \pm 0.015\, \GeV \ & \; &
M_{\PZ} = 91.1876 \pm 0.0021 \,\GeV \\
 \alpha(0) = 1/137.03599 &\; &  G_F = 1.16637 \, 10^{-5} \,\GeV^{-2}.  \label{eq:sminput}
\end{array}
\end{equation}

\noindent Plugging the above values into Eq.~\eqref{eq:deltar_def1} we get

\begin{equation}
 \Delta r_{\rm exp} = \frac{\sqrt{2}\,G_F}{\pi\alpha}\,M_{\PW}^2\,\left(1-\frac{M_{\PW}^2}{M_Z^2}\right) - 1 = 36.322\times 10^{-3}
\label{eq:deltarexp}.
\end{equation}

\noindent In Fig.\,\ref{fig:mw} the corresponding 2HDM and SM theory
predictions for  $M^{\rm th}_{W}$ are represented as a function of
$M_{\Azero}$ under the same conditions as in Fig.\,\ref{fig:deltaR}. One
can also appreciate the deviations of the theoretical $\PW$-boson mass
with respect to the latest experimental value $M^{\rm exp}_{\PW} = 80.385
\pm 0.015 \, \GeV$. Confidence levels for $M^{\rm exp}_{\PW}$ and $\Delta
r_{\rm exp}$ are also included in our plots as colored areas corresponding
to the $1\sigma$ (yellow) and $2\sigma$ (magenta) regions. In the case of
$\Delta r$,   we compute these uncertainties from those of $M_{W}^{\rm
exp}$ by standard error propagation methods. The region comprised between
the two parallel thin strips (in green) in Fig.~\ref{fig:mw} supplies an
estimate of the 2HDM theoretical uncertainty [$\Delta\,M_{\PW}^{\rm th}$].
The latter we quantify by means of Eq.~\eqref{eq:mwpred} upon substituting
$1 + \delta(\Delta r^{[1]}_{\rm 2HDM}) \to 1/(1-\delta(\Delta r^{[1]}_{\rm
2HDM}))$, which leads to a slightly modified prediction
[$\tilde{M}_{\PW}^{\rm th}$]. Such a shift operates an approximate
resummation of the 2HDM contributions, in such a way that the difference
relative to the central value, [$\Delta\,M_{\PW}^{\rm th}/M_{\PW}^{\rm th}
\equiv |\tilde{M}_{\PW}^{\rm th}-M_{\PW}^{\rm th}|/M_{\PW}^{\rm th}$],
should be indicative of the size of the pure 2HDM higher-order effects
neglected when truncating our perturbative expansion at
$\mathcal{O}(\aew)$ -- and that fall typically within the range of
$\sim 1-5$ MeV.  {As we will see later on, this estimate is well
in agreement with the leading 2HDM higher-order corrections computed in
Section~\ref{sec:higher}. Such 2HDM uncertainty is yet to be combined with
its SM counterpart $\Delta M_{\PW}^{\rm SM} \simeq 4$ MeV, as well as with
the parametric uncertainties -- dominated by the top mass measurement
$\Delta m_t = \pm 0.9$ GeV ~\cite{topmeasure} and entailing
approximately $\Delta M_{\PW}^{\rm param}\simeq 10$ MeV. Therefore, the
errors added in quadrature lead to a total uncertainty of roughly $12$ MeV.


Complementary vistas on the $\Delta r$ behavior across the 2HDM parameter
space are displayed in Fig.~\ref{fig:deltaR-newsets}. We compute once
again the full-fledged one-loop 2HDM prediction [$\Delta r^{\rm [1]}_{\rm
2HDM}$] and compare it to the SM result [$\Delta r_{\rm SM}$] and the
experimental value [$\Delta r_{\rm exp}$]. We keep a fixed $\tan\beta =
1.5$ but we now sweep all over the $\sin(\beta-\alpha)$ range. In doing
so, we effectively interpolate between the two corners with SM-like
Higgs/gauge boson couplings. In the light of the LHC findings, regions
away from $\sin(\beta-\alpha) \simeq 1$ -- or conversely
$\sin(\beta-\alpha) \simeq 0$ -- would be disfavored, if not simply
excluded, at a certain confidence level. We will refrain from considering
a more accurate treatment of this issue, as it is still subdued by large
statistical uncertainties and surely not relevant for our present
discussion. Fig.~\ref{fig:deltaR-newsets} displays a featureless $\Delta
r_{\rm 2HDM}$ profile as a function of $\sin(\beta - \alpha)$, which
follows from the approximate mass-degeneracy of the neutral \CP-even Higgs boson states.
Worth
noticing is also the influence of $\delta\rho$ (what is tantamount to say,
the size of $\Delta r^{[\delta\rho]}$ in the notation of
Eq.~\eqref{eq:DeltarGeneral}): the lighter the Higgs boson masses --and
the narrower their splittings -- the smaller becomes $\Delta
r^{[\delta\rho]}$. One can easily understand this behavior from equations
\eqref{drho2HDM} and \eqref{eq:deltarho_an}. One visible signature that
stands out in these plots is that the tension between the theoretical
prediction
[$\Delta r_{\rm 2HDM}$] and the experimental value
[$\Delta r_{\rm exp}$]
shrinks significantly when comparing the outcomes for
the different mass choices from Sets V to VIII. For Set V, for instance,
we have $|\delta\rho| \sim \mathcal{O}(10^{-3})$ -- at the very border of
the permitted custodial symmetry bounds. The predicted $\Delta r_{\rm
2HDM}$ is then pushed more than $2\sigma$ away from the experimental
result -- with an even larger tension than the SM prediction.
In contrast, for Set VIII the theoretical prediction [$\Delta r_{\rm 2HDM}$]
moves closer to [$\Delta r_{\rm exp}$], departing from it
roughly $\sim 0.6\,\%$ -- the SM prediction staying circa $\sim 4\%$ away.
All these
effects are qualitatively similar, although opposite in sign, when we swap
the mass hierarchy. Remarkably, we find that [$M_{\Azero} >
M_{\PHiggs^{\pm}}$] pulls our prediction for $M_{\PW}^{\rm th}$ slightly further
away from $M_{\PW}^{\rm exp}$, as compared to [$M_{\Azero} <
M_{\PHiggs^{\pm}}$]. For the EW observables under scope, therefore,
the scenarios we have considered here tend to favor 2HDM
realizations with the charged Higgs being heavier than the neutral,
\CP-odd one.

The predicted $M_{\PW}^{\rm th}$ values within the 2HDM are seen to follow
a smooth and monotonous variation, with changes of
$\mathcal{O}(10)\,\MeV$ -- equivalently, $\delta(\Delta r) \sim
\mathcal{O}(10^{-3})$ as retrieved by the previous Fig.~\ref{fig:deltaR}
-- when sweeping the Higgs boson mass range $190 < M_{\Azero} <  350$
[GeV]. The largest possible theoretical $\PW$-boson mass shift from
genuine 2HDM effects is substantial: $|\delta M_{\PW}^{\rm 2HDM}|\simeq
35$ MeV (more than twice the current experimental error on $M_{\PW}$).
Larger variations ensue from a trade-off between two different -- and
somewhat opposite -- conditions, which are the following: i) heavier Higgs
bosons, in correspondence with the mass suppression of the one-loop
Higgs-mediated contributions to $\Sigma_{\rm \PW,\PZ}$, yield relatively
tamed quantum effects; and ii) broader splittings among the different
Higgs boson masses lead to a stronger breaking of the approximate $SU(2)$
custodial symmetry, and hence to an augmented contribution $\Delta r^{\rm
[\delta\rho]}$. We will also identify a similar balance between both tendencies
when examining the leading higher order quantum corrections, as
discussed further down.

The SM theoretical prediction following from the full parametrization
\eqref{eq:cplxpar}-(\ref{eq: MWparameterization}) with  $M_{\PHiggs}=125$ GeV reads
\begin{equation}\label{eq:MWthSM}
M^{\rm SM}_{\PW}= 80.363 \, \GeV.
\end{equation}
The difference $M^{\rm SM}_{\PW}-M^{\rm exp}_{\PW}$
corresponds to the horizontal line marked as SM in all the
plots of Fig.\,\ref{fig:mw} -- except for Set II, for which
the identification $\hzero \equiv \PHiggs$ implies that
$M_{\PW}$ is to be evaluated at $M_{\PHiggs} = 115\,\GeV$, yielding
$M_{\PW} = 80.367\,\GeV $.
We see in these plots that the SM value is
systematically too low, it typically fails to match the central value
$M^{\rm exp}_{\PW}$ by more than $20$ MeV below it. This deviation is not
worrisome right now because it lies within the $1\sigma$ region, but this
could change in the future when the error in $M^{\rm exp}_{\PW}$ may
dramatically decrease at the $\sim 5$ MeV level. At this point the SM
prediction, whose intrinsic theoretical error is estimated to be $\sim 4$
MeV\,\cite{LEPEWWG}, could be in good shape or on the contrary it might
already clash with experiment. In contrast, the 2HDM numerical prediction
is seen to have the ability to easily cure the differences with respect to
the experimental value, should the conflict with the SM prediction become
confirmed. From Fig.\,\ref{fig:mw} it is evident that the 2HDM prediction
can stay very close to the central experimental value $M^{\rm exp}_{\PW}$
and moreover it can accommodate deviations of up to $20-35$ MeV with
respect to it, depending on the Higgs mass sets used from Table
\ref{tab:masses}.

\medskip{}

It is well-known that the MSSM has also the potential to shrink the gap
between $M^{\rm exp}_{\PW}$ and $M^{\rm SM}_{\PW}$. To illustrate it in
our framework, let us e.g. consider the SUSY-inspired Set III from
Table~\ref{tab:masses}, which features the \emph{maximal mixing} MSSM
scenario \cite{Arbey:2012dq}. The latter contains relatively light
$\mathcal{O}(100)$ GeV gaugino masses, along with heavy $\mathcal{O}(1)$
TeV squarks and sleptons and moderately massive stops in the
$\mathcal{O}(700)$ GeV ballpark. Using {\sc FeynHiggs} \cite{\citeFH} we
evaluate $M_{\PW}^{\rm MSSM}$ including all the presently known
contributions (up to leading two-loop order) and get: 

\begin{eqnarray}
 M_{\PW}^{\rm MSSM} = 80.373 \,\GeV \;  &\Rightarrow& M_{\PW}^{\rm MSSM} - M_{\PW}^{\rm SM} = 10 \, \MeV;\nonumber \\
&\phantom{\Rightarrow}& M_{\PW}^{\rm MSSM} - M_{\PW}^{\rm exp} = -12 \, \MeV\,. \nonumber \\
\label{eq:mw-mssm-compare}
\end{eqnarray}
\noindent For the same masses, the corresponding 2HDM prediction (taking e.g.
the decoupling limit $\alpha = \beta-\pi/2$, cf. also Table~\ref{tab:deltarho}) yields
\begin{eqnarray}
 M_{\PW}^{\rm 2HDM} = 80.394 \, \GeV \;  &\Rightarrow& M_{\PW}^{\rm 2HDM} - M_{\PW}^{\rm SM} = 31 \, \MeV;\nonumber \\
&\phantom{\rightarrow}& M_{\PW}^{\rm 2HDM} - M_{\PW}^{\rm exp} = 9 \, \MeV \,. \nonumber \\
\label{eq:mw-2hdm-compare}
\end{eqnarray}
Let us recall from Fig. \ref{fig:mw} that the SM prediction lies roughly $-22$ MeV below $M_{\PW}^{\rm exp}$.
In this example, both the 2HDM and the MSSM are comparably competitive in
softening the $M_{\PW}^{\rm th} - M_{\PW}^{\rm exp}$ tension.
Interestingly, this picture does not hold any longer if we decouple the
genuine SUSY degrees of freedom, i.e. all sparticle masses apart from the
MSSM Higgs bosons. Raising all SUSY particle masses up to $\sim
\mathcal{O}(1.5)$ TeV in this example -- it has essentially no impact here
on the resulting two-loop SUSY Higgs masses --, the MSSM prediction
converges to the SM,
\begin{eqnarray}
 M_{\PW}^{\rm MSSM} = 80.361 \,\GeV \;  &\Rightarrow& M_{\PW}^{\rm MSSM} - M_{\PW}^{\rm SM} = -2 \, \MeV;\nonumber \\
&\phantom{\rightarrow}& M_{\PW}^{\rm MSSM} - M_{\PW}^{\rm exp} = -24 \, \MeV\,, \nonumber \\
\label{eq:mw-mssm-compare-decoupling}
\end{eqnarray}
\noindent and therefore the $M_{\PW}^{\rm th} - M_{\PW}^{\rm exp}$ tension at the
level of $\sim 20$ MeV  is back once more between the MSSM and the
experimental measurement of the $W^{\pm}$ mass, as expected from
decoupling arguments. This example illustrates that in the MSSM, in
contrast to the general 2HDM, the bulk capability to reconcile
$M_{\PW}^{\rm exp}$ and $M_{\PW}^{\rm th}$ relies on the genuine SUSY
contributions, while the SUSY Higgs-mediated effects play a subdominant
role. This is after all a reflect of SUSY invariance, which links the
Higgs boson masses and the mixing angle $\alpha$ to the gauge couplings,
enforcing a very constrained Higgs boson sector -- with a naturally mild
custodial symmetry violation. Comprehensive surveys of the MSSM parameter
space, including a detailed discussion on the SUSY-decoupling scenarios,
can be found e.g. in Refs.~\cite{Heinemeyer:2006px}. Further discussion on
the 2HDM versus MSSM comparison is presented later on in
Section~\ref{sec:higher}.

Let us remind the reader at this point that, within the approach employed
in this section, we are not yet sensitive to the potential effects from
the $\lambda_5$ parameter. Neither the results show any direct sensitivity
to whether they are obtained within a type-I or a type-II 2HDM setup, as
this depends on the specific Yukawa couplings of each model. Still, we
should bear in mind that both features, namely the strength of the Higgs
self couplings and the pattern of Higgs/fermion Yukawa interactions are
indirectly present through their interplay with the theoretical and
phenomenological constraints -- which are duly taken into account in our
analysis (cf. Sec.\,\ref{subsec:phenorestrictions}). Let us recall in
particular the lower bound of $\sim 300$ GeV for the charged Higgs boson
mass for type II models.



We conclude this section by emphasizing what we believe is our most
important observation, hitherto unnoticed (to the best of our knowledge)
in the literature, to wit: the bulk presence of the term $\delta(\Delta
r_{\rm 2HDM}^{[1]})$ -- defined in Eq.\,\eqref{eq:deltar-def42} --
provides a significant shift on the $W$-mass genuinely caused by the 2HDM
heavy Higgs companions. This shift softens the influence of the SM(-like)
Higgs boson $[\hzero]$ on the W-boson mass prediction and therefore
relaxes the current $M_{\PW}^{\rm th}$-versus-$M_{\PW}^{\rm exp}$ tension,
which is well known to grow with $M_{\hzero}$ and constitutes one of the
main reasons why global fits to Electroweak precision data tend to favor
light Higgs bosons -- rendering the minimum $\chi^2$-values for masses
even below the LEP bounds (cf. e.g. \cite{Buchmueller:2009ng}). What our
results reflect is that such trend no longer holds in the context of the
general 2HDM. Here, and essentially for all the surveyed scenarios, the
neat effect of adding new scalar $SU_L(2)$ doublets leads to a systematic
downward shift of $\Delta r$, and so upward on $M_{\PW}^{\rm th}$, which
tends to improve the agreement with the corresponding experimental
measurement. Said differently, when compared to the SM expectations for a
$\sim 125$ GeV SM-like Higgs boson mass, the genuine 2HDM contribution
[$\delta(\Delta r^{[1]}_{\rm 2HDM})$] tends to pull down the overall
$\Delta r^{\rm [1]}_{\rm 2HDM}$ prediction. Accordingly, $M^{\rm
th}_{\PW}$ becomes larger, and hence closer to the corresponding
experimental measurement $M_{\PW}^{\rm exp}$.
This is the most relevant message to be conveyed at this
point of our analysis. In the next section we focus on the additional
influence of the 2HDM quantum effects through the $\lambda_5$ parameter.

\section{Higher order 2HDM effects from enhanced Higgs boson self-interactions}
\label{sec:higher}

\begin{figure}  \begin{center}
\includegraphics[scale=0.38]{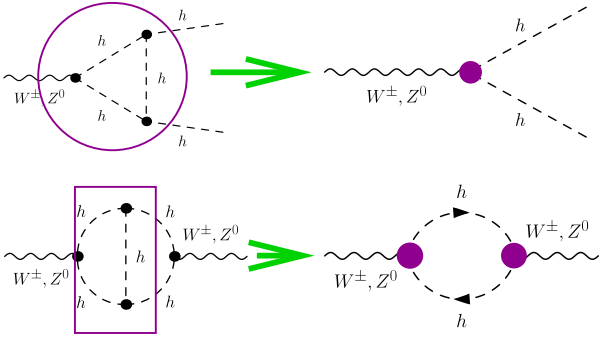}
\caption{Pictorial sketch of the form factors devised to
capture the leading 2HDM higher order corrections to $\delta\rho$.
In the limit of large Higgs boson self-interactions, these are
dominantly driven by the Higgs-mediated contributions to the weak
gauge boson self energies, as displayed in the lower part of the
figure (cf. Fig.\,\ref{fig:2self} for a specific sample). As a first
stage we compute the one-loop Higgs-mediated corrections to the
different Higgs/gauge (and Goldstone) boson interactions to order
$\mathcal{O}(\lambda^2_5)$; these correspond to a UV finite, gauge
invariant subset of contributions which encapsulates the (dominant)
effects triggered by the enhanced Higgs boson self-interactions
(upper-left corner). We then rewrite such corrections as one-loop
form factors and use them to effectively \emph{dress} the
leading-order Higgs/gauge boson couplings (upper-right corner). The
resulting set of Born-improved effective vertices (depicted as
colored blobs) can be finally plugged back in the calculation of the
one-loop gauge boson self energies (lower-right and lower-left
corners).\label{fig:composite}}
\end{center} \end{figure}

A number of recent studies
\cite{LopezVal:2009qy,LopezVal:2010yf}
have pointed out and exploited the possibility that the Higgs/gauge boson
couplings eventually undergo a drastic modification at the quantum level
in the limit of large Higgs boson self-interactions. These modifications
are in practice restrained by certain theoretical and phenomenological
bounds, but their net effect can still be quite sizeable in different
contexts. In the present case the potential enhancements can actually
be traded through the aforementioned $\mathcal{O}(\lambda^2_5)$ terms in the
Higgs boson self-coupling structure. Such scenarios are not only
phenomenologically viable, but also very well supported on theoretical
grounds. As a matter of fact, the rise of scalar resonances constitutes a
prominent manifestation of models with strong dynamics in the EWSB
sector\cite{composite,Evans:2009ga}. Interestingly, strong
self-interactions among Higgs, or Higgs-like states, are most often
identified with rather heavy mass spectra -- reflecting the relations
among masses and self-couplings settled through the Higgs potential.
Nevertheless, the 2HDM comprises a wider range of possibilities -- as not
all of the Higgs self-couplings are tied to the corresponding physical
Higgs masses. The fact that we can freely dial the parameter $\lambda_5$
enables to recreate genuine 2HDM regimes with relatively light, and yet
strongly coupled Higgs bosons.

As we have preliminarily discussed in
Sec.\,\ref{subsec:Deltardeltarho}, EW precision quantities -- most
significantly $\Delta r$ and $\delta \rho$ -- become sensitive to
the Higgs boson self-couplings via 2-loop effects on the gauge boson
self-energies. This is illustrated in Fig.~\ref{fig:composite}. Here
we do not target at an explicit 2-loop calculation. Instead, we aim
at getting insight into the foreseeable quantitative impact that
such Higgs boson self-couplings may exert on the theoretical
prediction for the W-boson mass $[M^{\rm th}_{\PW}]$. In fact, since
a fully-fledged 2-loop calculation is too cumbersome and probably
unnecessary at this point, we resort to a Born-improved
Lagrangian approach leading to a set of effective couplings or form
factors. The advantage of it is that we can focus more easily on
the main potential sources of enhancement and moreover we can even
have access to their influence beyond the 2-loop level.

In practice we devise our set of effective couplings as
follows:\footnote{Approaches along these lines are certainly not
foreign in the literature, see for example Ref.\cite{Lee:2012jn}. In
our opinion, however, scarce attention has been devoted to the
underlying assumptions and corresponding limitations.}:

\begin{enumerate}
\item{To start with, let us recall that after SSB of the EW gauge
    symmetry we can always trade the original Higgs self-couplings
    $\lambda_i$ (except $\lambda_5$) for the physical masses, mixing
    angles and gauge couplings. This is explicitly done e.g. in Sec.
    III of Ref.\,\cite{LopezVal:2009qy} -- cf. Eq.\,(12) there. From
    here we can easily implement the limit of large Higgs
    self-couplings by projecting out the $\lambda_5$-dependent part of
    each of the Higgs self-interactions in the model and neglecting
    the other terms. According to this procedure we can redefine
    $\lambda_i\to \tilde{\lambda}_i$ for all the original
    self-couplings $\lambda_i$ in the Higgs potential --cf.
    Eq.~\eqref{eq:potential} --  as follows:

\begin{eqnarray}
&& \tilde{\lambda_1} = \frac{1}{4}\,\lambda_5\,(1-\tan^2\beta); \quad  \tilde{\lambda_2} = \frac{1}{4}\,\lambda_5\,(1-\cot^2\beta);
\nonumber \\
&&  \tilde{\lambda_3} = -\frac{1}{4}\,\lambda_5;  \quad  \tilde{\lambda_4} = \lambda_6 = 0\,.
\label{eq:3Hproject}
\end{eqnarray}

\noindent The above approximation has its counterpart in the physical
basis. Following the splitting procedure \eqref{eq:splithhh}, we
collect the $\lambda_5$ terms associated to the Higgs self-couplings
in that basis. They render structures of the sort:
\begin{eqnarray}
\lambda_{\hzero\hzero\hzero} &\rightarrow&c_h\,\lambda_5= \nonumber \\
&& \left(i\,\frac{6\,M_W\,\sw}{e}\,\frac{\cos(\alpha+\beta) \cos^2(\beta-\alpha)}{\sin 2\beta}\right)\,\times\lambda_5\,,\nonumber \\
\label{eq:3Hproject2}
\end{eqnarray}
and analogously for all the self-interactions $\lambda_{\rm hhh}$ relating
the different Higgs boson fields to each other (we refer once more
e.g. to Table II of Ref.\cite{LopezVal:2009qy}). Notice that the
strength of all the $\lambda_{\rm hhh}$ couplings, and so also their
related potential enhancements, are now fully determined as functions
of $\lambda_5$ and $\tan\beta$ -- with no remaining dependence on the
Higgs nor the gauge boson masses. As it will become clear soon, this
is a necessary step in order to preserve gauge invariance within our
framework.
%
In Fig.~\ref{fig:values} we illustrate how this approximation performs
when compared to the complete analytical form of the triple (3h)
self-interations
 (see e.g. Table III of Ref.\cite{LopezVal:2009qy}).
All couplings in this figure are normalized to the value of the 3H
self-coupling in the SM, for a Higgs boson mass of $M_{\PHiggs} = 125$
GeV, see Eq.\eqref{eq:triplesm}. Noteworthy is that the 3h
enhancements within the 2HDM, as indicated in Fig.~\ref{fig:values}
for values of $|\lambda_5| \sim \mathcal{O}(10)$, are perfectly
consistent with the Lee-Quick-Thacker unitarity bound for the SM 3H
coupling \cite{lee}:
\begin{equation}\lambda_{\rm HHH} \simeq
\frac{3e M_{\rm H}^2}{2\sw\,M_{\PW}}\Big{]}_{M_{\rm H} = 1\,\TeV} \simeq 1.3 \times 10^{4} \,\GeV
 \label{eq:lqt}.
\end{equation}
The main virtue of Fig.~\ref{fig:values} is that it makes apparent how
both descriptions of the $\lambda_{hhh}$ couplings --- viz. the full
one versus the truncated approach $\lambda_{\rm hhh}\to c_h\lambda_5$
defined by \eqref{eq:splithhh}
--- nicely agree in the purported limit of large Higgs boson
self-interactions. At the end of the day, the goodness of this
approximation will depend on the extent to which this limit is
effectively realized by a given choice of $\lambda_5$ and $\tan\beta$.}

\begin{figure*} \begin{center}
\includegraphics[scale=0.55]{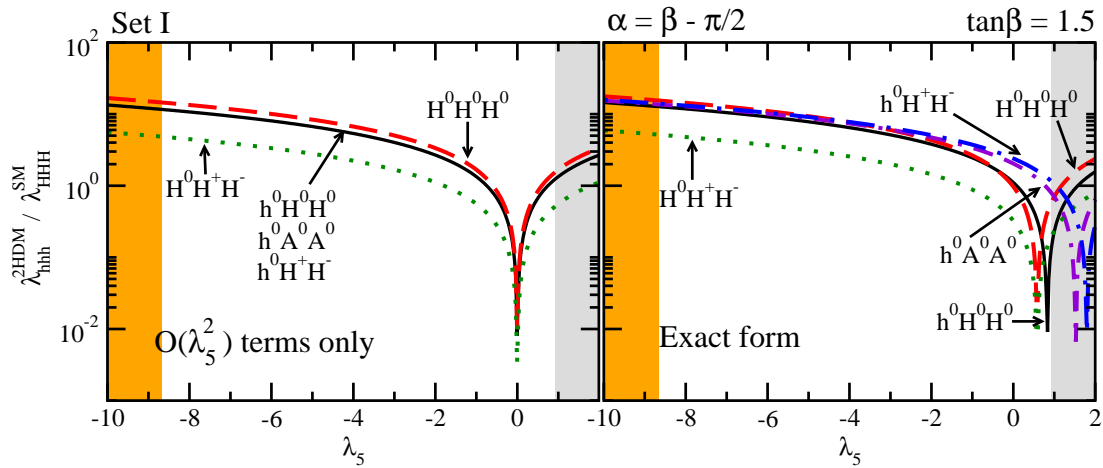}
\caption{Coupling strength [$\lambda_{\rm hhh}$] for a sample of
triple Higgs self-interactions, as a function of the parameter
$\lambda_5$. The couplings are normalized to the strength of the 3H
self-coupling in the SM, assuming that the SM Higgs boson mass yields
$M_{\PHiggs} = 125$ GeV. As a result we get $\lambda^{\rm SM}_{\rm
HHH}(M_{H} = 125\,\GeV) = 188.24\,\GeV$. We can check that the
obtained $\lambda_{\rm hhh}$ values lie systematically below the
Lee-Quick-Thacker bound $\lambda^{\rm SM}_{\rm HHH}(M_{H} \simeq
1\,\TeV) \simeq 1.3\times 10^4\,\GeV$ \cite{lee}.
The figure compares the size of these 3h self-couplings when
considering i) the projection of the $\mathcal{O}(\lambda_5)$
components only (left panel) and ii) its exact analytical form (right
panel) -- see the text for details. Without loss of generality, we
dwell here on set I of Higgs boson masses and we fix $\tan\beta =
1.5$. The shaded regions on the left (resp. right) side are excluded
by unitarity (resp. vacuum stability).\label{fig:values}} \end{center}\end{figure*}

\begin{figure*} \begin{center}
\includegraphics[scale=1.0]{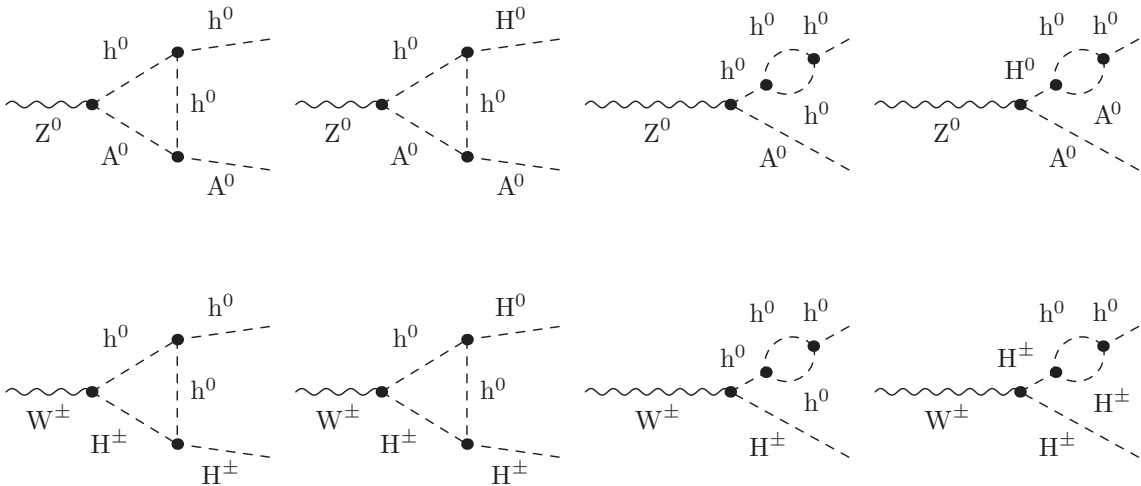}
 \caption{Sample of Feynman diagrams that account for the
$\mathcal{O}(\lambda^2_5)$ one-loop corrections to the
$\hzero\Azero\PZ^0$ and $\hzero \PW^{\pm} \PHiggs^{\pm}$ gauge
couplings in the 2HDM.} \label{fig:vert} \end{center} \end{figure*}

\item{Second, we examine the leading quantum effects on the different
    Higgs/gauge boson couplings in order to absorb them into effective
    vertices. These we can generically sort out into three categories:

\begin{tabular}{l}
 i) Higgs/Higgs/gauge boson couplings: $g_{\rm hhV}$  \\ ii) Higgs/gauge/gauge boson couplings: $g_{\rm hVV}$ \\
 iii) Higgs/gauge/Goldstone boson couplings: $g_{\rm hVG}$
\end{tabular}

\noindent with $[h = \hzero, \Hzero, \Azero,\PHiggs^{\pm}]$ standing
for a generic 2HDM Higgs field; $V = [\PW^{\pm}, \PZ^0]$ for a
weak gauge boson; and $G = [G^{\pm}, G^0]$ for the associated charged
and neutral Goldstone modes, namely the longitudinal components of the
gauge bosons in the 't Hooft-Feynman gauge. In practice, structures of
type iii) are formally equivalent to those of type i), as they involve
two scalar and one vector fields. In the regimes of interest here,
namely in the large Higgs self-interaction limit, the leading
corrections to the above couplings are thus driven by the interchange
of virtual Higgs bosons.
A sample of the corresponding Feynman diagrams is provided in
Fig.~\ref{fig:vert}, in which we single out the case of the
$g_{\hzero\Azero\PZ^0}$  and $g_{\hzero \PW^{\pm} \PH^{\pm}}$
effective couplings. The one-loop corrections to each of these
couplings involve both triangle diagrams and self-energy insertions;
the former account for the genuine $\mathcal{O}(\lambda_5^2)$ one-loop
vertex corrections, whereas the latter involve also
$\mathcal{O}(\lambda_5^2)$ pieces from the finite wave function
renormalization of the external Higgs boson legs. Starting from the
1-loop vertices of Fig.\,\ref{fig:vert} we can easily recognize them
as being part of some of the vertices in the 2-loop diagrams of
Fig.\,\ref{fig:2self}. When the vertices in the former collapse to a
point we obtain the kind of solid (colored) blobs indicated in
Fig.\ref{fig:composite}. These blobs represent static form factors
attached to the corresponding vertices, i.e. evaluated at zero
momentum, and are therefore appropriate to estimate the higher order
effects on the $\delta\rho$-parameter (even if they are actually not
suitable for addressing the corresponding effects on the full $\Delta
r$ parameter, see later on).

The practical recipe in our approach is now clear. Based solely on
power counting and dynamical considerations, the one-loop effects from
Fig.\,\ref{fig:vert} can be conveniently cast as
form factors of the guise:

\begin{eqnarray}
a_{\rm hhV} &\sim& \frac{1}{16\,\pi^2}\left(\frac{\lambda_{hhh}}{M_h}\right)^2\,
f_{\rm hhV}(p^2/M^2_h)\,(p\,\cdot \epsilon) \sim \mathcal{O}(\aew\,\lambda_5^2); \nonumber \\
b_{\rm hVV} &\sim& \frac{1}{16\,\pi^2}\left(\frac{\lambda_{hhh}}{M_h}\right)^2\,
f_{\rm hVV}(p^2/M^2_h)\,(\epsilon\,\cdot\, \epsilon) \sim \mathcal{O}(\aew\,\lambda_5^2); \nonumber \\
%
c_{\rm hGV} &\sim& \frac{1}{16\,\pi^2}\left(\frac{\lambda_{hhh}}{M_h}\right)^2\,
f_{\rm hGV}(p^2/M^2_h)\,(p\,\cdot\,\epsilon) \sim \mathcal{O}(\aew\,\lambda_5^2). \nonumber \\ \label{eq:formfactors}
\end{eqnarray}

%

These form factors carry the sought-for fingerprint of the Higgs boson
self-interactions, which we generically flag here as $[\lambda_{\rm
hhh}]$. The typical Higgs boson mass scale $M_{h}$ in
Eq.~\eqref{eq:formfactors} restores the corresponding dimensions and
accounts for the correct decoupling properties. At the same time, it
partially balances the enhancement effect of the Higgs
self-interactions in the numerator. By $f(p^2/m^2_h)$ we denote
a generic rational function, depending on the (ratios of the) relevant
scales involved in the Higgs boson-mediated one-loop diagrams, namely
the masses of the virtual particles and the transfered momentum. These
form factors we evaluate at zero momentum. Put another way, we reduce
the loop structure to a point-like interaction in which
$\lambda^2_{hhh}\to\lambda_5^2$ appears as the only effective coupling
in the large $\lambda_5$ limit. Finally, $1/16\pi^2$ stands for the
usual numerical factor from the one-loop integrals. We carry out this
calculation with the help of { \sc FormCalc}~\cite{formcalc}.
Explicit analytical details we provide in the Appendix, cf. Eqs.~\eqref{eq:formfactor1}-\eqref{eq:selfh0H0}.
Equivalent form
factors could of course be entertained for the MSSM case. However, in
stark contrast to the general 2HDM, they would show a pure ${\cal
O}(g^3)$ gauge structure and would therefore be inconspicuous from a
phenomenological viewpoint, in the sense that no characteristic MSSM
signature could easily emerge from them.

Noteworthy is also the following. The one-loop form factors
\eqref{eq:formfactors} computed this way are UV finite and gauge
invariant. The finiteness follows by simple power counting. The gauge
invariance ensues from the fact that we consistently retain all
$\mathcal{O}(\aew\lambda^2_5)$ contributions -- but not more. It
should be clear that it is because of our exclusive selection of the
characteristic $\lambda_5^2$ tag in these diagrams that we are granted
to successfully isolate a meaningful (gauge invariant) and finite
contribution. By the same token the sum of all the original 2-loop
diagrams as such carrying the $\lambda_5^2$ tag (cf.
Fig.\,\ref{fig:2self}\, for a representative sample) is gauge
invariant and finite.  Had we kept also the full dependence of the
trilinear couplings on the Higgs and gauge boson masses, we would have
been forced to take into account the remaining set of contributions
not only from the Higgs bosons, but also from the gauge sector (viz.
the gauge and Goldstone boson-mediated diagrams) in order to insure
the overall gauge invariance of the Higgs/gauge form factors
\eqref{eq:formfactors}. The fact that we single out these
$\mathcal{O}(\lambda_5^2)$ pieces from the 3h self-interactions could
be viewed as if we were effectively decoupling the dynamics of the
Higgs and the gauge sectors of the 2HDM. Of course this is not a
\emph{decoupling} in the usual sense, namely it is not due to the
presence of a hierarchy of masses -- notice, indeed, that the mass
spectra under examination typically feature $M_V \lesssim M_{\rm H}$.
Rather than ``decoupling'' we are ``detaching'' the
$\lambda_5$-effects from the rest of the quantum contributions in a
consistent way. There is actually some independence at the level of
QFT renormalizability between, say, the gauge and Yukawa sectors on
the one side, and the $\lambda_5$-structure of the Higgs boson
self-interactions on the other side. These sectors contain independent
parameters (gauge and Yukawa couplings in the the former versus
trilinear couplings in the latter). They all carry some characteristic
renormalizability ``flagpoles'', which can nevertheless be mixed. For
example, in the case of the Higgs self-interactions there is one part
that still carries the gauge flagpole, but there is another which does
not (the $\lambda_5$-part). It is precisely this part that has no
crosstalk with the others since it is gauge invariant and finite \emph{per
se}. As a result there can be, in principle, an arbitrary hierarchy of
strengths between the Higgs self-couplings and the (merely gauge)
Higgs/gauge and Higgs/fermion interactions. This is what allows to
detach one sector from the other without jeopardizing in any possible
way the QFT renormalizability  of the theory. In turn, this is what
authorizes the 2HDM Higgs bosons to couple to each other much stronger
than they do to the gauge bosons. Such property is impossible in the
MSSM, where the gauge flagpole is inherent (owing to the underlying
supersymmetry) to the entire structure of the Higgs self-interactions.

To summarize: in our calculation the $\lambda_5$ parameter plays the
role of a flagpole which marks an independent sector of the theory. By
selecting it we can factorize a gauge invariant and UV-finite piece
from the entire calculation. In the limit of large $\lambda_5$ such
well-defined piece encapsulates the dominant quantum effects on the
tree-level Higgs/gauge interactions.}

\item{Finally, we can fold these tree-level Higgs/gauge interaction
    Lagrangians with the form factors $[a_{\rm hhV}$, $b_{\rm hVV}$,
    $b_{\rm hGV}]$ derived above and build up a corresponding
    set of Born-improved Lagrangians:

\begin{eqnarray}
 \lag^{\rm eff}_{\rm hhV} &=& g_{\rm hhV}\,
 (1 + a_{\rm hhV})\,\left[ (\partial_\mu h)\,V^{\mu}\,h - h\,V^{\mu}\,\partial_\mu h \right]; \nonumber \\
 \lag^{\rm eff}_{\rm hVV} &=& g_{\rm hVV}\,(1 + b_{\rm hVV})\,h\,V_\mu V^\mu\, ; \nonumber \\
 \lag^{\rm eff}_{\rm hVG} &=& g_{\rm hVG}\,(1 + c_{\rm hVG})\,\left[ (\partial_\mu h)\,V^{\mu}\,G - h\,V^{\mu}\,\partial_\mu G \right]\nonumber\,.
 \\ \label{eq:efflag1}
\end{eqnarray}

The general structure
of the associated form factors can be illustrated e.g. in the case of the
effective $\hzero\Azero\PZ^0$ coupling,

 \begin{eqnarray}
 a_{\hzero\Azero\PZ^0}\Big{]}_{p^2=0} &=&
 {\Re e}\,V_{\hzero\Azero\PZ^0}\, - \frac{1}{2} {\Re e}\,\Sigma'_{\hzero} \nonumber \\
 &-& \frac{1}{2} {\Re e}\,\Sigma'_{\Azero} -
 \frac{\tan(\beta-\alpha)}{M^2_{\Hzero}}\,{\Re e}\,\retildehat_{\hzero\Hzero}\Big{]}_{p^2=0}, \nonumber \\
 \label{eq:formfactor1-prev}
 \end{eqnarray}

 \noindent where the
 explicit form of the two and three point functions is given in the
 Appendix.}


\end{enumerate}

\begin{figure*}
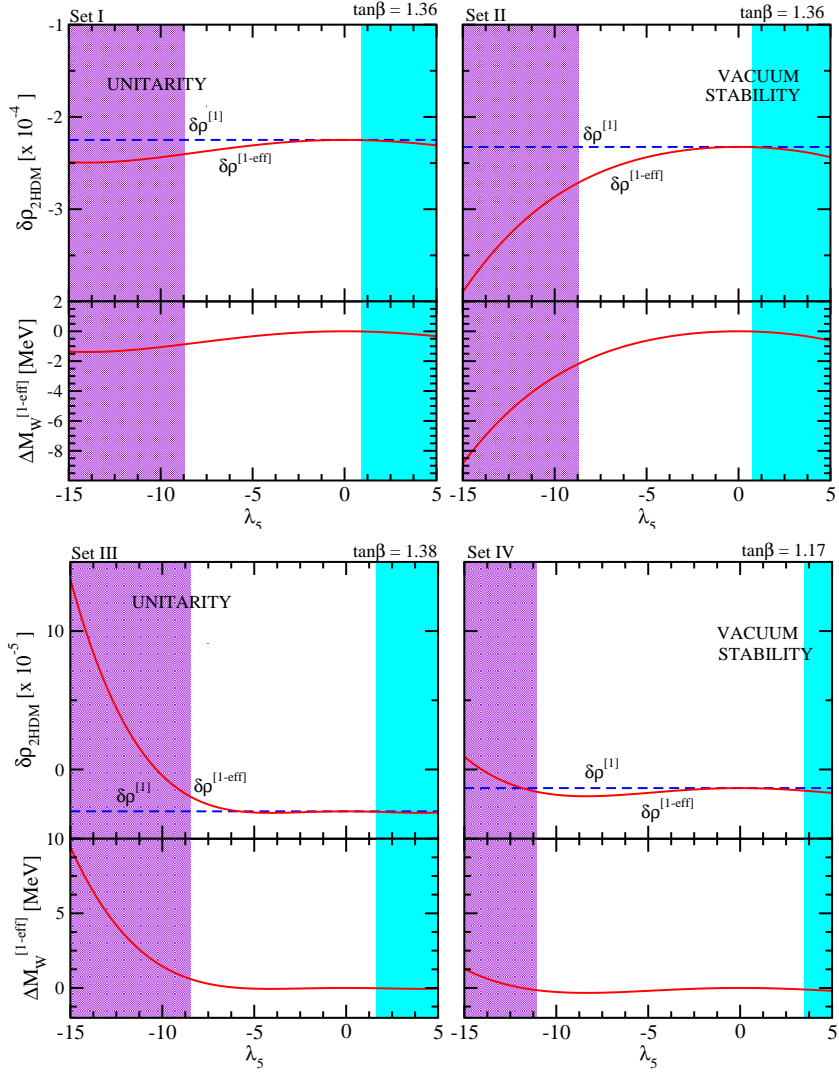
 \begin{center} \begin{tabular}{c}
 \includegraphics[scale=0.4]{overl5-one} \\
 \includegraphics[scale=0.4]{overl5-two}
\end{tabular}
 \caption{2HDM contribution to the parameter $\delta\rho$, as a function
of the self-coupling $\lambda_5$, for different sets of Higgs boson
masses. We superimpose the pure one-loop [$\delta\rho^{\rm [1]}$] and the
improved one-loop [$\delta\rho^{\rm [1-eff]}$] results. The latter trades
the dominant higher order effects from the enhanced Higgs boson
self-interactions. The quantitative impact of these higher order
corrections on the prediction of the W-boson mass is displayed in the
lower panels. The latter we compute via the contribution from
$\delta\rho^{\rm [1-eff]}$ to $\Delta r$, see Eq.\,(\ref{eq:mwestimate}).
For each set of Higgs boson masses, the value of $\tan\beta$ is chosen
such that the 3h self-couplings maximize for the largest allowed
$|\lambda_5|$ values. The shaded areas account for the unitarity (left)
and vacuum stability bounds (right). \label{fig:deltarho}} \end{center}
\end{figure*}

Additional momentum-dependent tensor structures
$\sim p_\mu\,p_\nu\,V^\mu\,V^\nu$ within the hVV form factor do not
include $\mathcal{O}(\lambda^2_5)$ contributions and are henceforth
disregarded in our treatment of $\delta\rho$ at higher orders. The idea
of absorbing the bulk of the quantum effects into \emph{improved}
leading-order effective interactions ~\eqref{eq:efflag1} is of course not
new.
A well known example is the enhanced
bottom/sbottom Yukawa coupling in the large $\tan\beta$ limit of the MSSM
\cite{Carena:1999py} -- see also \cite{Coarasa:1996qa}. The approximation
will be meaningful as long as we probe these effective couplings at
momentum scales significantly smaller than the typical Higgs boson masses.
For quantities which are momentum-dependent, though, such effective
coupling treatment may no longer be suitable. But in our case the approach
is fully justified, as we are using this approximation only to estimate
some higher order effects on the $\delta\rho$ parameter, a static
prominent piece of $\Delta r$, namely $\Delta r^{[\delta\rho]} \equiv
-(\cwd/\swd)\delta\rho$, for which the gauge boson self energies are to be
evaluated at zero momentum -- cf. Eq.~\eqref{eq:deltarho}. This is the
reason why the effective coupling treatment elaborated here is not
directly exportable to e.g. a full-fledged calculation of $\Delta r$. As
we will discuss in more detail, the $\Delta r^{\rm [\delta\rho]}$
component proves insufficient to completely describe the quantum effects from the
Higgs bosons, which means that the momentum-dependent part of the gauge
boson self energies (evaluated on-shell $p^2 = M^2_V$) cannot be
neglected, and in fact may be quantitatively relevant.

\begin{table*} \begin{center}
 \footnotesize{ \begin{tabular}{|l||c|c|c|c|c|c|} \hline
\multirow{2}{*}{ } & Set I & Set I & Set II & Set II & Set III & Set IV \\
  & $\alpha = \beta-\pi/2$ & $\alpha=0$ & $\alpha = \beta$ & $\alpha=\pi/2$ & $\alpha = \beta-\pi/2$ & $\alpha = \beta-\pi/2$ \\ \hline
$\Delta r_{\rm SM}\,[\times10^{-3}]$ & 37.744 & 37.744 & 37.443 & 37.443 &
37.744 & 37.744 \\ \hline
$\Delta r^{\rm [1]}_{\rm 2HDM}[\times 10^{-3}]$ & 36.701 & 36.650 & 36.352
& 36.409 & 35.845 & 35.603 \\ \hline
$\delta(\Delta r^{\rm [1]}_{\rm 2HDM})\,[\times 10^{-3}]$
 & -1.043 & -1.094 & -1.091 & -1.034 & -1.899 & -2.141 \\ \hline
$|\delta(\Delta r^{[1]}_{\rm 2HDM})/\Delta r^{\rm [1]}_{\rm 2HDM}|
\,[\%]$& 2.8 & 3.0 & 3.0 & 2.8 & 5.3 & 6.0 \\
\hline\hline
$\delta\rho^{\rm [1]}$ [$\times 10^{-4}$] & -2.249 & -2.192 & -2.325 &
-2.387 & -0.302 & -0.134 \\ \hline
$\delta\rho^{\rm [1-eff]}$ [$\times 10^{-4}$] & -2.403 & -2.348 & -2.709 &
-3.014 & -0.197 & -0.160 \\
\hline $|\Delta(\delta\rho^{\rm eff})/\delta\rho^{\rm [1]}|$ $[\%]$ & 6.8
& 7.1 & 16.5 & 26.3 & 35.0 & 19.0 \\ \hline $\delta M^{\rm
[1]}_{\PW}$ [MeV] & 17.257 & 18.078 & 18.021 & 17.102 & 31.071 & 34.967 \\
\hline
$\Delta M^{\rm [1-eff]}_{\PW}$ [MeV]& -0.865 & -0.880 & -2.166 & -3.537 &
0.596 & -0.144 \\ \hline \hline
$\tan\beta$ & 1.36 & 1.36 & 1.36 & 1.36 & 1.38 & 1.17 \\ \hline
$\lambda_5$ & -8.68 & -8.68 & -8.68 & -8.68 & -8.46 & -11.08 \\ \hline
  \end{tabular}}
 \caption{ Detailed numerical analysis of the different electroweak
quantities under survey, these are: $\Delta r$, $\delta\rho$ and
$M_{\PW}$. We examine different choices of Higgs boson masses (cf.
Table~\ref{tab:masses}) and trigonometric couplings. The values of
$\tan\beta$ and $\lambda_5$ (see the bottom rows of the table)
maximize the enhanced higher order effects induced by the Higgs
boson self-interactions. The notation $\delta(\Delta r^{[1]}_{\rm
2HDM})$ spells out the ``genuine'' one-loop 2HDM effects (i.e. after
consistent subtraction of the SM part) -- cf.
Eqs.~\eqref{eq:deltar-def4} - \eqref{eq:deltar-def42}. In turn,
$\delta\rho^{\rm [1]}$ and $\delta\,M^{[1]}_{\PW}$ denote the
one-loop shifts on these parameters from those genuine 2HDM one-loop
effects, while $\Delta(\delta\rho^{\rm eff})$ and $\Delta M^{\rm
[1-eff]}_{\PW}$ represent the corresponding higher order effects on
$\delta\rho$ and $M_W$ beyond one-loop, computed in our
approach --  cf. Eqs.
\eqref{eq:efflag1} and \eqref{eq:mwestimate}.} \label{tab:deltarho}
\end{center} \end{table*}

Following the above method, an estimate of $\delta\rho$
beyond one-loop -- viz. including the dominant
$\mathcal{O}(\aew\,\lambda^2_5)$ terms -- is now at reach with a moderate,
albeit still non-trivial, amount of work as compared to the full 2-loop
calculation.
The obtained numerical results are presented in the upper panels of
Fig.~\ref{fig:deltarho}. Here we plot the evolution of $\delta\rho$
versus the Higgs self-coupling $\lambda_5$ and for the Sets I-IV of
Higgs boson masses quoted in Table~\ref{tab:masses}. For each set,
we fix $\tan\beta$ so that the quantum effects governed by the
enhanced 3h self-interactions maximize for the largest attainable
$|\lambda_5|$ values within bounds. Alongside the pure 2HDM one-loop
effects [$\delta\rho^{[1]} \equiv \delta\rho^{[1]}_{\rm 2HDM}$] --
which correspond to the flat, dashed line -- we superimpose the
improved, $\lambda_5$-dependent, curves [$\delta\rho^{[1-\rm eff]}$]
derived via Eqs.~\eqref{eq:efflag1}. By dialing the
parameter $\lambda_5$ we portray the dependence of $\delta\rho^{[\rm
1 - eff]}$ on the Higgs boson self-interaction enhancements. With
decreasing $\lambda_5$ values, the difference
$\Delta(\delta\rho^{\rm eff}) \equiv (\delta\rho^{\rm
[1-eff]}-\delta\rho^{\rm [1]})$ obviously goes to zero -- and in
this case the curves $\delta\rho^{[1]}$ and $\delta\rho^{\rm
[1-eff]}(\lambda_5)$ tend to merge.

A complementary numerical account is provided in Table~\ref{tab:deltarho}.
Here we consider Sets I-IV of Higgs boson masses from
Table~\ref{tab:masses}, with corresponding SM-like couplings for the $\sim
125$ GeV \CP-even Higgs boson ($\hzero$ or $\Hzero$) state in all the
cases. Moreover, we also investigate the particular instances $\alpha=0$
(for Set I) and $\alpha=\pi/2$ (for Set II), these are the so-called
\emph{fermiophobic} Higgs limits of type-I 2HDM. Most significantly, in
this table we quantify the maximum attainable departures
$|\Delta(\delta\rho^{\rm eff})|$ from the plain one-loop predictions. These
optimal scenarios are realized for particular choices of $\tan\beta$ and
$\lambda_5$ (as quoted in the last rows of the table), saturating the 2HDM
unitarity bounds\cite{unitarity}. Such configurations are preferably
achieved for small $\tan\beta \sim 1-2$ and moderate (negative) $\lambda_5
\sim -5/-10$.


These results clearly spell out the physical meaning of $\delta\rho$ as a
measure of the $SU(2)$ custodial symmetry violation. The plain one-loop
prediction for $\delta\rho$ ranges from $\mathcal{O}(10^{-4})$ (for Sets I
and II, viz. for relatively light Higgs bosons with unconstrained mass
splittings) to $\mathcal{O}(10^{-5})$ (for Sets III and IV, wherein the
heavier Higgs bosons are tailored to mimic the mass splittings of
SUSY-like spectra). We may track down this same feature too from the right panels of
Fig.~\ref{fig:newsetsrho-overl5}, in which we again superimpose the mere
one-loop prediction [$\delta\rho^{\rm [1]}$] and the improved one
[$\delta\rho^{\rm [1-eff]}$], as a function of $\lambda_5$, for the
complementary mass Sets V-VIII. The sizable value of $\delta\rho$, at the
$\mathcal{O}(10^{-3})$ level for Set V (this means nearly overshooting the
custodial symmetry limit $|\delta\rho|\lesssim 10^{-3}$) depletes very
remarkably when the heavier Higgs boson masses -- and so the different
mass splittings -- are pulled down all the way from Set V to Set VIII
(cf. Table~\ref{tab:masses}).

\begin{figure*} \begin{center} \vspace{0.3cm}
\includegraphics[scale=0.5]{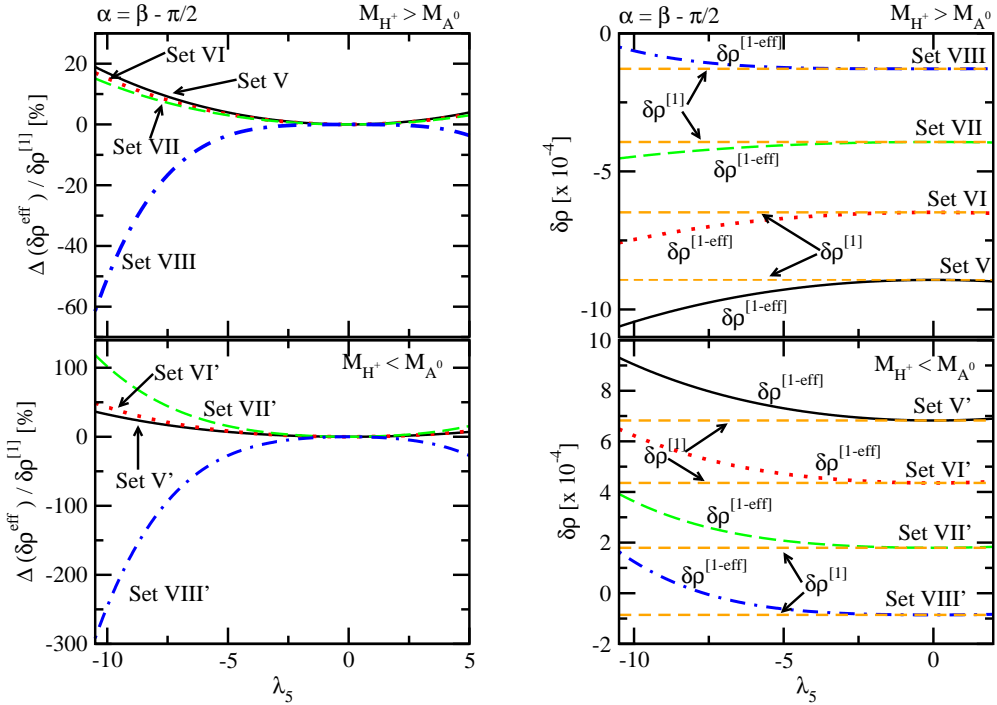}
\caption{Impact of the leading higher order corrections to $\delta\rho$.
On the left panels we evaluate their influence through the relative
departure [$\Delta(\delta^{\rm eff})/\delta\rho^{\rm [1]} \equiv
(\delta\rho^{\rm [1-eff]} - \delta\rho^{\rm [1]})/\delta\rho^{\rm [1]}\,[\%]$] as
a function of the self-coupling $\lambda_5$. We consider the Higgs boson
mass sets from Table~\ref{tab:masses} with both direct (Sets V-VIII, top
panel) and inverted (Sets V'-VIII', bottom panel) mass hierarchies between
the $\Azero$ and the $\PHiggs^{\pm}$ fields. We set $\alpha = \beta-\pi/2$
and fix $\tan\beta$ as quoted in Table~\ref{tab:masses}, so that the 3h
self-coupling enhancements maximize for a given $\lambda_5$ -- saturating
the unitarity bounds for each of the mass sets, cf. Table~\ref{tab:max}.
Complementarily, on the right panels we present the absolute value of
$\delta\rho$, again as a function of $\lambda_5$ and for the different
sets of masses. In each case we superimpose the mere one-loop results
[$\delta\rho^{\rm [1]}$] and those including the dominant higher-order
corrections [$\delta\rho^{\rm [1-eff]}(\lambda_5)$]. Let us notice that unitarity
and vacuum stability constraints are not explicitly included in the figure;
the phenomenologically viable $\lambda_5$ range for each of masses is
explicitly quoted in Table~\ref{tab:max}.} \label{fig:newsetsrho-overl5} \end{center}
\end{figure*}

\begin{table} \begin{center} \begin{tabular}{|l|cc|} \hline & Set V & Set
VIII \\ \hline
$a_{\Hzero\PHiggs^+\PW^-}$ & $8.06\times 10^{-2}$ & $1.56\times 10^{-1}$ \\
$a_{\Hzero\Azero\PZ^0}$ & $6.21\times 10^{-2}$ & $1.70\times 10^{-1}$ \\
$a_{\Azero\PHiggs^+\PW^-}$ & $1.47\times 10^{-3}$ & $1.16\times 10^{-2}$ \\
 $b_{\hzero\PZ^0\PZ^0}$ & $2.60\times 10^{-2}$ & $3.70\times 10^{-2}$ \\
$b_{\hzero\PW^+\PW^-}$ & $2.60\times 10^{-2}$ & $3.70\times 10^{-2}$ \\
\hline
\end{tabular}
\caption{Coupling strengths $a_{\rm hhV}$ and $b_{\rm hVV}$ -- cf.
Eq.\,\eqref{eq:formfactors} -- for a representative set of hhV and hVV
effective interactions. We compare them numerically for Sets V and VIII
(cf. Table~\ref{tab:masses}), assuming $\alpha=\beta-\pi/2$ and fixing
$\tan\beta$ and $\lambda_5$ such that the triple Higgs boson self-coupling
enhancements maximize -- in a way compatible with all the bounds. }
\label{tab:coupsize} \end{center} \end{table}
\smallskip{}

By the same token, the relative importance of the leading higher order
corrections to $\delta\rho$ becomes drastically promoted in some cases (at
the highest available values of $\lambda_5$), as we quantify explicitly on
the left panels of Fig.~\ref{fig:newsetsrho-overl5}. This is in part
because these higher order quantum effects are added to a decreasing
one-loop piece [$\delta\rho^{\rm [1]}$] -- which falls down by roughly one
order of magnitude if we compare again Sets V and VIII. At the same time, the boost in the
relative departure $\Delta(\delta\rho^{\rm eff})/\delta \rho^{\rm [1]}$ is
also partly explained due to the smaller mass suppression of the one-loop
Higgs-mediated effects. In Table~\ref{tab:coupsize} we settle this
statement quantitatively, by comparing the actual coupling strength of a
number of effective interactions for Sets V and VIII -- this is to say,
when we assume heavier (resp. lighter) Higgs bosons (cf.
Table~\ref{tab:masses}).

\begin{figure*} \begin{center}
\begin{tabular}{c}
  \includegraphics[scale=0.45]{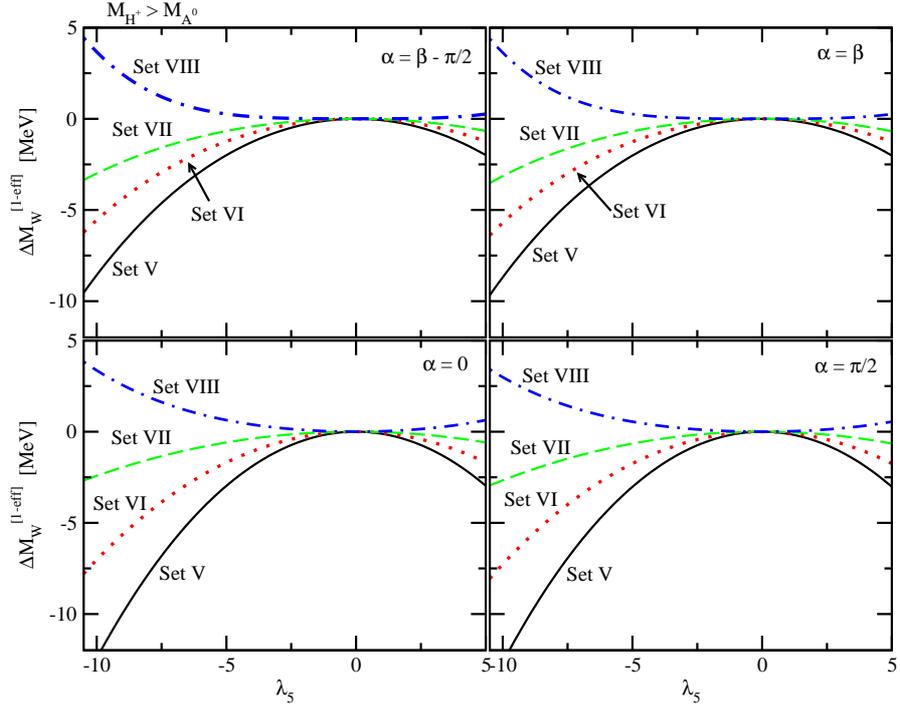}
\end{tabular}
\caption{Overall shift in the theoretical prediction for the W-boson mass
(cf. Eq.~\eqref{eq:mwestimate}), driven by the leading Higgs-mediated higher-order
corrections, as a function of the self-coupling $\lambda_5$. The results
are displayed separately for Sets V-VIII (cf. Table~\ref{tab:masses}) with a
\emph{direct} mass hierarchy [$M_{\PHiggs^{\pm}} - M_{\Azero} > 0$], and
for different choices of the mixing angle $\alpha$. We fix $\tan\beta$ as
quoted in Table~\ref{tab:max}, so that the 3H self-coupling enhancements
maximize at the largest allowed $|\lambda_5|$ values -- also given in the table.}
\label{fig:newsetsmass-overl5-one} \end{center} \end{figure*}

We can also rephrase these results as a relative contribution to $\Delta
r$ (cf. Eq.~\eqref{eq:DeltarGeneral}), namely via the term $ \Delta
r^{[\delta\rho]}$, of the order of $\sim 5\%$ to $\sim 50\%$ to the
overall 2HDM prediction [$\Delta r_{\rm 2HDM}$]. This quantity grows along
with the mass splittings between the different Higgs bosons. That explains
e.g. why, for Sets I and II, these $\delta\rho$-driven contributions are
more sizable than those for the SUSY-like Sets III and IV (that stagnate
at the $\mathcal{O}(10^{-5})$ level). In contrast, for the last two sets
the total $\Delta r$ values are larger. This simply means that the
\emph{on-shell} gauge boson self-energies or, in other words, the
non-static contributions to $\Delta r$, are more relevant here, as
compared to the self-energies evaluated at \emph{zero momentum}. The
induced shift (from strict one-loop 2HDM effects) on the W-boson mass
prediction [$\Delta M^{\rm [1]}_{\PW}$] is correspondingly larger for the
SUSY-inspired mass sets. Specifically, $\Delta M^{\rm [1]}_{\PW} \sim 15
\,\MeV$ for Sets I and II, versus $\Delta M^{\rm [1]}_{\PW} \sim 30
\,\MeV$ for Sets III and IV -- cf. Table~\ref{tab:deltarho} and
Fig.~\ref{fig:mw}. This is not surprising, after all. As we have
already analyzed in Section~\ref{sec:analysis}, the MSSM tends to relax
--sometimes very substantially -- the existing tension between the SM
theory predictions and the experimental measurements of electroweak
precision quantities \cite{Heinemeyer:2006px}. We may turn around this
observation and say -- emphasizing once more the claim made in Sec.
\ref{sec:analysis} -- that our study shows that the general 2HDM can
perform equally well than the MSSM in the task of fostering the agreement
between the theoretical prediction of the $\PW$-mass versus its
experimentally measured value.

In this respect, it is interesting to emphasize again that in the MSSM case the
Higgs bosons alone induce a rather moderate contribution to the EW
quantities $\Delta r$ and $\delta\rho$, as compared to the situation when
the squark, slepton and chargino-neutralino sectors are not very heavy.
Having already dwelled on $\Delta r$ and $M_{\PW}$ in
Section~\ref{sec:analysis}, let us now quantify the typical size of the
MSSM contributions to $\delta\rho$. Again, we start by considering Set III
of Higgs boson masses (cf. Table~\ref{tab:masses}), based on the
(\emph{maximal mixing}) MSSM benchmark -- which comprises
$\mathcal{O}(100)$ GeV charginos and neutralinos, $\sim 700$ GeV stops,
large Higgs-stop trilinear couplings, and the remaining sfermions above 1
TeV. Employing {\sc FeynHiggs} \cite{\citeFH} to extract the full MSSM
contribution to $\delta\rho$ within this scenario, we find
$\delta\rho_{\rm MSSM} = 2.186 \times 10^{-4}$. In addition we can
separately evaluate the contribution driven by the MSSM Higgs sector alone
by means of our 2HDM calculation, namely by plugging the mixing angle
value $\alpha_{\rm MSSM} = -0.188$ which corresponds to that specific MSSM
benchmark. Once more we can use {\sc FeynHiggs} for its numerical
evaluation, and we obtain $\delta\rho_{\rm MSSM}^{\rm Higgs} = -2.910
\times 10^{-5}$. This result clearly illustrates how the pure SUSY
Higgs-induced contribution to $\delta\rho$ falls roughly one order of
magnitude below the total MSSM budget. The SUSY embedding for Set IV, in
turn, features much heavier [viz. $\mathcal{O}(500)$ GeV] weak gauginos,
with $\sim 1 - 2 $ TeV sfermion masses and weaker Higgs-stop couplings. In
this case we find $\delta\rho_{\rm MSSM} = 5.211 \times 10^{-6}$ for the
total payoff of the MSSM,  whilst $\delta\rho_{\rm MSSM}^{\rm Higgs} =
-1.338 \times 10^{-5}$ for the specific SUSY Higgs part. As expected, the
effects from the more massive SUSY degrees of freedom are relatively
suppressed and the total contribution from sfermions and
chargino-neutralinos is, in this particular case, comparable (albeit
opposite in sign) to that of the Higgs bosons, so the overall MSSM yield
is significantly smaller than in the previously discussed case.

Let us now come back to the role of the leading higher-order corrections
governed by enhanced Higgs boson self-couplings of the 2HDM. We spotlight
significant effects which, in some particular corners of the parameter
space, may well trigger relative deviations with respect to the one-loop
prediction as large as $\Delta(\delta\rho^{\rm eff})/\delta\rho^{[\rm
1]}\sim \mathcal{O}(30)\%$ or above. Quite remarkably, these numerical
results could be foreseen from the rough analytical estimate
\begin{equation}
 \delta\rho^{\rm [1-eff]}_{\rm 2HDM} \simeq \delta\rho_{\rm 2HDM}^{[1]}\,
\left(1+ \frac{1}{16\,\pi^2}\frac{|\lambda_{\rm hhh}|^2}{M^2_{h}}\right)^2
\label{eq:estimate},
\end{equation}
which follows from the approximate form factors quoted in
Eq.~\eqref{eq:formfactors}.
On the one hand, the triple Higgs boson self-couplings may reach maximum
values of $\lambda_{\rm hhh} \sim \mathcal{O}(10^{3})$ GeV, as we can read
off Fig.~\ref{fig:values} and Eq.\,\eqref{eq:extrilinear}; and, on the
other, the typical Higgs boson masses considered in our analysis are of
$\mathcal{O}(100)\,\GeV$. Plugging these numbers into
Eq.~\eqref{eq:estimate} we indeed retrieve $\Delta(\delta\rho^{\rm eff})/
\delta\rho^{\rm [1]} \sim 50-100\%$, which agrees with the outcomes of our
full numerical calculation to a remarkable extent.

\begin{figure*} \begin{center}
\begin{tabular}{c}
  \includegraphics[scale=0.45]{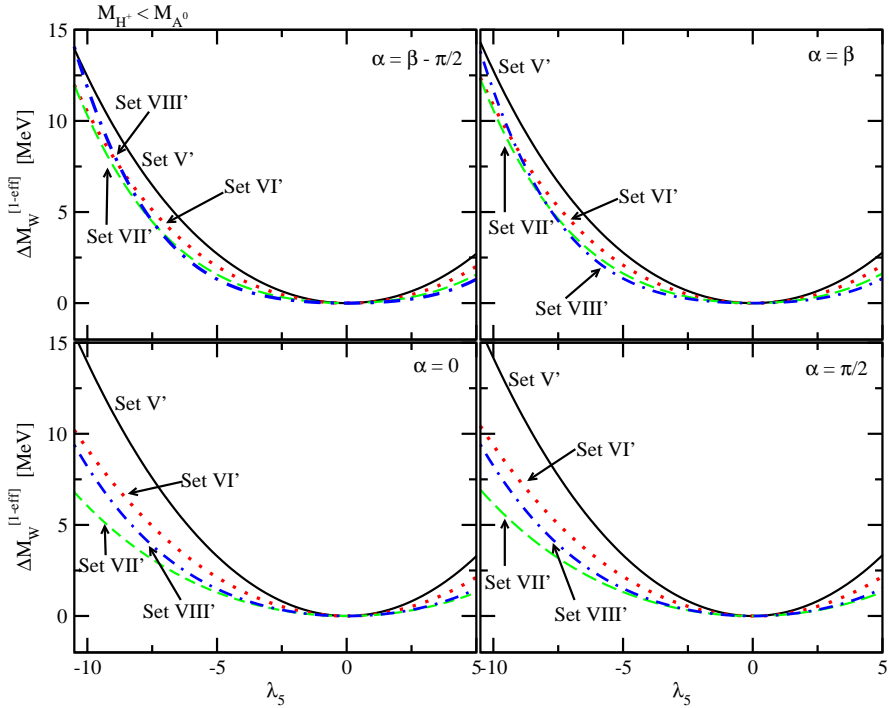}
\end{tabular}
\caption{Analogous setup and results as in
Fig.~\ref{fig:newsetsmass-overl5-one}, for Sets V'-VIII' of Higgs boson
masses with an \emph{inverted} mass hierarchy [$M_{\PHiggs^{\pm}} -
M_{\Azero} < 0$].} \label{fig:newsetsmass-overl5-two} \end{center} \end{figure*}

The parametric dependence on $\lambda_5$, as displayed in
Figs.~\ref{fig:deltarho}-\ref{fig:newsetsmass-overl5-two}, exhibits
the expected behavior $\delta\rho \sim (A\lambda_5^4 +
B\,\lambda_5^2 + C)$, which results from the structure of the
(Born-improved) Higgs/gauge boson couplings $g^{\rm eff}_{hhV} \sim
g\,(1 + \mathcal{O}(\lambda^2_5))$ -- entering of course square in
the gauge boson self energies. It is for this reason that the
leading correction to $\delta\rho$ is of ${\cal O}(\lambda_5^4)$.
Evidently this correction goes beyond the 2-loop level and therefore
it tests in  an effective way the largest possible effects that can
be expected from the $\lambda_5$ parameter at all orders. This kind
of effective approach is not new, let us note that it is fully
within the line of reasoning presented in the work of
Ref.\,\cite{LopezVal:2009qy}, which was devoted to compute not only
the full one-loop effects but also to test the largest possible
quantum corrections that the trilinear Higgs boson self-couplings
could produce at all orders in the pairwise production of neutral
2HDM Higgs bosons in a linear collider. We remark that the
one-loop diagrams used in that calculation are essentially identical
to those we are using here to construct the improved couplings
\eqref{eq:efflag1} for the present calculation, i.e. we are
proceeding along the same philosophy as in
Ref.\,\cite{LopezVal:2009qy}.

%

The leading large $\lambda_5$ corrections, if effectively realized in
nature, would also manifest as a shift  $\Delta M_W^{\rm [1-eff]}$ on the
W-boson mass prediction. This shift we can roughly estimate by means of
$\Delta r^{[\delta\rho]}$, whereby
\begin{eqnarray}
 \big{|}\Delta M_{\PW}^{\rm [1-eff]}\big{|} &\equiv& |\delta\,M_{\PW}^{\rm [1 - eff]}-\delta\,M_{\PW}^{\rm [1]}|
\nonumber \\
&\simeq& \frac{M_{\PW}}{2}\, \frac{\cwd}{\cwd-\swd}\,\times
\Big{\lvert}\delta\rho^{\rm [1-eff]} - \delta\rho^{[1]}
\Big{\lvert}\nonumber \\ &\simeq&  5.67\times 10^4\,[\MeV]\,
\Delta(\delta\rho^{\rm eff})
\label{eq:mwestimate}.
\end{eqnarray}

\begin{table*} \begin{center}
 \footnotesize{\begin{tabular}{|c|cc|c|cc|cc|cc|cc|} \hline
\multicolumn{4}{|c|}{\textbf{$M_{\PHiggs^{\pm}} > M_{\Azero}$}}   &
\multicolumn{2}{c|}{$\alpha=\beta-\pi/2$} & \multicolumn{2}{|c|}{$\alpha =
\beta$} & \multicolumn{2}{c|}{$\alpha = \frac{\pi}{2}$} &
\multicolumn{2}{c|}{$\alpha = 0$}\\ \hline &  $\lambda_5^{\rm min}$ &
$\lambda_5^{\rm max}$ & $\tan\beta$ & $\frac{\Delta(\delta\rho^{\rm
eff})}{\delta\rho^{\rm [1]}}$ & $\Delta M^{\rm 1-eff}_{\PW}$ &
$\frac{\Delta(\delta\rho^{\rm eff})}{\delta\rho^{\rm [1]}}$ & $\Delta
M^{\rm 1-eff}_{\PW}$ & $\frac{\Delta(\delta\rho^{\rm
eff})}{\delta\rho^{\rm [1]}}$ & $\Delta M^{\rm 1-eff}_{\PW}$
& $\frac{\Delta(\delta\rho^{\rm eff})}{\delta\rho^{\rm [1]}}$ & $\Delta M^{\rm 1-eff}_{\PW}$ \\
& &  &  & $[\%]$ & [MeV] & $[\%]$ & [MeV] & $[\%]$ & [MeV] & $[\%]$ &
[MeV]
\\ \hline
Set V & -10.82 & 0.93 & 1.19 & 20.22 & -10.19 & 20.51 & -10.34 & 29.76 & -15.00 & 29.15 & -14.70 \\
Set VI& -9.55& 0.87& 1.29  & 13.77 & -5.04 & 14.10 & -5.16 & 18.21 & -6.66 & 17.53 & -6.41 \\
Set VII & -8.04& 0.81& 1.42  & 8.30 & -1.84 & 8.57 & -1.91 & 7.76 & -1.72 & 6.97 & -1.55 \\
Set VIII& -6.55 & 0.76& 1.58 & -9.96 & 0.72 & -3.37 & 0.71 & -14.00 & 1.02
& -16.26 & 1.18 \\ \hline
\multicolumn{4}{|c|}{\textbf{$M_{\PHiggs^{\pm}} < M_{\Azero}$}}  &
\multicolumn{2}{c|}{}
 & \multicolumn{2}{c}{}  & \multicolumn{2}{c}{}  & \multicolumn{2}{c|}{} \\ \hline
Set V'  & -9.83 & 0.88& 1.27 & 31.46 & 12.07 & 31.86 & 12.38 & 35.64 & 13.78 & 35.08 & 13.52\\
Set VI'& -8.37& 0.83& 1.39 & 27.36 & 6.71 & 27.68 & 6.93 & 25.75 & 6.40 & 25.32 & 6.25\\
Set VII' & -6.85& 0.77& 1.55 & 34.93 & 3.53 & 24.65 & 6.17 & 23.39 & 5.82 & 25.00 & 2.57\\
Set VIII' & -5.40 & 0.72& 1.75 & -34.04 & 1.64 & -39.28 & 1.68 & -40.62 &
1.79 & -37.78 & 1.77 \\ \hline
\end{tabular}}
\caption{Parameter space survey for Sets V-VIII and V'-VIII' in
Table~\ref{tab:masses}. The left-most columns quote the allowed
$\lambda_5$ range, in agreement with the constraints from
perturbative unitarity and vacuum stability. Alongside we indicate
the corresponding value of $\tan\beta$ that maximizes the Higgs
boson self-interaction enhancements for the largest allowed
$|\lambda_5|$. Within these optimal scenarios, and for different
choices of $\alpha$, in the right part of the table we quantify i)
the relative size of the higher order effects on the
$\delta\rho$-parameter, $\Delta(\delta\rho^{\rm
eff})/\delta\rho^{\rm [1]}$; and ii) the induced shift on the
$\PW$-boson mass prediction [$\Delta M^{\rm 1-eff}_{\PW}$ ], as
defined by Eq.~\eqref{eq:mwestimate}. \label{tab:max}} \end{center} \end{table*}

Obviously this formula follows the same pattern as \eqref{eq:shiftMW} up
to the factor $\cwd/\swd$ inherent to the definition of $\Delta
r^{[\delta\rho]}$. The results for this estimate of $\Delta M_W^{\rm
[1-eff]}$ are displayed in the lower panels of Fig.~\ref{fig:deltarho}.
Notice that departures from the mere one-loop order by an amount of
$\Delta(\delta\rho^{\rm eff})/\delta\rho^{[1]}\sim \mathcal{O}(30)\%$
translate, for the mass sets under consideration, into a mass shift of up
to $\Delta(\delta M_{\PW}) \sim \mathcal{O}(3)$ MeV -- as we can confirm
from Table~\ref{tab:deltarho} and from the lower panels of
Fig.~\ref{fig:deltarho}. The largest absolute higher-order corrections to
$\delta\rho$  -- attained of course for the maximum allowed (negative)
values of $\lambda_5$ -- are possible within non-SUSY-like Higgs boson
mass spectra (Sets I and II). For the SUSY-like Sets III and IV, however,
these effects are much more tamed, and amount to barely ${\cal O}(1)$ MeV.
Recall that despite the inconspicuous $\lambda_5$-yield in this case, the
full one-loop 2HDM correction on $\Delta r$ is larger for those sets and
therefore the overall impact is more sizeable (cf. third row of Table 2
and Figs. 3-4).

We also study the complementary Sets V-VIII and V'-VIII' (i.e. considering
both the direct and the inverted mass hierarchies) in
Figs.~\ref{fig:newsetsmass-overl5-one} and
\ref{fig:newsetsmass-overl5-two} respectively, and also in
Table~\ref{tab:max}. For each of the mass sets we have different lower --
and upper constraints on $\lambda_5$,  that we explicitly account in the
left-most columns of the table. The parameter $\tan\beta$ we fix
accordingly, so that the Higgs self-coupling enhancements maximize for the
largest allowed $|\lambda_5|$. The maximum attainable relative deviations
on the $\delta\rho$-parameter [$\Delta(\delta\rho^{\rm
eff})/\delta\rho^{\rm [1]}$], alonside with the associated shift on the
predicted W-boson mass [$\Delta M_{\PW} ^{\rm [1-eff]}$], are documented
in the right columns. The results fall in the very same ballpark as for
sets I-IV that we have examined previously. Again, the relative impact of
the higher order effects can reach $\Delta(\delta\rho^{\rm
eff})/\delta\rho^{\rm [1]} \sim \mathcal{O}(30\%)$, but in this case
dragging the W-boson mass predictions up to $\Delta M_{\PW} ^{\rm
[1-eff]}\simeq 10-15$ MeV in those regimes (viz. Sets V and V') which
saturate the custodial symmetry bounds $|\delta\rho| \lesssim 10^{-3}$.
For this reason, within these complementary mass sets we identify
corrections to $M_{\PW}$ that can be up to three times bigger than in the
original Sets I-IV. We should nonetheless point out that such remarkable
effects beyond one-loop critically rely on a very large departure from the
custodial symmetry limit, and hence are not favored by the EW precision
data. Barring these borderline situations, a milder custodial symmetry
breaking -- e.g. at the level $|\delta\rho| \sim \mathcal{O}(10^{-4})$ --
typically renders one loop corrections of up to $\delta M^{\rm [1]}_{\PW}\simeq 40$ MeV combined with higher order mass shifts
[$\Delta M_{\PW} ^{\rm [1-eff]}$] in the ballpark of
$\mathcal{O}(1-5)$ MeV. This result is nevertheless quite significant, as
it implies that the characteristic higher order 2HDM corrections can be a
few times larger than the pure bosonic $\mathcal{O}(\alpha^2_{ew})$
two-loop corrections (which perform at the $\lesssim1$ MeV level), and
even comparable to most higher order corrections within the SM. Finally,
they can be as large as the estimated theoretical uncertainty ($\delta
M^{\rm th}_{\PW} \simeq 4 $ MeV)
of the SM effects beyond the 2-loop order \cite{LEPEWWG}. In all cases the
results are seen to be very much responsive to $\lambda_5$, while they
barely depend on the precise choice of the mixing angle $\alpha$.


The results illustrate once more the compromise between the two different
trends that govern the overall impact of the higher order effects, and
that somehow counterbalance each other: i) heavier -- versus lighter
masses: the heavier the Higgs bosons, the weaker will be the loop-induced
effects they generate -- these shall be typically suppressed by inverse
powers of such masses; ii) wider -- versus narrower mass splittings; as we
have discussed, these will determine the size of the pure one-loop piece
[$\delta\rho^{\rm [1]}$], and hence of the consequent mass shift
[$\delta M_{\PW}^{\rm [1]}$] and the corresponding higher-order contributions. This explains e.g. why Set V is the most
responsive one, as it combines a sizable mass splitting between the
\CP-even, the \CP-odd and the charged Higgs fields; and yet relatively light
(viz. $M_h \sim 125$ GeV) neutral, \CP-even states.
%

\section{Discussion and Conclusions}
\label{sec:conclusions}

In this work we have revisited the traditional electroweak (EW)
parameter $\Delta r$ from the viewpoint of the general (unconstrained)
Two-Higgs-Doublet Model
(2HDM)\,\cite{Frere:1982ma,Bertolini:1985ia,\hollikold,Froggatt:1991qw}.
That quantity, whose definition -- cf. Eq.~\eqref{eq:deltar_def1} --
was introduced more than thirty years ago
\,\cite{Sirlin:1980nh,Marciano:1980pb}-- spells out the quantum link
between the EW gauge boson masses ($M_{\PW}, M_Z$) and the Fermi constant
($G_F$), and thereby enables an accurate theoretical prediction for the
mass of the W-boson [$M^{\rm th}_{\PW}$], based on the precise knowledge
of the $Z$-boson mass [$M_Z$], the fine structure constant [$\alpha$] and
Fermi's constant [$G_F$] as experimental inputs. Analyses of $\Delta r$
have been instrumental in the past, both as precision tests of the SM and
as a strategy to seek for hints of new physics, as well as to derive
constraints on the parameter space of theories beyond the SM.
After carefully reexamining the one-loop 2HDM contributions
($\Delta r_{\rm 2HDM}$) to the parameter $\Delta r$ at one loop, namely
at $\mathcal{O}(\aew)$, we have folded these effects with all the known
higher order SM corrections to that quantity in a consistent way within
the allowed region of the 2HDM parameter space.  Of course the overlap of
the SM Higgs boson contribution and the lightest neutral \CP-even state of
the 2HDM has also been consistently removed. With this procedure we have
been able to bring to date the numerical analysis of $\Delta r$ and the
corresponding prediction of the $W$-mass within the 2HDM in a way
comparable to the SM case.

In the numerical analysis we have scrutinized a variety of regimes with
phenomenological relevance and in full compliance with a brought-to-date
set of restrictions that severely constrain the parameter space of the
model -- most significantly the unitarity and vacuum stability conditions,
together with the low energy flavor physics observables $\mathcal{B}(b \to
s\gamma)$ and $B^0_d - \overline{B}^0_d$. Needless to say, we have included
as well the important phenomenological implications associated to the
recent $\sim 125$ GeV Higgs-like state observed at the ATLAS and CMS
detectors. We have examined various sets of Higgs boson masses, and used
them to test the different response to considering: i) relatively light
versus relatively heavy spectra;  ii) relatively broad versus relatively
narrow spectra; iii) a fully unconstrained choice versus a supersymmetric
MSSM-like one. In all cases the genuine 2HDM effects on $\Delta r$ exhibit
a characteristic dependence on the Higgs boson masses, which manifests as
a sort of antagonism between the following two tendencies: 1) the boosting
of the 2HDM contributions to $\Delta r$  from  a significant mass
splitting among the 2HDM Higgs states (caused by a stronger violation of
the underlying $SU(2)$ custodial symmetry); 2) the weakening of the 2HDM
contributions to $\Delta r$ when employing heavier sets of Higgs boson
masses -- obviously related to the suppression of the corresponding loop
corrections.

Let us also note that while our final results do of course depend on the
details of the Higgs mass spectra, they are essentially unresponsive to
the precise setup of Yukawa-coupling structures, i.e. they are insensitive
to type-I or type-II 2HDM. This is not surprising at first sight, as no
Higgs/fermion couplings enter the leading (one-loop) evaluation of the
gauge boson self-energies. What is less obvious is that at higher orders,
where the 2HDM Yukawa couplings could, in principle, play a non-negligible
role, they are finally impotent, too, owing to the stringent theoretical
and phenomenological constraints currently in force, in particular the
condition that the regime of $\tan\beta={\cal O}(1)$ is highly preferred.
As a result the strength of the 2HDM Yukawa couplings is essentially the
same as in the SM case within the physically allowed region. The different
impact of type-I or type-II 2HDM on the evaluation of $\Delta r$ enters
our calculation in a completely indirect way, namely only through the
specific (and significantly unequal) parameter space restrictions
affecting each one of these models.

Doubtlessly an expected but most salient feature of our analysis of
$\Delta r$ within the general 2HDM is the following: the sole addition of
a second $SU_L(2)$ Higgs boson doublet significantly relaxes the
persistent $M^{\rm th}_{\PW}-M^{\rm exp}_{\PW}$ tension existing between
theory and experiment. With only one Higgs doublet the global fits to EW
precision data favor a Higgs boson lighter than expected (even below the
LEP bounds) -- precisely due to the dependence of $\Delta r$, and so of
$M^{\rm th}_{\PW}$, on the Higgs boson mass(es). With an additional
doublet the SM-like Higgs boson can stay comfortably at a mass value
closer to the present experimental measurements since the rest of the
(heavier) Higgs bosons provide the mechanism to compensate for the
difference. All the explored scenarios feature very similar trends
concerning both the obtained results and parameter space dependence for
the quantities under scrutiny. We identify a variety of regions within the 2HDM
parameter space leading to typical variations of $\Delta r$ in the
ballpark of $\delta\,(\Delta r)\sim \mathcal{O}(10^{-3})$, which translate
into $\PW$-mass shifts of $\delta M_{\PW}^{\rm 2HDM}\sim 20 -40$
MeV. Being the current experimental error on the $\PW$-mass of $\delta
M_{\PW}^{\rm exp}=\pm 15$ MeV (and the current discrepancy with the SM
value of the same order of magnitude: $M_{\PW}^{\rm SM}-M_{\PW}^{\rm{exp}}
\simeq -20 \,\MeV$), it is pretty obvious that the 2HDM effects are
potentially very important to help the SM theoretical prediction (which is
in deficit) to better match the experimental value. This is the more true
if we recall that the future measurements may reduce the mass uncertainty
further (hopefully at the $\sim 5$ MeV level), which means that we might
be able to eventually confirm or exclude the SM value at a rather
significant confidence level.

%
%

While the above feature has been previously emphasized in the literature
within the MSSM case, we confirm here for the first time that the general
2HDM shares this virtue. This may perhaps be viewed as not too surprising
a posteriori since this fact basically depends (at one-loop) on the bulk
contribution of the interactions of the gauge bosons with the new 2HDM
Higgs states as compared to the single Higgs state in the SM. Let us note
that in the context of the MSSM the Higgs sector is highly constrained and
in general it provides a rather modest contribution to $\Delta r$.
However, there we still have the additional one-loop contributions from
the genuine supersymmetric particles (e.g. from squarks, sleptons, and
chargino-neutralinos). At one loop all these genuine SUSY effects are
gauge-like interactions (since no Yukawa-like enhancement is possible at
all at this order), and therefore they may amply compensate for the meager
output from the MSSM Higgs sector, especially if the masses of some of the
sfermions and/or chargino-neutralinos remain relatively light, i.e. at the
few hundred GeV level. It is thanks to this feature that the overall
theoretical predictions on $\Delta r$ (and $M_{\PW}$) can be more in
accordance with experiment within the MSSM than in the SM.

What we have shown in this work is that this is also the case for the
general non-supersymmetric 2HDM, albeit for a completely different reason.
Here the Higgs sector is less constrained and we have found that upon
enforcing all the known restrictions from perturbative unitarity, vacuum
stability, flavor and custodial symmetry, and direct search limits, the
Higgs sector alone is indeed
able to render corresponding $\Delta r$ and $M_W$ predictions perfectly
comparable to the total payoff of the MSSM. We believe it is useful and
remarkable to have verified this fact explicitly, as it was not obvious a
priori that the stringent current constraints on the 2HDM parameter space
would still permit such possibility.

No less remarkable in our study is the possible (and distinctive)
quantum effects that the general 2HDM can provide beyond one-loop. This
has been explored in the second part of our study, where we have dwelled
upon the role of the higher order effects induced by the Higgs boson
self-interactions on these EW precision quantities, specifically on the
bearing they may have on the parameter $\delta\rho$ and hence on the
characteristic $\Delta r$ part carrying its influence (viz.
$-\cwd\,\delta\rho/\swd$). Owing to the various constraints, the Higgs
boson self-interactions are most efficiently enhanced in the limit of large values
of the $\lambda_5$-parameter in the general 2HDM Higgs potential. In turn
the weak gauge boson self energies are responsive to them via Higgs
boson-mediated corrections only at the two-loop level
and beyond, which obviously makes the full
task of computing them rather cumbersome. We have worked out a numerical estimate of
such higher order effects by means of a Born-improved Lagrangian
approach
which tracks the leading $\lambda_5$ self-coupling effects in the
limit of large values of this parameter (within the allowed bounds).
First, we have computed the dominant radiative corrections to all
Higgs/gauge boson interactions in this limit. These corrections can
be accounted for via a gauge invariant, UV-finite subset of
(Higgs-mediated) one-loop diagrams
from which we retain the $\mathcal{O}(\aew\,\lambda_5^2)$ contributions.
Second, we have rewritten these one-loop amplitudes as static form factors
and used them to effectively \emph{dress} the bare Higgs/gauge boson
interactions.

Quantitatively speaking, these higher order quantum effects on the
$\delta\rho$ parameter may entail characteristic corrections to the 2HDM
one-loop result as large as $\sim 30\,\%$ -- for Higgs boson
self-interactions already bordering the unitarity limit. The corresponding
impact on the theoretical prediction of the W-boson mass is typically of a
few MeV.
%
%
%
This is already a noteworthy result, since these typical effects
are of the order of the present theoretical uncertainty within the SM
prediction. Let us however mention that in limiting regions of the
parameter space the higher-order 2HDM corrections could shift the $W$-mass
predictions up to $\sim 10-15$ MeV. Overall we find that the estimated
higher-order effects within the 2HDM are small enough to remain beneath
the limited precision of the current EW data, but at the same time they
could be at reach of the expected experimental precision attainable at the
LHC and future TeV-range linear colliders. Finally, to better appreciate
the potential significance of these effects, it is instructive to note
that they can be not only substantially larger than the pure bosonic
${\cal O}(\aew^2)$ two-loop contribution within the SM (which is long
known to be of order of $1$ MeV, at most) but they could even be of the
same order than the remaining set of higher order pure QCD, EW and mixed
QCD-EW effects.

The upshot of our study is that the general 2HDM possesses, under the full
set of currently known theoretical and phenomenological constraints, both
the capability to improve the bulk theoretical prediction of the
$\PW$-mass as compared to the SM, and also the possibility to provide
distinctive quantum effects that could reveal its underlying dynamics
through the role played by the 2HDM Higgs boson self-interactions. We have
demonstrated that their characteristic impact on $\Delta r$ might not be
inconspicuous at all, and could in fact be quite relevant for an accurate
theoretical determination of the EW precision quantities. Ultimately, they
could emerge above the experimental precision and signal a smoking gun of
physics beyond the SM and the MSSM\,\footnote{The potential importance of
these distinctive quantum effects on $\Delta r$, as a trademark structure
of (non-supersymmetric) extended Higgs sectors, was first suggested to the
best of our knowledge in Ref.\,\cite{LopezVal:2009qy}.}. In this
respect it is interesting to remark the apparently detected excess at the
LHC in the $H\to\gamma\gamma$ decay mode with respect to the SM
prediction. Very recently, some studies have suggested that such excess
could be explained from the general 2HDM, specifically owing to the additional
one-loop contribution from the charged Higgs bosons coupling to the
photon, see e.g.\,\cite{Altmannshofer:2012ar,Pich13} 
and references therein. 
In point of fact, this kind of possible enhancement was previously noticed in
the first detailed studies of single Higgs boson photoproduction and decay in
the context of the 2HDM ~\cite{Bernal:2009rk}.
If so, relatively light
charged Higgs bosons could be around the corner, maybe even within the LHC
reach. If confirmed, the excess in the diphoton decay at the LHC combined
with the detailed measurement of $\Delta r$ and the $W^{\pm}$ mass could
strengthen the case for the general 2HDM as a powerful source of new Higgs
boson physics beyond the SM. Of course an eventual direct detection of the
charged Higgs bosons would round off the job.
%
Bearing in mind the remarkable improvement in the experimental accuracy
which is expected for the forthcoming W-mass measurements at the LHC [viz.
$\delta M^{\rm exp}_{\PW} \sim 10\,\MeV$], the analysis of the
quantum effects on $\Delta r$ combined with the valuable information from
the direct searches may become instrumental in the near future. Now that
the LHC seems to be finally closing in on the Higgs issue, efforts on the
theory side are of foremost importance to broaden the assortment of
strategies that will be necessary to bring this research program to a most
successful completion.

\begin{acknowledgement}
The authors are very grateful to Wolfgang Hollik for enlightening
conversations on this topic and also for providing useful
references. The work of JS has been supported in part by the
research Grant PA-2010-20807; by the Consolider CPAN project; and
also by DIUE/CUR Generalitat de Catalunya under project 2009SGR502.
\end{acknowledgement}

\vspace{0.4cm}

\paragraph{Appendix}

\vspace{0.3cm}

For the sake of completeness, we provide herewith a more detailed
analytical account on selected aspects of our calculation.
All UV divergences we handle by means of conventional dimensional
regularization in the 't Hooft-Veltman scheme, setting the number of
dimensions to $d = 4-\varepsilon$. As usual, we introduce an (arbitrary)
mass scale $\mu$ in front of the loop integrals in order not to alter the
dimension of the result in $d$ dimensions with respect to $d=4$. After
renormalization (in the on-shell scheme, in our case) the results for the
physical quantities are finite in the limit $d\to 4$. Furthermore, in the
practical aspect of the calculation all one-loop structures are reduced in
terms of standard Passarino-Veltman coefficients in the conventions of
Ref.\cite{formcalc}.

%
\paragraph{\textbullet \,One loop functions at zero momentum:}
the one-loop vacuum integrals that enter the evaluation of the
parameter $\delta\rho$, which is built upon the weak gauge boson self
energies at vanishing momenta, cf. Eq.~\eqref{eq:deltarho},  read
as follows:

\begin{eqnarray}
&&\mu^{4-d}\,\int\,\frac{d^d\,q}{(2\pi)^d}\,\frac{g^{\mu\nu}}{q^2-A + i\epsilon}\,=
\frac{i\,g^{\mu\nu}}{16\,\pi^2}\,A\,\left(\Delta_\epsilon - \log(A) \right) \nonumber \\
&=&
\frac{i\,g^{\mu\nu}}{16\,\pi^2}\,A_0(A);
\end{eqnarray}
\begin{eqnarray}
&&\mu^{4-d}\,\int\,\frac{d^d\,q}{(2\pi)^d}\,\int^1_0\,dx\,
\frac{4\,k^\mu\,k^\nu}{\left[q^2 - Ax - B(1-x) + i\epsilon \right]^2} \nonumber \\ \qquad
&=& \frac{i\,g^{\mu\nu}}{16\,\pi^2}\,\left[
A\,(\Delta_\epsilon - \log\,A) + B\,(\Delta_\epsilon - \log\,B)
+ F(A,B)  \right] \nonumber \\
&=& \frac{i\,g^{\mu\nu}}{4\,\pi^2}\,\tilde{B}_{00}\,(A,B);
\end{eqnarray}

\begin{eqnarray}
&& \mu^{4-d}\,\int\,\frac{d^d\,q}{(2\pi)^d}\,\int^1_0\,dx\,
\frac{g^{\mu\nu}}{\left[q^2 - Ax - B(1-x) + i\epsilon \right]^2} \nonumber \\
&& \quad = \frac{i\,g^{\mu\nu}}{16\,\pi^2\,A}\,\left[
A\,(\Delta_\epsilon - \log\,A)
- \frac{A+B}{2} + F(A,B) \right] \nonumber \\
&=& \frac{i\,g^{\mu\nu}}{16\,\pi^2}\,\tilde{B}_{0}\,(A,B);
\end{eqnarray}

\noindent where $\Delta_\epsilon = 2/\epsilon +1 -\gamma_E +
\log(4\pi\,\mu^2)$ and the function $F(x,y)$ is defined as follows:

\begin{equation}
F(x,y) = F(y,x) = \left\{ \begin{array}{lr}
\frac{x+y}{2}- \frac{xy}{x-y}\,\log(\frac{x}{y}) & \qquad x \neq y \\
 & \\
 0 & \qquad x=y
\end{array} \right.
\label{eq:fxy}.
\end{equation}

\noindent The tilded notation for the Passarino-Veltman functions, e.g.

\begin{align}
\tilde{B}(m_1^2,m^2_2)\equiv B(0,m_1^2,m^2_2)\; ,
\end{align}

\noindent indicates that these integrals are evaluated at zero
momentum. The parameters $A,B$ can be identified with the (squared of
the) masses of the virtual particles propagating in the loop, $A\equiv
m_1^2$, $B\equiv m_2^2$.

With these expressions at hand, it is straightforward to write down a
compact analytical form for $\delta \rho_{\rm 2HDM}$ at one-loop in
the 't Hooft-Feynman gauge, starting from the definition of
Eq.~\eqref{eq:deltarho}:

\begin{eqnarray}
\delta\rho_{\rm 2HDM} &=& \frac{-\alpha}{16\,\pi\,\sw^2\,M_{\PW}^2}\,
\Bigg\{\cos^2(\beta-\alpha)\left[\,F(M^2_{\hzero},M^2_{\PHiggs^{\pm}}) \right. \nonumber \\
&& \left. -F(M^2_{\hzero},M_{\Azero}^2)\right] \nonumber \\
&&+\sin^2(\beta-\alpha)\left[\, F(M^2_{\Hzero},M^2_{\PHiggs^{\pm}}) -
F(M^2_{\Hzero},M^2_{\Azero})\right] \nonumber \\ && + F(M^2_{\Azero},M^2_{\PHiggs^{\pm}})
\nonumber \\
&&-  3\cos^2(\beta-\alpha)\left[F(M^2_{\Hzero},M_{\PW}^2) + F(M^2_{\hzero},M_Z^2) \right. \nonumber \\
&& \left. - F(M^2_{\Hzero},M_Z^2) - F(M^2_{\hzero},M_{\PW}^2)\right] \Bigg\}. \nonumber \\
\label{eq:deltarho_an}
\end{eqnarray}

From the above equation we can explicitly read off how the size of
$\delta\rho$ depends on the mass splitting between the different Higgs
bosons, as well as on the strength of the Higgs/gauge boson couplings
-- which is modulated by $\tan\beta$ and the mixing angle $\alpha$.
The first two lines of the full expression (\ref{eq:deltarho_an}) is
the part that we have denoted $\delta\rho_{\rm 2HDM}^{*}$ in
Sec.\,\ref{subsec:Deltardeltarho}, see Eq.\,\eqref{drho2HDM}. We
remark that for $M_{\Azero}\to M_{H^{\pm}}$ \emph{and}
$|\beta-\alpha|\to\pi/2$ (in which the $h^0$ field behaves SM-like) the full
$\delta\rho_{{\rm 2HDM}}\to 0$. This is the precise formulation of the
decoupling regime for the unconstrained 2HDM.

In the case of the SM the Higgs contribution to the
$\delta\rho$-parameter \eqref{eq:deltarho} is not finite if taken in
an isolated form. The complete bosonic contribution to $\Delta r$ is
of course finite and gauge invariant, and therefore unambiguous. To
define a Higgs part of it is then a bit a matter of convention. What
is important is that the complete $M_H$-dependence is exhibited
correctly and coincides in all conventions. After removing the
UV-parts which cancel against other bosonic contributions one arrives
at
\begin{eqnarray}\label{eq:deltarhoHSM2}
\delta\rho_{\rm SM}^H&=&-\frac{3\sqrt{2}\,G_F}{16\,\pi^2}\left[M_Z^2\ln{\frac{M_Z^2}{\mu^2}}-M_W^2\ln{\frac{M_W^2}{\mu^2}} \right. \nonumber\\
&& -\frac43\,\left(M_Z^2-M_W^2\right)
+\left.F(M_{\PHiggs},M_{\PZ})-F(M_{\PHiggs},M_{\PW})\right]\,. \nonumber \\
\end{eqnarray}
The explicit dependence on the scale $\mu$ is unavoidable in
quantities which are not UV-finite by themselves.  It is however
natural to set e.g. the EW scale choice $\mu=M_W$. In the limit
$M_{\PHiggs}^2\gg M_{\PW}^2$ we can see Veltman's screening theorem at
work in the SM, as there remain no $M_{\PHiggs}^2$ terms but a logarithmic
Higgs mass dependence. Indeed, in that limit the expression
(\ref{eq:deltarhoHSM2}) reduces to
\begin{equation}\label{eq:deltarhoHSM2b}
\delta\rho^{H}_{\rm SM}\simeq -\frac{3\sqrt{2}\,G_F\,M_{\PW}^2}{16\,\pi^2}\,\frac{\swd}{\cwd}\,\left\{\ln\frac{M_H^2}{M_W^2}-\frac56\right\}\,,
\end{equation}
which coincides with the result quoted in Eq.\eqref{eq:deltarhoHSM} of
Sec.\,\ref{Sec:Deltarandallthat}.

The SM Higgs boson contribution to $\delta\rho$ can also be retrieved
from the 2HDM result (\ref{eq:deltarho_an}) by selecting the $\hzero$
parts of the contributions involving the $\hzero$ and the gauge
bosons, namely in the last line of that equation. By performing the
identification $\rm \PHiggs \equiv \hzero$ and removing the
trigonometric factors we are led to
\begin{eqnarray}
\delta\rho_{\rm SM}^H &=& \frac{3\alpha}{16\,\pi\,\sw^2\,M_{\PW}^2}\,
\left[F(M^2_{\PHiggs},M_Z^2) - F(M^2_{\PHiggs},M_{\PW}^2)\right]\nonumber\\
&=& -\frac{3\sqrt{2}\,G_F\,M_{\PW}^2}{16\,\pi^2}\,\frac{\swd}{\cwd}\,\ln\frac{M_{\PHiggs}^2}{M_W^2}+...
\label{eq:deltarho_SM}
\end{eqnarray}
We see that the last expression coincides with
Eq.(\ref{eq:deltarhoHSM2}) up to finite additive parts, which of
course reflects the arbitrariness of the scale setting $\mu$. As we
said, this is not important because the full bosonic contribution to
$\Delta r$ is finite and unambiguous. The fact that we can recover the
SM result from (\ref{eq:deltarho_an}) in such a way suggests that the
expression in the first line of (\ref{eq:deltarho_SM}) should be
subtracted from (\ref{eq:deltarho_an}) in order to compute the genuine
2HDM effects on $\delta\rho$, i.e. the Higgs boson quantum effects
beyond those associated to the Higgs sector of the SM. This is in fact
the practical recipe that we follow in this paper. Finally, let us
notice that the $\delta\rho_{\rm 2HDM}^{*}$ part of
(\ref{eq:deltarho_an}), i.e. the one which is completely unrelated to
the SM Higgs contribution, is precisely the part of the full
$\delta\rho_{\rm 2HDM}$ that violates the screening theorem in the
2HDM, as it is manifest from Eq.\,\eqref{drho2HDM} of
Sec.\,\ref{subsec:Deltardeltarho}.
\paragraph{\textbullet\, 2HDM contributions to the gauge boson self-energies:}
We quote herewith their complete analytical form, in terms of the
standard Passarino-Veltman coefficients and following the conventions
of Ref.\cite{formcalc}. The self-energies are evaluated for
on-shell gauge bosons, e.g. $p^2 = M^2_V\, [V = \PW^{\pm}, \PZ^0]$, in
the way they enter the calculation of $\Delta r$.

\begin{itemize}
\item{Two Higgs-boson contributions: \jump

\begin{eqnarray}
\self_{\PW}\big{]}_{\rm 2HDM}^{\rm Higgs} &=& \frac{\alpha}{16\,\pi\,\sw^2}\,
\left[-A_0(M^2_{\hzero})-A_0(M^2_{\Hzero}) \right. \nonumber \\ && \left. -A_0(M^2_{\Azero})
-2\,A_0(M^2_{\PHiggs^{\pm}}) \right.
\nonumber \\
&&
+ 4\,\cos^2(\beta-\alpha)\,B_{00}\,(M_{\PW}^2,M^2_{\hzero},M^2_{\PHiggs^{\pm}})
\nonumber \\
&& \left. + 4\,\sin^2(\beta-\alpha)\,B_{00}\,(M_{\PW}^2,M^2_{\Hzero},M^2_{\PHiggs^{\pm}})
\right. \nonumber \\ && \left. +
4\,B_{00}\,(M_{\PW}^2,M^2_{\Azero},M^2_{\PHiggs^{\pm}}) \right].
\label{eq:self_full_higgsWW}
\end{eqnarray}
\begin{eqnarray}
\self_{\PZ}\big{]}_{\rm 2HDM}^{\rm Higgs} &=& \frac{\alpha}{16\,\pi\,\sw^2\,\cw^2}\,
\left[-A_0(M^2_{\hzero})-A_0(M^2_{\Hzero}) \right. \nonumber \\
 && \left.-A_0(M^2_{\Azero}) -2\,(\cw^2 - \sw^2)^2 \,A_0(M^2_{\PHiggs^{\pm}})\right.
\nonumber \\
&&
+ 4\,\cos^2(\beta-\alpha)\,B_{00}\,(M_Z^2,M^2_{\hzero},M^2_{\Azero})
\nonumber \\
&& \left. + 4\,\sin^2(\beta-\alpha)\,B_{00}\,(M_Z^2,M^2_{\Hzero},M^2_{\Azero})
 \right. \nonumber \\ &&+ \left.
4\,(\cw^2 - \sw^2)^2\,
B_{00}\,(M_Z^2,M^2_{\PHiggs^{\pm}},M^2_{\PHiggs^{\pm}}) \right]. \nonumber \\
\label{eq:self_full_higgsZZ}
\end{eqnarray}
}

\item{Higgs/gauge boson and Higgs/Goldstone boson contributions:
    \jump

\begin{eqnarray}
&& \self_{\PW}\big{]}_{\rm 2HDM}^{\rm Higgs/gauge} = \nonumber \\
&&  \frac{\alpha}{4\,\pi\,\sw^2}\,
\left\{
\cos^2(\beta-\alpha)\,\left[B_{00}\,(M_{\PW}^2,M^2_{\Hzero},M^2_{\PW})
\right. \right. \nonumber \\ && - \left. \left.
B_{00}\,(M_{\PW}^2,M^2_{\hzero},M^2_{\PW})\right] \right. \nonumber \\
&& \left. - \cos^2(\beta-\alpha)\,M^2_{\PW}\,\left[
B_{0}\,(M^2_{\PW},M^2_{\Hzero},M^2_{\PW}) \right. \right. \nonumber \\
&& \left. \left. -  B_{0}\,(M^2_{\PW},M^2_{\hzero},M^2_{\PW})\right]
\right\}. \nonumber \\
\label{eq:self_full_gaugeWW}
\end{eqnarray}
\begin{eqnarray}
&&\self_{\PZ}\big{]}_{\rm 2HDM}^{\rm Higgs/gauge} = \nonumber \\
&&   \frac{\alpha}{4\,\pi\,\sw^2\,\cw^2}\,
\left\{
\cos^2(\beta-\alpha)\,\left[B_{00}\,(M_{\PZ}^2,M^2_{\Hzero},M^2_{\PZ})
\right. \right. \nonumber \\
&& - \left. \left.
 B_{00}\,(M_{\PZ}^2,M^2_{\hzero},M^2_{\PZ})\right] \right. \nonumber \\ && \left. -
\cos^2(\beta-\alpha)\,M^2_{\PZ}\,\left[
B_{0}\,(M^2_{\PZ},M^2_{\Hzero},M^2_{\PZ}) \right. \right. \nonumber \\
&& \left. \left. - B_{0}\,(M^2_{\PZ},M^2_{\hzero},M^2_{\PZ})\right]
\right\}
\label{eq:self_full_gaugeZZ}.
\end{eqnarray}}
\end{itemize}

\noindent Let us notice that, in the last two expressions, we have
explicitly removed the overlap with the SM Higgs boson contribution,
to wit:

\begin{eqnarray}
&& \self_{V}\big{]}_{\rm 2HDM} \equiv \self_{V} -  \self_{V}\big{]}_{\rm SM} \propto \nonumber \\
&& \left[ \cos^2(\beta-\alpha)\,B_{00}\,(M_{V}^2,M^2_{\Hzero},M^2_{V}) \right. \nonumber \\ && +
\sin^2(\beta-\alpha)\,B_{00}\,(M_{V}^2,M^2_{\hzero},M^2_{V}) \nonumber \\
&& - M_V^2\,\cos^2(\beta-\alpha)\,B_{0}\,(M_{V}^2,M^2_{\Hzero},M^2_{V}) \nonumber \\ && -
 M_V^2\,\sin^2(\beta-\alpha)\,B_{0}\,(M_{V}^2,M^2_{\hzero},M^2_{V}) \nonumber \\
&&  -\left. B_{00}\,(M_{V}^2,M^2_{\PHiggs}, M^2_{V})\big{|}_{\rm \PHiggs \equiv \hzero} \right. \nonumber \\
&& \left. + M_V^2\,B_{0}\,(M_{V}^2,M^2_{\PHiggs},M^2_{V})\big{|}_{\rm \PHiggs \equiv \hzero} \right]. \nonumber \\
\label{eq:smsubtract}
\end{eqnarray}

\paragraph{\textbullet\, Effective Higgs/gauge boson interactions} To better
    illustrate how we build up in practice the effective Higgs/gauge
    boson couplings employed in this study, herewith we provide
    explicit analytical details for the construction of one of such
    Born-improved interactions. We carry out the calculation with the
    help of the standard algebraic packages {\sc FeynArts} and {\sc
    FormCalc} \cite{feynarts,formcalc}. Without loss
    of generality, let us take the concrete case of the Z boson
    coupling to the \CP-odd and the light \CP-even neutral Higgs
    bosons $[g_{\hzero\Azero\PZ^0}]$. A sample of the Feynman diagrams
    describing the $\mathcal{O}(\lambda^2_5)$ corrections to this
    coupling is displayed in the upper row of Fig.~\ref{fig:vert}. The
    general structure of the associated form factor
    $a_{\hzero\Azero\PZ^0}$ may be cast as:

\begin{eqnarray}
a_{\hzero\Azero\PZ^0}\Big{]}_{p^2=0} &=&
\underbrace{{\Re e}\,V_{\hzero\Azero\PZ^0}}_{\rm vertex}\, - \underbrace{\frac{1}{2} {\Re e}\,\Sigma'_{\hzero}
- \frac{1}{2} {\Re e}\,\Sigma'_{\Azero}}_{\rm wave-function} \nonumber \\ && -
\underbrace{\frac{\tan(\beta-\alpha)}{M^2_{\Hzero}}\,{\Re e}\,\retildehat_{\hzero\Hzero}\Big{]}_{p^2=0}}_{\rm mixing}
\label{eq:formfactor1}.
\end{eqnarray}

\noindent Notice that we define our form factors to be real, in order
to preserve the hermiticity of the Born-improved Lagrangians derived
from them. The different building blocks of Eq.~\eqref{eq:formfactor1}
correspond to:

a)  $V_{\hzero\Azero\PZ^0}$, the genuine vertex corrections (cf. e.g.
the first two diagrams in the upper row of Fig.~\ref{fig:vert}). For
illustration purposes, we provide its complete analytical form:

\begin{eqnarray}
&& V_{\hzero\Azero\PZ^0}\,(0)
= \frac{1}{16\,\pi^2}\,\left\{
\lambda_{\Hzero\Azero\Azero}\lambda_{\hzero\hzero\Hzero}\,
\tilde{C}_2(M^2_{\Azero},M^2_{\hzero},M^2_{\Hzero}) \right. \nonumber \\
&& \quad + \left.
\lambda_{\hzero\Azero\Azero}\lambda_{\hzero\Azero\Azero}\,
\tilde{C}_1(M^2_{\Azero},M^2_{\Azero},M^2_{\hzero})  \right. \nonumber \\
&& \quad - \lambda_{\hzero\Azero\Azero}\lambda_{\hzero\hzero\hzero}\,
\tilde{C}_2(M^2_{\Azero},M^2_{\hzero},M^2_{\hzero})  \nonumber \\
&& \quad + \tan(\beta-\alpha)\,\left[
\lambda_{\hzero\Azero\Azero}\lambda_{\hzero\hzero\Hzero}\,
\tilde{C}_1(M^2_{\Azero},M^2_{\hzero},M^2_{\Hzero}) \right. \nonumber \\
&& \quad \left. -\lambda_{\Hzero\Azero\Azero}\,\lambda_{\hzero\Azero\Azero}\,\tilde{C}_1(M^2_{\Azero},M^2_{\Azero},M^2_{\Hzero}) \right.
\nonumber  \\
&& \quad \left. \left. +\, \lambda_{\Hzero\Azero\Azero}\,\lambda_{\hzero\Hzero\Hzero}\,\tilde{C}_2(M^2_{\Azero},M^2_{\Hzero},M^2_{\Hzero})\right]
 \right\}\,. \nonumber \\ \label{eq:vert-h0A0Z0}
\end{eqnarray}

b) the wave-function corrections associated to each of the external
Higgs boson legs (including, as we single out in the last
term of Eq.~\eqref{eq:formfactor1}, the $\hzero-\Hzero$ mixing one-loop diagrams):

\begin{eqnarray}
&& \Sigma'_{\hzero}(0) = \frac{1}{16\,\pi^2}\,\left\{
\lambda^2_{\hzero\hzero\Hzero} \tilde{B}'_0(\mhzero^2,\mHzero^2) \right. \nonumber \\
&& \quad + \left. \lambda^2_{\hzero\Hpm\Hpm} \tilde{B}'_0(\mHp^2,\mHp^2) + \right. \nonumber \\
&& \quad +\left. \frac{1}{2}\left[
\lambda^2_{\hzero\Hzero\Hzero}\,\tilde{B}'_0(\mHzero^2,\mHzero^2)
+\lambda^2_{\hzero\Azero\Azero}\,\tilde{B}'_0(\mAzero^2,\mAzero^2)
\right. \right. \nonumber \\
&& \quad + \left.\left.
\lambda^2_{\hzero\hzero\hzero}\,\tilde{B}'_0(\mhzero^2,\mhzero^2) \right]
\right\}. \nonumber \\
 \label{eq:selfh0h0}
\end{eqnarray}


\begin{eqnarray}
&& \Sigma'_{\Azero}(0) = -\frac{1}{16\,\pi^2}\,\left[
\lambda^2_{\hzero\Azero\Azero} \tilde{B}'_0(\mAzero^2,\mhzero^2) \right.
\nonumber \\
&& + \left. \lambda^2_{\Hzero\Azero\Azero} \tilde{B}'_0(\mHzero^2,\mAzero^2)
\right]
 \label{eq:selfA0A0}.
\end{eqnarray}

\begin{eqnarray}
&& \retildehat_{\hzero\Hzero}(0) = \frac{1}{32\,\pi^2}\,\left\{
\lambda_{\Hzero\Azero\Azero}\,\lambda_{\hzero\Azero\Azero}\,[\tilde{B}_0(\mAzero^2,\mAzero^2)
\right. \nonumber \\
&& \left. \qquad -{\Re e}\,B_0(q^2,\mAzero^2,\mAzero^2)] \right. \nonumber \\
&&  + \lambda_{\hzero\hzero\hzero}\,\lambda_{\hzero\hzero\Hzero}\,[\tilde{B}_0(\mhzero^2,\mhzero^2)
\nonumber \\
&& \qquad-{\Re e}\,B_0(q^2,\mAzero^2,\mAzero^2)] \nonumber \\
&& + \lambda_{\Hzero\Hzero\Hzero}\,\lambda_{\hzero\Hzero\Hzero}\,[B_0(0,\mHzero^2,\mHzero^2) \nonumber \\
&&\qquad-{\Re e}\,B_0(q^2,\mAzero^2,\mAzero^2)]  \nonumber \\
&& + 2\left(\lambda_{\hzero\Hzero\Hzero}\,\lambda_{\hzero\hzero\Hzero}\,[\tilde{B}_0(\mhzero^2,\mHzero^2)
\right. \nonumber \\
&& \qquad-{\Re e}\,B_0(q^2,\mhzero^2,\mHzero^2)] \nonumber \\
&& \left. \left. + \lambda_{\hzero\PHiggs^+\PHiggs^-}\,\lambda_{\Hzero\PHiggs^+\PHiggs^-}\,[\tilde{B}_0(\mHp^2,\mHp^2)
\right. \right. \nonumber \\
 &&\left. \left. \qquad-{\Re e}\,B_0(q^2,\mHp^2,\mHp^2)]\right)
 \right\}. \nonumber \\
 \label{eq:selfh0H0}
\end{eqnarray}

\noindent In the last equation, the $\hzero-\Hzero$ mixing self-energy
$\retildehat_{\hzero\Hzero}(q^2) = \Sigma_{\hzero\Hzero}(q^2) +
\delta\,Z_{\hzero\Hzero}(q^2-M^2_{\hzero})/2 +
\delta\,Z_{\hzero\Hzero}(q^2-M^2_{\Hzero})/2
 - \delta M^2_{\hzero\Hzero}$ involves the renormalization of the
mixing angle $\alpha$, that we anchore via the relation
${\Re e}\retildehat_{\hzero\Hzero}(q^2) = 0$
 according to \cite{LopezVal:2009qy}, with the renormalization scale
chosen at the average mass $q^2 \equiv (M^2_{\hzero}+M^2_{\Hzero})/2$.
As mentioned
above, the tilded Passarino-Veltman functions are evaluated at vanishing external momentum.

Let us note in passing that, for the case of the $g_{\rm hVV}$-type
couplings, and due to he fact that just one single scalar leg is
present there, only pieces of type b) shall give rise to
$\mathcal{O}(\lambda^2_5)$ contributions. The same holds as well for
the Higgs/gauge/Goldstone boson couplings [$g_{\rm hVG}$].


\end{document}